\documentclass[12pt]{article}
\usepackage{amsmath,amssymb,hyperref,url,color,dsfont}
\usepackage{amscd,fancyhdr,makeidx}
\usepackage{feynmp}
\usepackage{mathtools}
\usepackage{hhline,footnote,caption,wrapfig}
\usepackage{fullpage}
\usepackage{verbatim}
\usepackage{amsfonts}
\usepackage{gensymb}
\usepackage{nicefrac} 
\usepackage{epsfig}
\usepackage{slashed}
\usepackage{braket}
\usepackage{tikz}
\usepackage{tikz-feynman}
\usepackage{float}

\newcommand{\half}{\frac{1}{2}}
\newcommand{\vk}{{\vec{k}}}
\newcommand{\beq}{\begin{equation}}
\newcommand{\eeq}{\end{equation}}
\newcommand{\be}{\begin{equation}}
\newcommand{\ee}{\end{equation}}
\newcommand{\bdm}{\begin{displaymath}}
\newcommand{\edm}{\end{displaymath}}
\newcommand{\bea}{\begin{eqnarray}}
\newcommand{\eea}{\end{eqnarray}}
\newcommand{\nn}{}
\newcommand{\del}{\partial}
\newcommand{\shalf}{{\textstyle\frac12}}
\newcommand\sfrac[2]{{\textstyle\frac{#1}{#2}}}
\newcommand{\vx}{\vec{x}}

\newcommand{\vp}{\vec{p}}
\newcommand{\vcp}{\vec{p}}
\newcommand{\vpp}{{\vec p}\,'}

\newcommand{\cL}{{\cal L}}

\newcommand{\wV}{\widetilde V}

\newcommand{\cM}{{\cal M}}

\newcommand{\cS}{{\cal S}}

\newcommand{\cT}{{\cal T}}
\newcommand{\cbar}{{\bar c}}

\newcommand{\ubar}{{\overline u}}

\newcommand{\psibar}{{\bar\psi}}
\newcommand{\tA}{{\tilde A}}

\newcommand{\exercise}[1]{\bigskip\medskip
\hrule height 0.5pt\kern 2pt {\em Exercise:} #1\\[-3mm] \hrule height 0.5pt \bigskip}

\def\d{\displaystyle}
\def\da{\dagger}
\newcommand{\sint}{{\textstyle\int}}
\newcommand\dsl{{\del\kern-6.5pt/}}
\newcommand\Dsl{{D\kern-6.5pt/}}
\def\psl{{p\kern-5pt/}}
\def\ksl{{k\!\!\!/}}
\def\epsl{{\epsilon\kern-4pt/}}

\def\lcol{{:}}
\def\rcol{{:}}
\def\tr{{\rm \,tr\,}}

\def\gsim{\:\raisebox{-0.5ex}{$\stackrel{\textstyle>}{\sim}$}\:}

\newcommand\bpm{\begin{pmatrix}}
\newcommand\epm{\end{pmatrix}}
\newcommand\bsp{\begin{split}}
\def\esp{\end{split}}
\newcommand{\eref}[1]{Eq.(\ref{#1})}
\newcommand{\Comment}[1]{}

\author{{\bf Sunil Mukhi}\\[2mm] 
\em Indian Institute of Science Education and Research,\\
\em Homi Bhabha Rd, Pashan\\
\em Pune 411 008, India\\[6mm] 
\large An introductory review, based on a series of lectures\\ delivered at the XXXI SERC School,\\[1mm]
Kalyani University, 9--18 January 2017.\\[6mm]
Lecture Notes by {\bf Shibasis Roy} and {\bf Anirban Karan}
}
\title{{\Large\bf Renormalisation in Quantum Field Theory}}

\date{}

\makeindex

\numberwithin{equation}{section}

\begin{document}

\clearpage
\maketitle

\thispagestyle{empty}

\newpage

\setcounter{page}{1}

\tableofcontents

\newpage

\advance\baselineskip by 3pt
\advance\parskip by 5pt


\section*{Introduction and Acknowledgements}
\label{intro}

These notes are based on a series of 9 lectures delivered at the 31st SERC School in Theoretical High-Energy Physics, held at Kalyani University, West Bengal, India in January 2017. At the outset I would like to mention that the SERC School series in Theoretical High Energy Physics is among the longest uninterrupted advanced physics school series in the world, having been initiated in 1985 and held annually (with a few minor glitches in schedule) since then. Since 1996 it actually split into a Preparatory and an Advanced School, each lasting around 3 weeks and held separately. It is aimed primarily at Indian students working for a Ph.D. in particle physics, and postdoctoral fellows in the field. The IISER system, in which I work, also has undergraduates who are active in research and some of them too have participated in SERC schools. 

One of the most remarkable features of these Schools is that they are aimed at both the hep-ph and hep-th communities, whose focus is respectively on phenomenologically motivated and formal theoretical approaches to particle physics. That the Schools have continued in this spirit despite all kinds of frictions between the two sub-fields (all over the world) is a credit to the Indian particle physics community. In particular, they have gone some way in removing the ignorance and misunderstandings that divide the two sub-communities, which ought to count as a major service to Physics by itself. It must also be mentioned that a sizable fraction of particle physicists currently holding faculty positions in various Indian universities and institutes are alumni of one or more SERC schools.

The Preparatory School provides basics that are taught only in some universities in India at the Master's level, such as perturbative Quantum Field Theory, the Standard Model and advanced Mathematical Methods. Some students who do not receive this training in their home institution, or feel the need to deepen their understanding, attend this school. A larger number participate in the Advanced School, which has lecture courses that either describe a topic of current research interest, or (as with these notes) describe established techniques that are important in contemporary research. I am proud to have lectured and tutored at numerous SERC schools, starting with the third one held at Shantiniketan in 1987. Through some periods I was also associated with the organisation of this activity. This brought with it the joy of interacting with the Department of Science and Technology, the funding agency for this activity which has been extremely generous and supportive. Time erases the few negative memories such as budget cuts, bureaucratic rules and the building catching fire once while we were in a meeting.

In retrospect, teaching at SERC schools has been one of the greatest joys of my life. This is true even without discussing the merits of the food, which both at Shantiniketan in 1987 and at Kalyani in 2017 was of a sublime quality, creating the perfect spiritual mood in which to design lectures. I will skip over side-effects such as weight gain and the urge to take a nap in the middle of one's own lecture. If any of the current organisers chance to read these notes, let me sneakily embed here a request to be invited to an SERC school again.

I am particularly thankful to Shamik Banerjee and Santosh Kumar Rai, the two tutors for the course who spent many hours helping the students solve exercises and understand the lectures better, and Shibasis Roy and Anirban Karan who, as students at the School, duly took notes of my lectures and typed up a preliminary version. I have edited their version thoroughly and expanded it in parts, so any remaining errors or misunderstandings are my responsibility alone. The school was very well-organised, for which credit goes to Jyoti Prasad Saha of Kalyani University together with many local volunteers, and Debashis Ghoshal as the Chair of the organising committee. I would also like to thank Rohini Godbole whom I consulted on several fine points of the exposition.

Coming now to the technical part, the purpose of the present notes is to introduce Ph.D. students working in high energy physics, who have already had a first course in perturbative Quantum Field Theory, to perturbative renormalisation. I am aware from experience that most students who use renormalisation in their research (and dare I say, many faculty members too) think of it as a mere prescription. Which in a sense it is, but it has genuine physical motivations and I have tried to highlight these. I have consistently emphasised that the extraordinary predictive power of quantum field theory comes into play only after we calibrate the parameters of the theory using a small set of experiments. The theory then makes a variety of predictions which can be tested in a larger class of experiments. On keeping this firmly in mind, one finds the renormalisation procedure much less arbitrary. I sensed that the students at the School seemed particularly gratified to appreciate this point. 

In a relatively short course like this, it became necessary to pick and choose topics and approaches. These notes start with a rather lengthy review of quantum field theory, not all of which actually featured in the lectures, mainly to set the notation and establish basic concepts. I expect a large number of readers of these notes to skip directly to Chapter \ref{qftoneloop}, where I start to work out the perturbative expansion. For explicit calculations, I restrict myself to one-loop perturbation theory. Also I work mostly with dimensional regularisation, though only after phrasing the basic ideas in the language of a momentum cutoff. While the discussion of loop diagrams starts out with scalar field theory, we quickly switch to gauge theory and thereafter carry out most of the work in the context of quantum electrodynamics. Later in the notes I move on to non-Abelian gauge theory where I have necessarily had to be a little more brief because of the increased technical complexity. Parts of these chapters will, moreover, only be accessible to students who have studied path integral quantisation, for which I am unfortunately unable to provide a self-contained introduction here. Nonetheless, the reader will find in these notes a fairly detailed guide to the calculation of the one-loop $\beta$-function of Yang-Mills theory. This is one of the most beautiful examples of the power of quantum field theory (for anyone who needs convincing): a straightforward computation in renormalised quantum field theory predicts a striking (and for its time, novel) type of physical behaviour that is visible in experiments, namely asymptotic freedom. The notes conclude with a short chapter on the renormalisation of Higgsed, or ``spontaneously broken'', gauge theory. This brief chapter should facilitate the student's deeper study of the subject from a textbook or by reading the original papers, which are listed at the end.

I make no claim to originality in these notes, but I have tried to present all the results in a logical sequence that is sometimes (even to me) not visible in the various references I have used. I hope the end result will be useful to some.

Among major omissions: Ward and Ward-Takahashi identities are used in the Abelian context, but their fundamental role in proving renormalisability in the non-Abelian case is glossed over. Throughout the text I have not been able to provide detailed references to the original literature, preferring to refer the student to two excellent textbooks: ``Introduction to Quantum Field Theory'' by Peskin and Schroeder \cite{Peskin:1995ev}, and ``Quantum Field Theory and the Standard Model'' by Matthew Schwartz \cite{Schwartz:2013pla}. But the biggest omission is that there is no mention of Ken Wilson and his transformative ideas about renormalisation, and in this respect the notes are shamelessly old-fashioned. This is not because I would dare to deny the greatness of Wilson's contribution but simply because my lectures were mostly directed to practical particle physicists working in areas like Beyond Standard Model physics, Precision Tests, QCD and so on. And also, because the Wilsonian approach merits its own series of lectures.

\section{Review: Tree-level quantum field theory}
\label{sfttree}

\subsection{$\cS$-matrix and cross-sections}
\label{smatcros}

One of the principal goals of quantum field theory is to compute scattering cross-sections for elementary particles. These are numerical quantities that can be compared with the results of experiments. The spectacular agreement between calculations in quantum field theory and experiments at particle colliders tells us that we understand something profound and eternal about nature. This is all the more true given that quantum field theories have limited numbers of parameters. For example, all elementary particle processes involving electrodynamics have to be explained in terms of the electric charge of the particles involved and their masses. No more arbitrary parameters can be introduced. Indeed, at very high energies one can even ignore particle masses. It is truly amazing that one can then explain the results of multiple experiments involving very different processes from an essentially simple theoretical starting point.

In scattering experiments, particles come in from far away where they originated in the far past, enter an interaction region where they collide with each other, and then again move apart in the far future. In the past and future, the particles can be taken to be non-interacting. The amplitude for the scattering process is the overlap of the incoming state and the outgoing state. This overlap defines a matrix called the $\cS$-matrix.

A typical experiment involves the collision of two particles, since it is hard to make three or more particles collide at the same place. The number of outgoing particles, $n$, is arbitrary. We refer to the process as $2\to n$ scattering. The ``in'' state of interest is denoted $|\vec p_1, \vec p_2\rangle_{\rm in}$ and is associated to two particles of 4-momenta
$(E_1, \vec p_1)$ and $(E_2,\vec p_2)$.  The ``out'' state is similarly labeled $_{\rm out}\langle p'_1, \cdots, p'_n|$ and corresponds to $n$ particles of 4-momenta $(E'_i,\vpp_i)$.
In an interacting theory it is not easy to rigorously define these states, but we will present a pragmatic definition and then develop approximate methods to study the $\cS$-matrix.

In the above notation the $\cS$-matrix is defined as the overlap, or inner product, of the
``in'' state with the ``out'' state:
\be
\cS_{(\vpp_1,\cdots,\vpp_n|\vp_1,\vp_2)}=
{}_{\rm out}\langle \vpp_1,\cdots,\vpp_n|\vec p_1,\vec p_2\rangle_{\rm in}
\ee
Clearly, all possible ``in'' states of a given quantum field theory form a complete basis of quantum states of that theory. The same must be true of the ``out'' states. So physically speaking, all the $\cS$-matrix does is to convert one basis to another. One can define an operator (also called $\cS$) such that its matrix elements give us the $\cS$-matrix:
\be
\cS_{(\vpp_1,\cdots,\vpp_n|\vp_1,\vp_2)}=
\langle \vpp_1,\cdots,\vpp_n|\cS|\vec p_1,\vec p_2\rangle
\ee
where the states on the RHS are defined at a common time rather than in the far past and far future. The above expression gives us the probability amplitude for given out-states to be created for a given in-state. We will see very shortly that this can be used to find the probability and scattering cross-section for any scattering process in the theory. 

The fact that it converts one complete basis into another implies a very important property of the $\cS$-matrix, namely that it is unitary:
\be
\cS^\dagger \cS = 1
\ee

Certain contributions to the $\cS$-matrix are trivial. For example in the case of elastic $2\to 2$ scattering (``elastic''
 means the initial particle species are identical to the final ones), it is possible for the incoming particles to simply pass through the interaction region and become outgoing particles without undergoing any interaction. Such ``disconnected processes'' are not of physical interest.
Therefore we define a new matrix called ${\cal T}$ to be the nontrivial part of $\cS$. Schematically we write:
\be
\cS={\bf 1}+i{\cal T}
\ee
where ${\bf 1}$ represents the trivial contribution from disconnected processes while ${\cal T}$ is the transition amplitude between distinct initial and final states.

Because of momentum conservation, we can be sure that ${\cal T}$ will contain a $\delta$-function equating incoming and outgoing momenta. Therefore we pull this factor out explicitly and write:
\be
{\cal T} = (2\pi)^4\delta^4(p_1+p_2-\sum_{i=1}^n p'_i)\, {\cal M}
\label{matel}
\ee
This defines the so-called ``matrix element'' ${\cal M}$. Note that this relation involves a certain amount of short-hand: the object $\cT$ on the LHS should really be labelled $\cT_{(\vpp_1,\cdots,\vpp_n|\vp_1,\vp_2)}$ and correspondingly for $\cM$, so as to keep track of the incoming and outgoing 3-momenta.

The probability $P_T$ of a transition between initial and final states is the square of the transition amplitude. Here we must take care of several points. One is that because ${\cal T}$ contains a momentum-conserving $\delta$-function, its square will contain a factor $(2\pi)^4\delta^4(0)$ which is divergent. This factor is interpreted as the volume of space-time and we drop it to get the transition probability per unit volume per unit time. The second point is that due to the standard normalisation of one-particle states, we must divide the square of ${\cal T}$ by $\prod_{i=1}^2 (2E_i)\prod_{j=1}^N (2E'_j)$. The last point is that 
since no experiment measures momenta with arbitrary precision, we must allow for some resolution in the final-state momenta. Therefore we include a factor $d^3\vpp_i/(2\pi)^3$ for each outgoing particle and label the corresponding object as the differential transition probability $dP_T$. Thus we find that:
\be
dP_T = \frac{(2\pi)^4\delta^4(p_1+p_2-\sum_{i=1}^n p'_i)\,|{\cal M}|^2}{\prod_{i=1}^2 (2E_i)\prod_{j=1}^n (2E'_j)}\prod_{i=1}^n \frac{d^3\vpp_i}{(2\pi)^3}
\label{peetee}
\ee

\exercise{Show that the differential transition probability has dimensions of area.}

The physical observable we want to define is called the scattering cross-section. This is a familiar concept in classical physics. Suppose we fire a particle at a hard classical target such that the line of incidence passes the target at a distance $b$. We call this the {\em impact parameter} of the incident particle. Let the target be a hard sphere of radius $b_0$. If $b<b_0$ then the target is hit, while if $b>b_0$ then it is missed. The area that the target presents to the incident particle, known as the {\em cross-section}, is $\sigma=\pi b_0^2$. 

We would like to define, by analogy, a set of corresponding concepts for elementary particles. In this context there is really no such thing as a hard disc or hard sphere -- all interactions take place via fundamental forces, which in a classical picture cause the incident particle to smoothly scatter off the target. In a quantum-mechanical picture one has to think of scattering of wave-packets that are well-localised in position and momentum space. Either way, the radius $b_0$ of the target can be defined by saying that for impact parameter $b>b_0$ there is essentially no scattering. For long-range interactions like the Coulomb interaction $b_0$ can be infinite, so what we would call the total cross-section will also be infinite. It therefore makes sense to limit our attention to outgoing particles with definite energies and emerging in definite directions. The corresponding object is the {\em differential cross-section}, and corresponds closely to what detectors actually measure. It carries more information than the total cross-section, which in turn can be found from the differential one by integrating it over angles and energies (whenever it is finite).

We now present a somewhat simplified derivation of the relation between the cross-section and the matrix element ${\cal M}$. More rigorous definitions, involving the construction of wave-packets and putting the system in a box of finite volume, can be found in textbooks on quantum field theory. 

We start by working in the laboratory frame. Then we have a beam of particles incident on a stationary target. The cross-section must be proportional to the differential transition probability, which already has dimensions of area. However, it should also be inversely proportional to the velocity of the beam (a slower beam has a greater probability of interaction). Therefore we divide $dP_T$ by $|{\vec v}|$ (in general units, by $|{\vec v}|/c$) to get:
\be
d\sigma = \frac{1}{|\vec v|}\,dP_T
\ee
From \eref{peetee} we find:
\be
d\sigma = \frac{|{\cal M}|^2}{|{\vec v}|}\,\frac{1}{4E_1E_2}\left[\prod_{i=1}^n\frac{d^3p_i}{(2\pi)^3 2E'_i}\right]\,(2\pi)^4\delta^4(p_1+p_2-\sum_{i=1}^n p'_i)
\label{colinform}
\ee
If now we integrate over all final-state 3-momenta $\vpp_i$ we get the total cross-section for the given process. If we also sum over all numbers $n$ of particles in the final state, we get the total {\em inclusive} cross-section. In practice we decide whether to integrate over final-state momenta, and then whether to sum over the number of particles, depending on the experiment with which we wish to compare.

The above formula actually holds not just in the laboratory frame, but in any frame where the 3-momenta of the two incoming particles are parallel. Such frames are obtained by boosting the system along the incoming beam direction. In these frames $\vec v$ must be replaced by ${\vec v}_{12}$, the relative velocity of the two incoming beams. Note that the centre-of-mass (CM) frame is an example of such a frame.

There is an important correction to this formula if the final state contains identical particles. Suppose there is a pair of identical particles that are measured to have 4-momenta $p'_1,p'_2$. Because the two particles cannot be distinguished, it is physically meaningless to ask which of them has momentum $p'_1$ and which has $p'_2$. But our derivation treated the two possibilities as distinct and therefore overcounted by a factor of 2. We compensate for this by including a symmetry factor of $\half$. More generally, we insert a factor $\frac{1}{j!}$  for every set of $j$ identical particles in the final state. We will not insert this factor explicitly into our formulae, but it is essential to take care of it whenever appropriate.

\eref{colinform}, valid in both laboratory and centre-of-mass frames, is sufficient for most purposes since in particle physics we usually work in one of these two frames. However, it is more satisfying to extend this to a Lorentz-invariant formula. The factors $d^3\vpp_i/(2\pi)^3 2E'_i$ are already Lorentz-invariant, so the non-invariance comes only from the denominator factor $E_1E_2 v_{12}$. We claim the general formula for the differential cross-section, valid for any incoming 4-momenta $p_1,p_2$, is: 
\be
d\sigma = |{\cal M}|^2 \frac{1}{4\sqrt{(p_1\cdot p_2)^2-(m_1m_2)^2}}\left[\prod_{i=1}^n 
\frac{d^3\vpp_i}{(2\pi)^3 2E'_i}\right] (2\pi)^4 \delta^4(p_1^\mu + p_2^\mu - \sum_{i=1}^n p'^\mu_i)
\label{genform}
\ee
It is easy to show that:
\be
(p_1\cdot p_2)^2-(m_1m_2)^2=(E_1E_2|v_{12}|)^2 + \Delta
\label{non-collin}
\ee
where $\Delta = (\vp_1\cdot \vp_2)^2-\vp_1^{\,2}\vp_2^{\,2}$ is a measure of the non-collinearity of the incoming beams. For collinear beams $\Delta=0$ and the square-root in the denominator of the above formula reduces to $E_1E_2|{\vec v}_{12}|$ as desired. This proves our claim.

\exercise{Verify \eref{non-collin} above.}

\subsection{$2\to 2$ scattering}
\label{twototwo}

Let us now consider the special case of two particles scattering into two particles. A certain parametrisation of the Lorentz invariant combinations of 4-momenta turns out to be very useful in understanding this situation. Given incoming momenta $p_1,p_2$ and outgoing momenta $p'_1,p'_2$ as in the previous sub-sections, we define the scalar variables:
\be
s=(p_1+p_2)^2,\quad t=(p_1-p'_1)^2,\quad u=(p_1-p'_2)^2
\label{manddef}
\ee
These are known as Mandelstam variables.
If the incoming particles have masses $m_1,m_2$ and the outgoing ones have masses $m_1',m_2'$ then we easily see that:
\be
s=m_1^2+m_2^2+2p_1\cdot p_2,\quad
t=m_1^2+{m_1'}^2-2p_1\cdot p'_1,\quad
u=m_1^2+{m_2'}^2-2p_1\cdot p'_2
\label{stuexp}
\ee
Adding up the three invariants and using momentum conservation: $p_1+p_2=p'_1+p'_2$ we see that:
\be
s+t+u=\sum_{i=1}^4 m_i^2
\ee
Notice that $t$ and $u$ can be interchanged if we exchange our labelling of outgoing momenta. The scattering angle is defined as the angle between the incoming particle and whichever outgoing particle has been conventionally chosen to have momentum $p_1'$. Hence the variable $t$ encodes the scattering angle:
\be
t= m_1^2 + {m'_1}^2 -2E_1E'_1+2|\vp_1||\vpp_1|\cos\theta
\label{tscat}
\ee
Also note that in the CM frame, $s=E_{\rm CM}^2$.

Finally, we note that the Lorentz-invariant kinematical factor on the LHS of \eref{non-collin} has an elegant expression in terms of the Mandelstam variable $s$. Define a function of three variables:
\be
\lambda(a,b,c)\equiv a^2+b^2+c^2-2ab-2bc-2ac
\label{lamfn}
\ee
It is an easy exercise to show that:
\be
\lambda(s,m_1^2,m_2^2) = 4\left[(p_1\cdot p_2)^2-(m_1m_2)^2\right]
\label{lamdenom}
\ee
It follows that the factor:
\be
\frac{1}{4\sqrt{(p_1\cdot p_2)^2-(m_1 m_2)^2}}
\ee
that appears in \eref{genform} can be replaced by:
\be
\frac{1}{2\sqrt{\lambda(s,m_1^2,m_2^2)}}
\ee
The $\lambda$-function defined above occurs frequently in the study of $2\to 2$ particle scattering.

\subsection{Perturbative scalar field theory}
\label{PertQFT}

Now we turn to the main purpose of this course, to learn how to compute the matrix element $\cM$ for various quantum field theories in perturbation theory, and deal with the puzzling infinities that arise when we undertake this computation.

We start by studying a real scalar field, denoted $\phi(x)$, with a quartic self-interaction. We will work in units in which $\hbar=1$ and $c=1$. The Lagrangian density is:
\be
{\cal L} = {1\over2} \partial_\mu \phi \partial^\mu \phi - {1\over2} 
m^2\phi^2 - \frac{\lambda}{4!}\phi^4
\label{Lscalar}
\ee
The corresponding equation of motion is:
\be
(\partial_\mu \partial^\mu + m^2)\phi + {\lambda \over 3!} \phi^3 = 0
\ee

To study the resulting quantum theory, we are going to assume that the
effect of interactions is small and answers can be expressed as a
power series in $\lambda$, with $\lambda = 0$ giving the ``free'' theory. This is an extremely strong assumption and in some situations very unreasonable!

\exercise{Think about why this assumption might be reasonable or unreasonable. Consider simple quantum-mechanical models.}

The free theory is quantised in the standard way via canonical commutators:
\be
\bsp
&[\phi(t,\vec x\,),\pi(t,\vec x\,')] = i\hbar\, \delta^3 (\vec x - \vec x\,')\\
&[\phi(t,\vec x\,), \phi(t,\vec x\,')] = [\pi(t,\vec x\,),\pi(t,\vec x\,')] = 0
\end{split}
\ee
Note that the field commutators are taken at equal times.

The most important thing we need to know about the free theory is the Feynman propagator:
\be
D_F(x_1-x_2)=\langle 0|T\big(\phi(\vx_1,t_1)\phi(\vx_2,t_2)\big)|0\rangle
\ee
where the $T$ symbol represents time-ordering. An integral representation of the Feynman propagator is given by:
\be
D_F(x_1-x_2)=\int\frac{d^4k}{(2\pi)^4}\frac{i}{k^2-m^2+i\epsilon}e^{-ik\cdot(x_1-x_2)}
\ee

Now let us turn to the interacting theory. Its Hamiltonian can be written:
\be
\bsp
H &= \int d^3 x \left({1\over2} \pi^2 + {1\over2} (\vec\nabla \phi)^2 + 
{1\over2} m^2 \phi^2 + {\lambda \over 4!} \phi^4\right) \\ 
&= H_0 + H_{\rm int}
\end{split}
\ee
where $H_{\rm int} = \d{\lambda \over 4!} \int d^3x\, \phi(x)^4$.

Because of the nonlinearities in $H_{\rm int}$, the state $\Omega$ of minimum energy or the ``true vacuum''
\be
(H_0 + H_{\rm int})|\Omega\rangle = E_{\rm min}|\Omega\rangle
\ee
is very complicated and cannot be calculated
exactly. However, in terms of this state we can defined a quantity that we would like to
compute: the probability for a particle to propagate from one point to
another at a later time in the true vacuum of the interacting theory:
\be
\langle \Omega|T(\phi(x) \phi(y))|\Omega\rangle
\label{intcorr}
\ee
The approximation scheme we will develop, perturbation theory, gives an expression for this quantity (suitably normalised) in terms of only free fields and the free-field vacuum. Of course the expression will depend on the interaction term in the Hamiltonian. 

Intuitively, the procedure is as follows. Consider a field configuration at an initial time $t=0$, call it $\phi(0,\vx)$. This evolves with time in the usual way, by sandwiching it between $e^{iHt}$ and $e^{-iHt}$. Now $H$ is the sum of a free and an interacting part. The free part $H_0$ evolves $\phi(0,\vx)$ as a free field. By suitable manipulations, we can express the propagator \eref{intcorr} of the interacting theory in terms of the time-evolved free field along with corrections due to the interaction Hamiltonian. There are some important subtleties because the free and interaction Hamiltonians do not commute. The detailed derivation can be found in textbooks. Here we simply present the result and explain its various features.

The answer, which forms the basis of perturbative quantum field theory, is:
\be
\frac{\langle \Omega|T\left(\phi(x)
\phi(y)\right)|\Omega\rangle}{\langle
\Omega|\Omega\rangle} =
\lim_{T \rightarrow \infty(1-i\epsilon)} {\langle 0|T\left(\phi_0(x) \phi_0(y) 
e^{-i\int^{T}_{-T} dt' H_I(t')}\right)|0\rangle \over \langle 0|T\left(
e^{-i\int_{-T}^T dt' H_I(t')}\right)|0\rangle}
\label{inttwopt}
\ee
The limit on the RHS just corresponds to taking the time $T$ to infinity, but to ensure good convergence we have assigned it a small negative imaginary part. Here, $H_I(t)$ is just the same function of the free field that $H_{int}$ was of the interacting field. Thus, in our example of a scalar field with a quartic interaction, we have:
\be
H_I(t) = \frac{\lambda}{4!}\int d^3x\, \phi_0(x)^4
\ee

Although \eref{inttwopt} looks a little intimidating at first sight, it is physically very accessible. We learn that the propagation amplitude for an interacting particle is equal to the propagation amplitude for a free particle plus a series of corrections obtained by expanding the exponential in a power series. These corrections are induced by the interaction Hamiltonian, and the number of powers of this Hamiltonian determines the order of the correction. The denominator factor just normalises the resulting expression appropriately. 

Since the Hamiltonian $H$ is the spatial integral of the Hamiltonian density ${\cal H}$, and since $T\to\infty$, we can roughly think of the exponent in the above formula as:
\be
-i\int^{T}_{-T} dt' H_I(t')\simeq -i\int d^4 x\,{\cal H}_I
\ee
where ${\cal H}_I$ is the interaction Hamiltonian density. Thus the physical picture is that, up to numerical factors, the $n$th order correction to particle propagation due to interactions involves the free particle ``colliding'' with the interaction Hamiltonian density ${\cal H}_I$ $n$ times. The location of this ``collision'' (the space-time argument of the Hamiltonian density) can be anywhere, so we integrate over all space and time.

As will be seen from the Appendix, the formula in \eref{inttwopt} generalises to multi-point amplitudes (also referred to as Green's functions, correlation functions, or $n$-point functions) of the interacting theory in a straightforward way. Thus:
\be
\frac{\langle \Omega|T\left(\phi(x_1)
\cdots \phi(x_n)\right)|\Omega\rangle}{\langle
\Omega|\Omega\rangle} =
\lim_{T \rightarrow \infty(1-i\epsilon)} {\langle 0|T\left(\phi_0(x_1)
\cdots \phi_0(x_n) 
e^{-i\int^{T}_{-T} dt' H_I(t')}\right)|0\rangle \over \langle 0|T\left(
e^{-i\int_{-T}^T dt' H_I(t')}\right)|0\rangle}
\label{intnpt}
\ee
In the next section we will see how $n$-point functions can be systematically evaluated order-by-order in powers of the interaction Hamiltonian. As we have explained at the outset, these are the scattering amplitudes for interacting particles, which can be converted by a series of manipulations into experimentally measurable quantities.

\subsection{Wick's theorem}
\label{wicksth}

The general result \eref{intnpt} expresses all $n$-point functions of the interacting theory in terms of the free theory. By expanding the exponential in powers, the right hand side becomes a sum of $m$-point functions of a free theory where $m\ge n$, the additional powers of $\phi_0$ coming from $H_I$. The fields in these $m$-point functions will not all be located at distinct space-time points. The $\phi_0$'s that come from each $H_I$ insertion are at coincident points, which moreover are integrated over. So in the end, the answer will only depend on the spacetime points $x_1,x_2,\cdots,x_n$ as expected.

We therefore need two distinct procedures: one to evaluate general $m$-point functions for all possible $m$ in free field theory, the other to make some sets of points coincident and integrate over them. Together these will constitute the Feynman diagram technique. For those who already have a passing familiarity with this technique, the first procedure leads to propagators, while the second leads to vertices and (after going to Fourier space) momentum integrals. In this section we focus on the first procedure, namely a rule to express:
\be
\langle 0|T\left(\phi_0(x_1) \cdots \phi_0(x_m)\right)|0\rangle 
\ee
entirely in terms of products of the Feynman propagator $D_F(x-y)$ of the free theory:
\be
D_F(x-y)=\int \frac{d^4k}{(2\pi)^4}\frac{i}{k^2-m^2+i\epsilon}e^{-ik\cdot(x-y)}
\label{feynpropint}
\ee
The rule is called ``Wick's theorem''. Since we will work entirely with free field theory in the rest of this section, we will drop the ``0'' index on $\phi$ but the reader should keep in mind that these results are specific to free fields. 

Instead of embarking on a derivation of Wick's theorem, which is quite technical, let us offer a physical explanation in terms of particles. A free field is associated to a free particle, and this cannot do anything but propagate! So the expectation value of a product of free fields, which is like a scattering amplitude for many free particles, must be made up entirely of free propagation of each particle from one point to another. Thus the free $n$-point function decomposes as a product of 2-point functions in which one particle propagates from $x_1$ to $x_2$, another from $x_3$ to $x_4$ etc, plus all permutations thereof. From this viewpoint we can guess an answer for the 4-point function:
\be
\begin{split}
\langle 0| T(\phi_1\phi_2\phi_3\phi_4)|0\rangle = &\langle 0| T(\phi_1\phi_2)|0\rangle\langle T(\phi_3\phi_4)|0\rangle  \\
&+\langle 0| T(\phi_1\phi_3)|0\rangle\langle T(\phi_2\phi_4)|0\rangle \\
&+\langle 0| T(\phi_1\phi_4)|0\rangle\langle T(\phi_2\phi_3)|0\rangle
\end{split}
\ee
where we have simplified our notation by writing $\phi(x_i)$ as $\phi_i$.

The above guess turns out to be precisely correct. Moreover, the objects on the RHS are just Feynman propagators, which we have already evaluated. So we can write the above equation as:
\be
\begin{split}
\langle 0| T(\phi_1\phi_2\phi_3\phi_4)|0\rangle = D_F(x_{12})D_F(x_{34})+
D_F(x_{13})D_F(x_{24})+
D_F(x_{14})D_F(x_{23})
\label{fourptwick}
\end{split}
\ee
where we have introduced another piece of shorthand: $x_{ij}=x_i-x_j$. This provides a precise technique to evaluate 4-point functions in free field theory. 

The above reasoning is easily extended to all even-point functions, which reduce to products of two-point functions in
all possible distinct ways. Thus one has:
\be
\begin{split}
\langle 0| T(\phi_1\phi_2\cdots\phi_m)|0\rangle = \sum D_F(x_{i_1i_2})D_F(x_{i_3i_4})\cdots
D_F(x_{i_{m-1}i_m})
\end{split}
\label{nptwick}
\ee
where the $i_1,i_2,\cdots,i_m$ are a permutation of $1,2,\cdots,m$ and the sum is over all distinct ways of dividing the numbers $1,2,\cdots,m$ into pairs.

This formula is best remembered in words. An $m$-point function in free field theory is a sum over all possible ways of breaking up the numbers from 1 to $m$ into pairs. For each such way, the contribution is the product of Feynman propagators for every pair. We sometimes refer to the process of pairing off points as a ``contraction'' of the fields at those points. Every contraction is represented by a Feynman propagator. The total number of ways of contracting $m$ points pairwise is easily seen to be
$(m-1)!!=(m-1)(m-3)\cdots$. For the 4-point function we have 3
independent contractions, as above, while for the 6-point function we
get 15 independent contractions. Odd $m$-point functions vanish, since in that case there is no way to break up the $m$ points into pairs.

\subsection{Interactions via Wick's theorem}

Now we turn to the purpose for which we introduced Wick's theorem, namely the study of $n$-point functions in the interacting theory. Here we must reintroduce the subscript 0 on free fields. As a first example, consider the leading correction to the propagator in the theory with $H_{\rm int} = \d{\lambda \over 4!} \int d^3x\, \phi_0^4$. We find:
\be
\begin{split}
&\langle \Omega|T(\phi(x_1) \phi(x_2))|\Omega\rangle
=\\
& \qquad\qquad \langle 0|T(\phi_0(x_1) \phi_0(x_2))|0\rangle -  i\,{\lambda \over 4!} 
\int d^4 y\,\langle 0|T(\phi_0(x_1) \phi_0(x_2) \phi_0^4(y))|0\rangle + {\cal O}(\lambda^2)
\end{split}
\ee
The  second term on the RHS is a 6-point function of free fields, 
which we now know how to
compute using Wick's theorem. The new feature is that four of the
fields are at the {\em same} space-time point, and there is an
integral over all possible locations of that point.

\begin{figure}[H]
\begin{center}
\includegraphics[height=4cm]{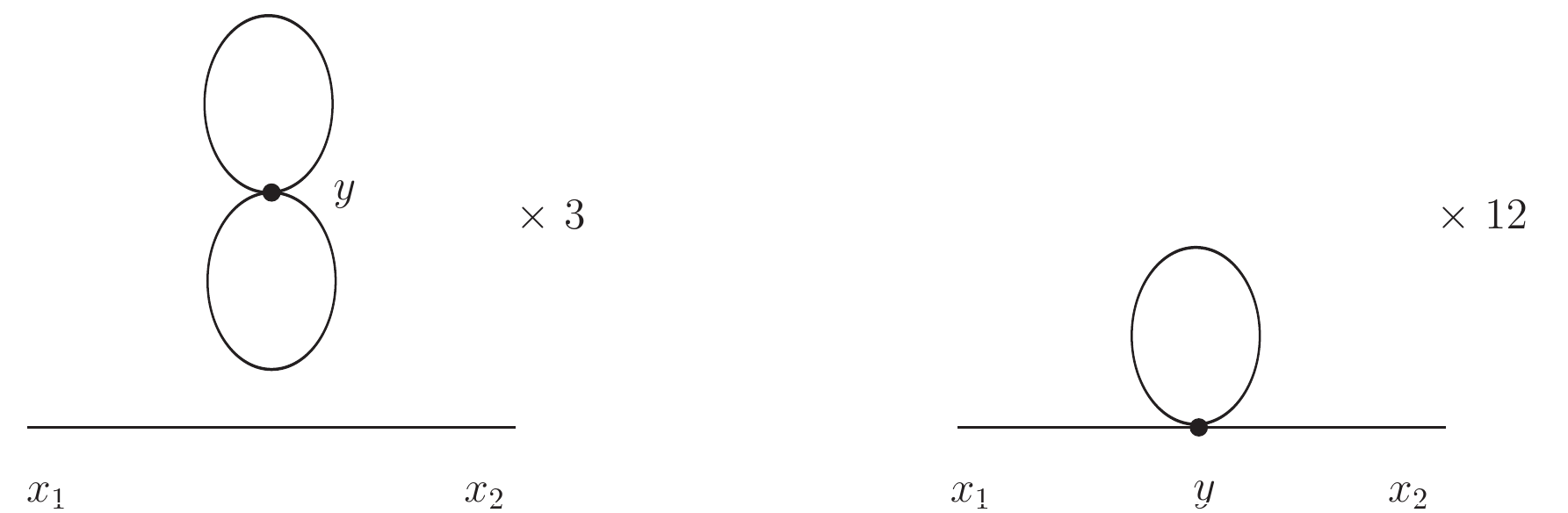}
\caption{Feynman diagrams for first propagator correction}
\label{firstfeyn}
\end{center}
\end{figure}

Wick's theorem applied to this 6-point function gives:
\be
\begin{split}
\langle 0|T(\phi(x_1) \phi(x_2) \phi^4(y))|0\rangle &=
3\, D_F(x_1-x_2)\,D_F(y-y)\,D_F(y-y)\\ &+~ 
12\, D_F(x_1-y)\,D_F(x_2-y)\,D_F(y-y)
\end{split}
\ee
Such expressions are conveniently represented in terms of diagrams. For example, the expression above can be represented as in Fig.\ref{firstfeyn}.

In this diagram, each line represents a Feynman propagator from the starting point to the end-point of the line.  Since we are working to first order in the interaction, the first term in the expansion of
$e^{-i\int H_I}$ is represented by a single four-point vertex. Correspondingly the entire term is multiplied by a single power of $\lambda$, which parametrises the interaction strength or ``coupling constant'' of the theory. In general the power of $\lambda$, or equivalently the number of four-point vertices in the diagram, counts the
order of perturbation theory. The numerical factors 3 and 12 in the above expression are ``statistical factors'' that arise from counting the number of different Wick contractions that lead to the same expression. Diagrams with propagators and vertices that depict the possible terms arising from Wick contraction in perturbation theory are called {\em Feynman diagrams}.

The simple example we have worked out already exhibits a variety of potential problems. In the
second term we get a 
factor of $D_F(y-y) = D_F(0)$. Now, 
$$
D_F(0) = \int d^4p\, \d{i \over p^2 - m^2}
$$ 
which is divergent. The divergence comes from the region of large
$|\vec p|$, so it is called an {\em ultraviolet divergence}.

The first term is even more problematic. Diagrammatically it corresponds to a loop disconnected from the freely propagating particle. The integrand
is $y$-independent and therefore we appear to have a
divergence proportional to the (infinite) volume of space-time. This is
multiplied by a factor $(D_F(0))^2$ which as we have seen above is
also divergent.

In general, we should always imagine cutting off ultra-violet
divergences using a large momentum cutoff, for example by specifying that in the integral, $|\vec p| < \Lambda$. This
is called a UV cutoff. For the volume divergence, we can put the system
in a finite box in space-time, which is called an infra-red or IR cutoff. The problem is then how to remove and/or makes sense of the cutoffs.  We will discuss this problem in some generality at a later stage.

Fortunately, we can easily dispose of the problem of disconnected
diagrams. Our intuition suggests that the one in the example above is rather unphysical: a ``virtual'' interaction takes place at an arbitrary point $y$ while the external particle propagates freely from $x_1$ to $x_2$. The disconnected bubble is unrelated to the physical process taking place. This intuition is correct: such a term is cancelled by a similar term coming from the denominator of \eref{intnpt}, where we have to expand:
\be
\langle 0|T(e^{-i\int H_I})|0\rangle = 
1 - 3{i\lambda \over 4!} \int d^4y\, D_F(y-y) D_F(y-y) +\cdots
\ee
Thus we have, to order $\lambda$,
\be
\begin{split}
&\d{\langle \Omega|T(\phi(x)\phi(y))|\Omega\rangle \over 
\langle \Omega|\Omega\rangle}\\
&=~\frac{D_F(x_1-x_2)(1-3\frac{i\lambda}{4!}\int\! d^4 y\, D_F(0)^2)
 -12 \frac{i\lambda}{4!}\int\! d^4 y\,
D_F(0)D_F(x_1-y)D_F(x_2-y) +\cdots}{1-3\frac{i\lambda}{4!}\int\! d^4 y\,
D_F(0)^2+\cdots}\\[3mm]
&=~  D_F(x_1-x_2)\,\left(1-3\sfrac{i\lambda}{4!}\sint d^4 y\, D_F(0)^2\right)
\left(1+3\sfrac{i\lambda}{4!}\sint d^4 y\, D_F(0)^2\right)\\
&\qquad\qquad\qquad\qquad
 -~12\sfrac{i\lambda}{4!}\sint d^4 y\, D_F(x_1-y)\,D_F(x_2-y)\,D_F(0)
+{\cal O}(\lambda^2)
\end{split}
\ee

Since we are working to order $\lambda$, we find that:
\be
\left(1-3\sfrac{i\lambda}{4!}\sint d^4 y\, D_F(0)^2\right)
\left(1+3\sfrac{i\lambda}{4!}\sint d^4 y\, D_F(0)^2\right)=1+{\cal
O}(\lambda^2)\sim 1
\ee
and the bubble term has cancelled! Recall that we are working to order $\lambda$ and therefore cannot retain terms of order $\lambda^2$. If we actually work to the next order and keep all terms of that order, the disconnected terms cancel again. Indeed this is a general phenomenon: all ``bubble'' diagrams can be shown to
be cancelled by denominator contributions. Moreover this uses up all
denominator contributions, so we can consistently ignore bubble
diagrams as well as denominators. We will not give a precise derivation of this result, the interested student should try it on her own or look up a textbook. We  will use the result repeatedly to simplify our calculations.

\begin{figure}[H]
\begin{center}
\includegraphics[height=10cm]{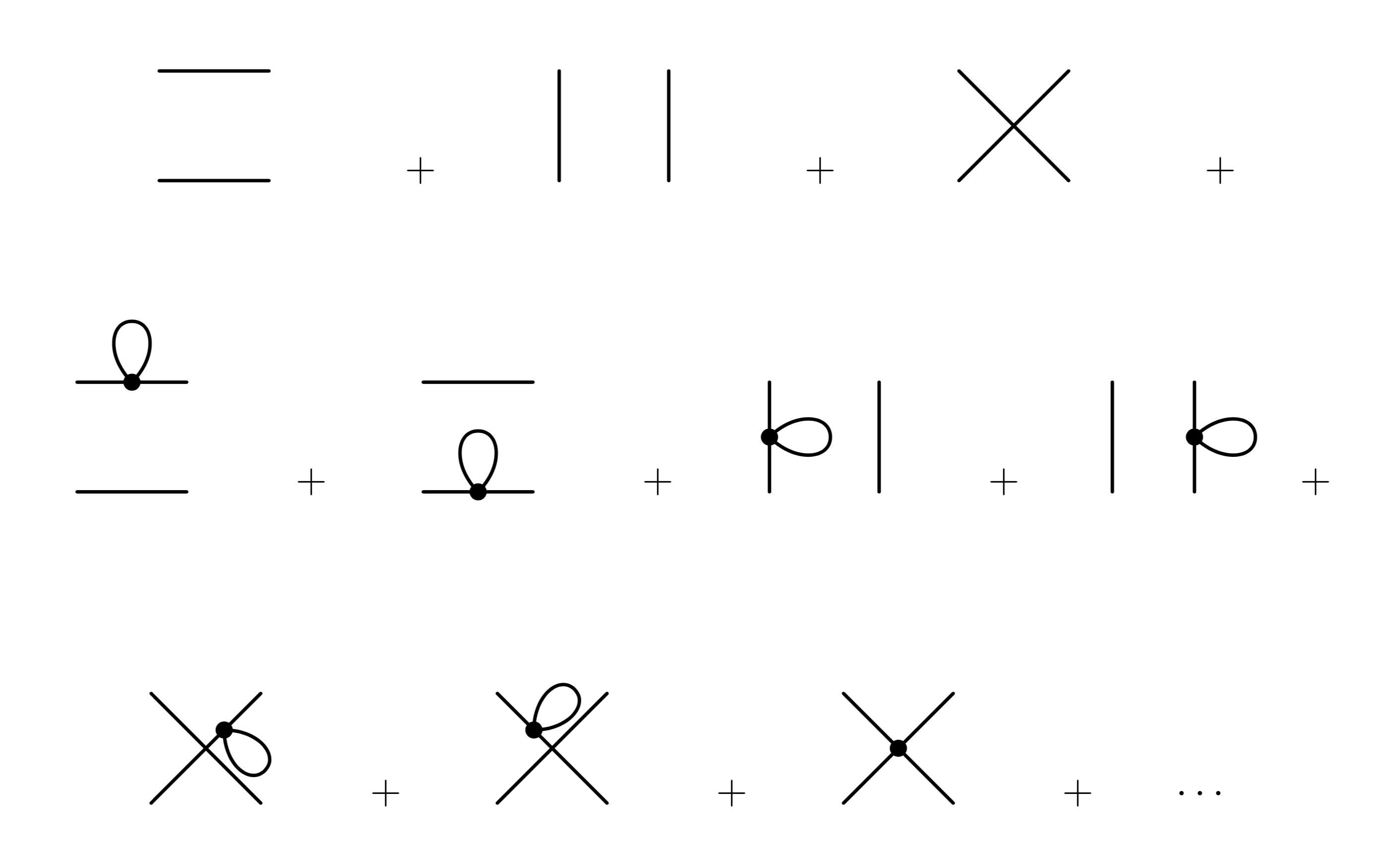}
\caption{Feynman diagrams for 4-point function to order $\lambda$}
\label{feynfourpt}
\end{center}
\end{figure}

The rule of ignoring bubble diagrams means that when we calculate any $n$-point function:
\be
\langle 0|T\Big(\phi(x_1)\phi(x_2)\cdots \phi(x_n)\Big)|0\rangle
\ee
we only include those diagrams where all parts are connected to at least one of the external
points $x_1,x_2,\cdots,x_n$. As an example, the diagrams for the four-point function
\be
\langle 0|T\Big(\phi(x_1)\phi(x_2)\phi(x_3)\phi(x_4)\Big)|0\rangle
\ee
are given to order $\lambda$ by Figure \ref{feynfourpt}.

This figure once more exemplifies the power of Feynman diagrams in providing physical intuition. The first three diagrams involve no scattering at all\footnote{For clarity we explicitly denote interaction vertices with a dot. Thus in the third figure of the first line, there are two propagators crossing each other. But in the third figure of the last line, there are four propagators meeting at a common vertex.}. External particles (interpreted as being either in the initial or final state) connect to each other pairwise. These terms would be present even in the free theory, consistent with the fact that they involve no scattering. The next six terms, though they are of order $\lambda$ (as highlighted by the dot at the interaction vertex) also do not involve a true scattering process. In each of them, one of the incoming particles undergoes a sort of ``self-interaction'' through a virtual loop before going out. What characterises each of the first 9 diagrams is that they are disconnected, in the sense that not all external points are connected to {\em each other} through a series of lines. Thus to order $\lambda$, the only connected diagram is the last one, in which two particles undergo a contact interaction and scatter into other particles. Only this diagram will contribute to the $\cS$-matrix. These intuitive observations will be made more precise in subsequent sections.

\subsection{Momentum space Feynman rules}
\label{momspacefeyn}

Feynman diagrams are shorthand symbols for numbers, namely interacting $n$-point functions. They depend on
external positions as well as the coupling constant $\lambda$ (from the expansion of the Hamiltonian) and the mass
$m$ (on which the propagator depends). But in practice we never create particles at fixed positions! They are created at fixed momenta so that they can scatter off each other. After scattering, new particles again emerge with fixed momenta. Therefore it is more appropriate to consider the Fourier transform of the position-space $n$-point function that we have studied up to now. In fact, we will see that not only experimentally but also theoretically, momentum space is more natural and simpler than position space.

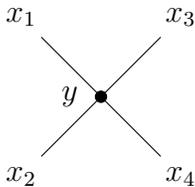
\begin{figure}[h!]
\begin{center}
\begin{tikzpicture}
\begin{feynman}
\vertex [dot] (c) {};
\vertex [above left=of c] (a) {$x_1$};
\vertex [below left=of c] (b) {$x_2$};
\vertex [above right=of c] (k) {$x_3$};
\vertex [below right=of c] (l) {$x_4$};
\vertex [left=1em of c] {\(y\)};
\diagram* {
(a) -- [] (c);
(b) -- [] (c);
(k) -- [] (c);
(l) -- [] (c);
};
\end{feynman}
\end{tikzpicture}
\caption{Connected part of 4-point function at order $\lambda$}
\label{fourptconn}
\end{center}
\end{figure}

Let us start by defining:
\be
\tilde D_F(k) = \d{i \over k^2 - m^2 + i\epsilon}
\label{momprop}
\ee
This is the Fourier transform of the Feynman propagator \eref{feynpropint}. Now consider the diagram in Fig.\ref{fourptconn}. It can be evaluated by putting together all the ingredients of which it is made up. We find:
\be
\begin{split}
&-i\lambda
\int d^4y\, D_F(x_1-y) D_F(x_2-y) D_F(x_3-y) D_F(x_4-y)  \\
&= -i\lambda\int d^4y \prod_{i=1}^4 \left(\int {d^4k_i \over (2\pi)^4}
{i\over (k^2_i - m^2 + i\epsilon)}\right) e^{-i\sum_i k_i\cdot(x_i-y)} 
\end{split}
\ee
The factor $-i$ comes from the expansion of the exponential of $H_I$. This also brings down a factor of $\lambda/4!$ but the denominator is cancelled because there are $4!$ ways to construct this graph (i.e. ways to contract one of the external fields with one of the four $\phi$'s in the interaction Hamiltonian). Finally, one multiplies a free propagator for each of the four ``legs'' of the diagram.

The $y$-integral gives:
\be
\int d^4y \ e^{iy\cdot\sum_ik_i}
= (2\pi)^4 \delta^4\Big(\sum_i k_i\Big)
\ee
Thus the above expression is equal to:
\be
-i\lambda\int \prod_{i=1}^4\left( {d^4k_i \over (2\pi)^4}  {i \over
(k^2_i - m^2 + i\epsilon)}\right)  
(2\pi)^4\delta\Big(\sum_i k_i\Big) 
e^{-i\sum_i k_i\cdot x_i}
\ee
As we saw above, this diagram is one contribution to 
\be
G(x_1,\cdots,x_4) = \langle \Omega|T(\phi(x_1) \cdots \phi(x_4))|\Omega\rangle
\ee
We now write it as: 
\be
\int \prod_i {d^4 k_i \over (2\pi)^4}\, e^{-i\sum k_i \cdot x_i}\, 
(2\pi)^4\delta\Big(\sum_i k_i\Big)\,{\tilde G}(k_1,\cdots,k_4)
\ee
where:
\be
{\tilde G}(k_1,\cdots,k_4)= -i\lambda\,\prod_{i=1}^4
{i \over k^2_i - m^2 + i\epsilon}
\ee
We refer to ${\tilde G}(k_1,\cdots,k_4)$ as the ``momentum space
correlation function'' although it is not $\tilde G(k_1,\cdots,k_4)$
but rather $(2\pi)^4\delta(\Sigma k_i) \tilde G (k_1, \cdots, k_4)$ that is the
Fourier transform of $G(x_1 \cdots x_4)$. We simply keep in mind that every ${\tilde G}$ is accompanied by a momentum-conserving delta-function which also brings along its own factor of $(2\pi)^4$.

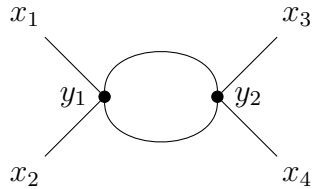
\begin{figure}[H]
\begin{center}
\begin{tikzpicture}
\begin{feynman}
\vertex [dot] (c) {};
\vertex [above left=of c] (a) {$x_1$};
\vertex [below left=of c] (b) {$x_2$};
\vertex [right=of c, dot] (f) {};
\vertex [above right=of f] (k) {$x_3$};
\vertex [below right=of f] (l) {$x_4$};
\vertex [left=1em of c] {\(y_1\)};
\vertex [right=1em of f] {\(y_2\)};
\diagram* {
(a) -- [] (c);
(b) -- [] (c);
(f) -- [] (k);
(f) -- [] (l);
(c) -- [half left, looseness=1.2] (f);
(f) -- [half left, looseness=1.2] (c);
};
\end{feynman}
\end{tikzpicture}
\caption{Second order contribution to 4-point function.} 
\label{secord}
\end{center}
\end{figure}

Let us now try to get some insight into the structure of a general Feynman diagram. For this, let us look at a particular contribution to the 4-point function that arises by expanding the exponential to second order. The diagram is shown in Figure \ref{secord}.

Let us first focus on the propagators. There are six in all, four connecting an ``external'' point $x_i$ to an ``internal'' point $y_a$ while two connect the internal points $y_1$ and $y_2$. Writing out each of the six propagators and integrating over $y_1,y_2$ as required, we have (apart from numerical factors):
\be
\begin{split}
&\sim \lambda^2\int d^4y_1\,d^4y_2 \prod_{i=1}^6 \left(\int\frac{d^4k_i}{(2\pi)^4}\frac{1}{k_i^2-m^2+i\epsilon}\right)\times\\
&\qquad\qquad e^{-ik_1\cdot(x_1-y_1)}e^{-ik_2\cdot(x_2-y_2)}e^{-ik_3\cdot(x_3-y_1)}e^{-ik_4\cdot(x_4-y_2)}e^{-ik_5\cdot(y_1-y_2)}e^{-ik_6\cdot(y_1-y_2)}
\end{split}
\ee
The $y_1,y_2$ integrals are easily performed, leading to delta-functions:
\be
(2\pi)^8 \,\delta^4(k_1+k_3-k_5-k_6)\,\delta^4(k_2+k_4+k_5+k_6)
\ee
From this we learn that there is a momentum-conserving delta function at every vertex. 

Since the second delta function vanishes when its argument is nonzero, we can substitute $k_5+k_6=-k_2-k_4$ into the first delta function, to get:
\be
(2\pi)^8 \,\delta^4(k_1+k_2+k_3+k_4)\,\delta^4(k_1+k_3-k_5-k_6)
\ee
The first delta-function conserves overall momentum. We remove this, together with a factor of $(2\pi)^4$, as well as a factor $\prod_{i=1}^4 e^{-ik_i\cdot x_i}$, to get the momentum-space version of the diagram:
\be
\begin{split}
\sim \lambda^2 \left(\prod_{i=1}^4 \frac{1}{k_i^2-m^2+i\epsilon}\right)
&\int \frac{d^4k_5}{(2\pi)^4}\frac{d^4k_6}{(2\pi)^4} \, \frac{1}{k_5^2-m^2+i\epsilon}\frac{1}{k_6^2-m^2+i\epsilon}\times\\
&\qquad\qquad(2\pi)^4\delta^4(k_1+k_3-k_5-k_6)
\end{split}
\label{momspacediag}
\ee
Finally we perform the integral over $k_6$ using the delta-function. This leaves just one remaining integration variable, $k_5$ which we re-label as $q$. Then we have:
\be
\begin{split}
\sim \lambda^2 \left(\prod_{i=1}^4 \frac{1}{k_i^2-m^2}\right)
&\int \frac{d^4q}{(2\pi)^4} \, \frac{1}{q^2-m^2}\frac{1}{(k_1+k_3-q)^2-m^2}
\end{split}
\ee
Note that we have dropped the $i\epsilon$ term in every denominator. It is always understood to be present, and will be re-introduced as needed.

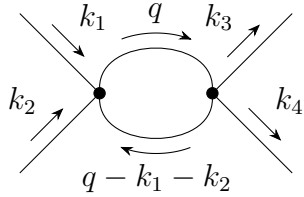
\begin{figure}[H]
\begin{center}
\begin{tikzpicture}
\begin{feynman}
\vertex [dot] (c) {};
\vertex [above left=of c] (a);
\vertex [below left=of c] (b);
\vertex [dot, right=of c] (f) {};
\vertex [above right=of f] (k);
\vertex [below right=of f] (l);
\diagram* {
(a) -- [momentum={[arrow shorten=0.1mm, arrow distance=2mm]\(k_1\)}] (c);
(b) -- [momentum={[arrow shorten=0.1mm, arrow distance=2mm]\(k_2\)}] (c);
(f) -- [momentum={[arrow shorten=0.1mm, arrow distance=2mm]\(k_3\)}] (k);
(f) -- [momentum={[arrow shorten=0.1mm, arrow distance=2mm]\(k_4\)}] (l);
(c) -- [half left, looseness=1.2, momentum={[arrow shorten=0.11mm, arrow distance=2mm]\(q\)}] (f);
(f) -- [half left, looseness=1.2, momentum={[arrow shorten=0.11mm, arrow distance=2mm]\(q-k_1-k_2\)}] (c);
};
\end{feynman}
\end{tikzpicture}
\caption{Second order contribution to 4-point function in momentum space.} 
\label{secordmom}
\end{center}
\end{figure}

The above manipulations illustrate a number of general principles. A Feynman diagram with multiple vertices has a momentum-conserving $\delta$-function at {\em each} vertex of the diagram. 
In momentum space, these diagrams can be evaluated by 
giving each leg an independent momentum, then assigning a propagator: 
\be
\tilde D_F(k) = \d{i \over k^2-m^2+i\epsilon}
\ee 
to each line, putting in a momentum-conserving $\delta$-function at
each vertex and finally integrating over each momentum except the ones on legs that are
directly connected to external points. Thus we label diagrams with momenta flowing along the propagators, rather than by the positions of the vertices. The momenta must have arrows
describing their orientation. The arrows are arbitrary but must be
assigned once and for all at the start. The signs of the momenta appearing in
momentum-conserving $\delta$-functions will be determined by these
arrows. In Figure \ref{secordmom} we exhibit the same diagram as in Figure \ref{secord} but appropriately labeled with momenta rather than positions. We have introduced arrows to depict the direction of momentum flow.

From this example it is easy to see that there is one momentum integral for
every closed loop. The physical interpretation is that there is a
``virtual particle'' circulating in the loop. This particle does not
need to satisfy $k^2=m^2$, nor does it need to have a positive value
for $k^0$. These properties are evident from the above example where the loop momentum is $q$ and we integrate each component independently from $-\infty$ to $+\infty$. We know that the  momenta of external (physical) particles must satisfy $k^2 = m^2$, a property referred to as being ``on-shell''. Thus we see that internal or loop momenta, by contrast, are ``off-shell''.

A puzzling feature of \eref{momspacediag} is that if the external momenta $k_1,\cdots,k_4$ are
placed on-shell then the external propagators all diverge! Therefore
temporarily we will allow even external momenta to be
off-shell. Keeping the external legs of correlation functions
off-shell can be very useful to develop recursive formulae among them.
When we relate the $n$-point function to physically measurable
quantities, specifically the scattering matrix (``S-matrix''), we will
argue that the external propagator factors should be dropped and {\em then} the
external momenta put on-shell.

Returning to the diagram we drew above, let us briefly look at the last remaining integral that needs to be performed, the one over the ``loop momentum'' $q$. This
integral is:
\be
\int \frac{d^4q}{(2\pi)^4}\,
\frac{i}{q^2-m^2}\,\frac{i}{(k_1+k_3-q)^2-m^2}
\ee
We see that if all components of $q$ are scaled to infinity, the numerator and denominator both scale
like $|q|^4$. Thus the integral is {\em logarithmically divergent}. Because this arises from the upper end of the integration (large momentum), it is called an ultraviolet divergence. How to deal with such divergent
integrals will be the subject of the rest of these lecture notes.

\subsection{Computation of the $\cS$-matrix}
\label{compsmat}

We have developed a considerable amount of machinery to compute $n$-point correlation functions $\langle
\Omega|T(\phi(x_1)\phi(x_2)\cdots\phi(x_n))|\Omega\rangle$ in an interacting field theory. The final step is to relate these to quantities that are physical, i.e. experimentally measurable. As we have already explained, it is the $\cS$-matrix, or the reduced matrix element ${\cal M}$, that we must compute using quantum field theory.

As we saw, the $\cS$-matrix is the overlap of ``in'' and ``out'' states of the interacting theory. To understand these states better, recall how we defined the interacting in-vacuum:
\be
|\Omega\rangle = \lim_{T \rightarrow \infty(1-i\epsilon)} e^{-i HT}|0\rangle
\ee
and the out-vacuum:
\be
\langle \Omega| = \lim_{T \rightarrow \infty(1-i\epsilon)} \langle 0|
e^{-iHT}
\ee
(both upto normalisation).
We computed the overlap between the two vacua in perturbation theory and derived the formula:
\be
\langle \Omega|\Omega\rangle = \lim_{T\rightarrow\infty(1-i\epsilon)} 
\langle 0|T\left(e^{-i \int^{T}_{-T} H_I(t')dt'}\right)|0\rangle
\ee
The physical interpretation of this expression is that the free vacuum in the far past is evolved to the free vacuum in the far future by the unitary operator in the middle. The $\cS$-matrix will be a
similar unitary operator, but it will take us from a state containing a number of non-interacting particles in the far past to another one in the far future.

As indicated earlier, the ``in'' state of interest denoted $|\vec p_1,
\vec p_2\rangle_{\rm in}$ is hard to define precisely. However in the free theory we know that the corresponding state is:
\be
|\vec p_1, \vec p_2\rangle_0=
\sqrt{2\omega_{\vec p_1}} \sqrt{2\omega_{\vec p_2}}~ a^\da_{\vec p_1} 
a^\da_{\vec p_2} |0\rangle
\ee
We define the state of the interacting theory through the same operator that we used to evolve the free in-vacuum to the interacting one. Thus:
\be
|\vec p_1,\vec p_2\rangle_{\rm in} = \lim_{T \rightarrow 
\infty(1 - i\epsilon)} e^{-iHT} |\vec p_1,\vec p_2\rangle_0
\ee
For the ``out'' state the procedure is similar except that we allow for an arbitrary number $n$ of outgoing particles:
\be
_{\rm out}\langle \vpp_1, \cdots, \vpp_n| = \lim_{T \rightarrow \infty
(1 - i\epsilon)} \ _0\langle \vpp_1,\cdots,\vpp_n|e^{-iHT}
\ee

Now recalling the definition of the $\cS$-matrix as the overlap of the
``in'' state with the ``out'' state:
\be
\cS_{(\vpp_1,\cdots,\vpp_n|\vp_1,\vp_2)}=
{}_{\rm out}\langle \vpp_1,\cdots,\vpp_n|\vec p_1,\vec p_2\rangle_{\rm in}
\ee
as well as the preceding manipulations involving correlation functions, we may guess a formula to evaluate the $\cS$-matrix (upto a multiplicative constant) in terms of free fields in perturbation theory:
\be
\cS_{(\vpp_1,\cdots,\vpp_n|\vp_1,\vp_2)}=
 \lim_{T \rightarrow \infty(1 - i\epsilon)} \ _0\langle \vpp_1,\cdots,
\vpp_n |T\left(e^{-i \int^T_{-T} H_I(t')dt'}\right) |\vec p_1,
\vec p_2\rangle_0
\ee
Earlier we proposed an analogous formula for correlation functions in the vacuum, and just as in that case, we will assume the above formula to be true. Note that although we have not provided a derivation, let alone a rigorous one, the formula itself is extremely precise.

\exercise{Derive the above equation (or look it up).}

Consider the special case of $2 \rightarrow 2$ scattering, which means
the final state has two particles in it. Expanding the above formula, we
have:
\be
\begin{split}
\cS_{(\vpp_1,\vpp_2|\vp_1,\vp_2)}
&= \lim_{T \rightarrow \infty(1 - i\epsilon)} \ _0\langle \vpp_1, \vpp_2 |
T\left(e^{-i \int^T_{-T} H_I(t')dt'}\right)|\vec p_1,\vec p_2\rangle_0 \\
&=\quad {}_0\langle \vpp_1,\vpp_2|\vec p_1,\vec p_2\rangle_0 + {\cal O}(\lambda) 
\end{split}
\label{twototwoamp}
\ee
where for definiteness we continue to consider an interaction of the form $H_{\rm
int} = {\lambda \over 4!} \int \phi^4$ as we have been doing in the previous sections. The first term in the last line above just sets the initial state equal to the final state and does not take into account interactions. This is what we have earlier called a disconnected process and we would like to drop it. It can be considered as part of the ${\bf 1}$ in the decomposition:
\be
\cS={\bf 1}+i{\cal T}
\ee

Just as we did for $n$-point amplitudes, we can use Feynman diagrams to depict $\cS$-matrix elements. However, before doing so we must note an important point. In the diagrams we drew for $n$-point correlation functions, for example those in Fig.(\ref{feynfourpt}), there was no specification of any ``incoming'' or ``outgoing'' particles. When talking about the $\cS$-matrix, however, this notion is fundamental and we must fix a definite direction for the flow of time. Accordingly, in the Feynman diagrams to follow, time will be taken to flow from left to right. Lines coming from the left will represent incoming particles and those going to the right will represent outgoing ones.

Now consider the first three diagrams in Fig.(\ref{feynfourpt}), but interpreted as $\cS$-matrix contributions. Now the second one is no longer permitted since it would represent incoming particles propagating into each other. That leaves the first and third diagrams, which do contribute to $\cS$ and in fact give precisely the terms describing no scattering. As we saw, these diagrams
are ``disconnected'', not in the sense of the bubble diagrams that we encountered earlier,
but in the sense that particles pass through without interacting with each other. Thus, in order to compute the $\cT$-matrix we simply drop them.

Next consider the terms in \eref{twototwoamp} of order $\lambda$:
\be
-i {\lambda \over 4!} \ _0\langle \vpp_1,\vpp_2 |\, T \Big(\int \phi^4 
(y) d^4y\Big) |\vec p_1,\vec p_2\rangle_0 
\label{ordlambda}
\ee
(here $\phi$ refers to the free field in the interaction picture,
which should really be denoted $\phi_0$ but we drop the $0$ hoping no
confusion will arise). Notice that this time we do not have a vacuum
to vacuum amplitude so we cannot just expand the time-ordered product
using Wick's theorem. However we can use a more general result that expresses the time-ordered product in terms of normal-ordered products and Feynman propagators. To define normal-ordering we decompose the field as $\phi=\phi_++\phi_-$, with:
\be
\begin{split}
\phi_+(t,\vx) &= \int
\frac{d^3k}{(2\pi)^3}\frac{1}{\sqrt{2\omega_\vk}}\, a_\vk\,e^{-ik\cdot
x}\\
\phi_-(t,\vx) &= \int
\frac{d^3k}{(2\pi)^3}\frac{1}{\sqrt{2\omega_\vk}}\, a^\da_\vk\,e^{ik\cdot
x}
\end{split}
\ee
Normal-ordering consists of placing $\phi^+$ to the right of $\phi^-$ in every monomial. It is denoted by $\lcol ()\rcol$. 
Now the relevant result is:
\be
T(\phi^4(y))=\lcol\phi^4(y)\rcol + 6D_F(0)\, \lcol \phi^2(y) \rcol + 3
D_F(0)^2 
\label{phifournorm}
\ee

\exercise{By writing $\phi=\phi_++\phi_-$ and rearranging terms, derive the above equation. Also verify that if we sandwich both sides of the above equation between vacuum states, we get the same result as predicted by Wick's theorem. For this you will need the result $\phi_+|0\rangle=0$, which follows easily from the definition of $\phi_+$ and $|0\rangle$.}

Although $\phi_+$ annihilates the vacuum, one easily sees that it does not annihilate a one-particle state, i.e.
$\phi_+ (x) |\vec p_1\rangle$ is nonzero:
\be
\phi_+ (x) |\vec p_1\rangle =
\int {d^3k \over (2\pi)^3} {1 \over \sqrt{2\omega_{\vec k}}}\, a_{\vec k}\, 
e^{-i{k\cdot x}} \sqrt{2\omega_{\vec p_1}}\, a^\da_{\vec p_1} |0\rangle
= e^{-i{p_1\cdot x}}|0\rangle
\label{phipone}
\ee
Next we use this to compute $\langle \vpp_1,\vpp_2| \lcol \phi^4(y) \rcol|\vec p_1,\vec p_2\rangle$.
From a calculation similar to \eref{phipone} it is easy to show that the application of more than two $\phi_+$ factors on the right, or more than two $\phi_-$ factors on the left, gives a vanishing result. Therefore the only term we need to keep
from $\lcol\phi(y)^4\rcol $ is $6\,\phi_-^2(y) \phi_+^2(y)$.
Thus we have:
\be
\begin{split}
&-i {\lambda \over 4!} \ _0\langle \vpp_1,\vpp_2 |\, T \Big(\int \phi^4 
(y) d^4y\Big) |\vec p_1,\vec p_2\rangle_0\\
&\qquad\qquad=
6\cdot -{i \over 4!} \lambda \int d^4y 
\langle \vpp_1,\vpp_2|\, \phi_-^2(y)\phi_+^2(y)\, |\vp_1,\vp_2\rangle_0
\end{split}
\ee
It is easily seen that:
\be
\phi_+(y)\, |\vp_1,\vp_2\rangle_0
= e^{-i{p_1\cdot y}}|\vp_2\rangle_0 + e^{-i{p_2\cdot y}} |\vp_1\rangle_0 
\ee
Repeating this, we find:
\be
\phi_+^2(y)\, |\vp_1,\vp_2\rangle_0 
= 2\, e^{-i{(p_1+p_2)\cdot y}}|0\rangle_0 
\ee
Similarly,
\be
\langle \vpp_1,\vpp_2|\,\phi_-^2(y) = 2\, \langle 0|e^{i{(p'_1+p'_2)\cdot y}}
\ee
Thus we finally get:
\be
- i\lambda \int d^4y\, e^{i{(p'_1+p'_2-p_1-p_2)\cdot y}}
= -i\lambda\, (2\pi)^4 \delta^4(p'_1 + p'_2 - p_1 - p_2)
\ee
This term describes genuine scattering: it is of order $\lambda$ and it allows individual particles to exchange their momenta while only conserving the total incoming momentum. Thus it is a valid contribution to $\langle {\cal T}\rangle$, the non-trivial part of the $\cS$-matrix. The corresponding diagram is the same as Fig.(\ref{fourptconn}) but with momentum-space labels.
Recalling the definition \eref{matel} of the matrix element ${\cal M}$, we conclude that the term coming from $\lcol \phi^4(y)\rcol$ contributes ${\cal M} = -\lambda$. 

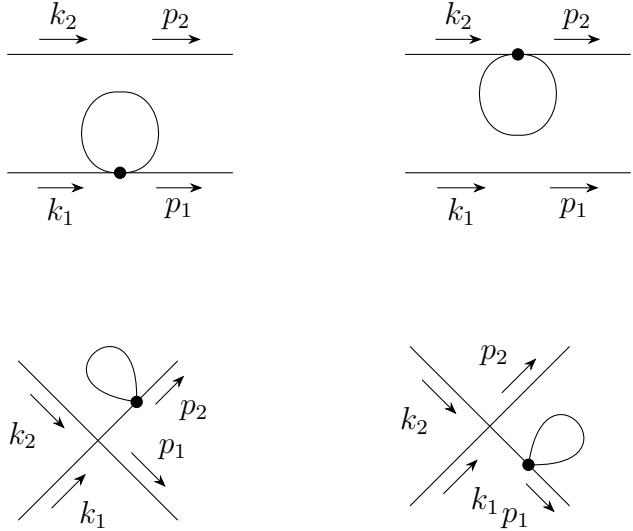
\begin{figure}[H]
\begin{center}
\begin{tikzpicture}
\begin{feynman}
\vertex (a);
\vertex [above=of a, yshift=-10mm] (b);
\vertex [left=of b] (c);
\vertex [right=of b] (d);
\vertex [dot, below=of a, yshift=5mm] (e) {}; 
\vertex [left=of e] (f);
\vertex [right=of e] (g);
\diagram* {
(c) -- [momentum={[arrow shorten=0.1mm, arrow distance=2mm]\(k_2\)}] (b);
(b) -- [momentum={[arrow shorten=0.1mm, arrow distance=2mm]\(p_2\)}] (d);
(f) -- [momentum'={[arrow shorten=0.1mm, arrow distance=2mm]\(k_1\)}] (e);
(e) -- [momentum'={[arrow shorten=0.1mm, arrow distance=2mm]\(p_1\)}] (g);
(e) -- [half left] (a) --[half left] (e);
};
\end{feynman}
\end{tikzpicture}
\hspace*{2cm}
\begin{tikzpicture}
\begin{feynman}
\vertex (a);
\vertex [dot, above=of a, yshift=-5mm] (b) {};
\vertex [left=of b] (c);
\vertex [right=of b] (d);
\vertex [below=of a, yshift=10mm] (e); 
\vertex [left=of e] (f);
\vertex [right=of e] (g);
\diagram* {
(c) -- [momentum={[arrow shorten=0.1mm, arrow distance=2mm]\(k_2\)}] (b);
(b) -- [momentum={[arrow shorten=0.1mm, arrow distance=2mm]\(p_2\)}] (d);
(f) -- [momentum'={[arrow shorten=0.1mm, arrow distance=2mm]\(k_1\)}] (e);
(e) -- [momentum'={[arrow shorten=0.1mm, arrow distance=2mm]\(p_1\)}] (g);
(b) -- [half left] (a) --[half left] (b);
};
\end{feynman}
\end{tikzpicture}\\[7mm]
\begin{tikzpicture}
\begin{feynman}
\vertex (c);
\vertex [above left=of c] (a);
\vertex [below left=of c] (b);
\vertex [above right=of c] (k);
\vertex [below right=of c] (l);
\vertex [dot, above right=of c, xshift=-6mm, yshift=-6mm] (m) {};
\diagram* {
(a) -- [momentum'={[arrow shorten=0.1mm, arrow distance=2mm]\(k_2\)}] (c);
(m) -- [out=170, in=90, loop, min distance=1.4cm] (m)
(b) -- [momentum'={[arrow shorten=0.1mm, arrow distance=2mm]\(k_1\)}] (c);
(c) -- [] (m) --  [momentum'={[arrow shorten=0.04mm, arrow distance=2mm]\(p_2\)}] (k);
(c) -- [momentum={[arrow shorten=0.1mm, arrow distance=2mm]\(p_1\)}] (l);
};
\end{feynman}
\end{tikzpicture}
\hspace*{2cm}
\begin{tikzpicture}
\begin{feynman}
\vertex (c);
\vertex [above left=of c] (a);
\vertex [below left=of c] (b);
\vertex [above right=of c] (k);
\vertex [below right=of c] (l);
\vertex [dot, below right=of c, xshift=-6mm, yshift=6mm] (m) {};
\diagram* {
(a) -- [momentum'={[arrow shorten=0.1mm, arrow distance=2mm]\(k_2\)}] (c);
(m) -- [out=80, in=0, loop, min distance=1.4cm] (m)
(b) -- [momentum'={[arrow shorten=0.1mm, arrow distance=2mm]\(k_1\)}] (c);
(c) -- [momentum={[arrow shorten=0.1mm, arrow distance=2mm]\(p_2\)}] (k);
(c) -- [] (m) -- [momentum'={[arrow shorten=0.04mm, arrow distance=2mm]\(p_1\)}] (l);
};
\end{feynman}
\end{tikzpicture}
\caption{Four disconnected diagrams: there are self-interactions at the dotted vertex, but the two particles do not interact with each other.} 
\label{sfourterms}
\end{center}
\end{figure}

Now we consider the other terms in \eref{phifournorm}. For example, $D_F(0)\, \lcol
\phi^2(y)\rcol $ gives:
\be
D_F(0)\, \langle \vpp_1, \vpp_2 | \phi_-(y) \phi_+(y) |\vec p_1, 
\vec p_2\rangle 
\ee
This can be represented as the sum of the 4 diagrams in Fig.\ref{sfourterms}. 

These diagrams do contain interactions, as represented by the dotted vertices, but those interactions merely convert one of the free propagators to an interacting propagator. The two incoming particles never interact with each other! Thus these diagrams are also included in the  $\bf 1$ part of $\cS = {\bf 1} + i {\cal T}$. Finally, terms with $(D_F(0))^2$ give bubble diagrams which involve none of the incoming or outgoing particles, and therefore these too are dropped.
Thus at the end of this rather lengthy discussion, we have found that:
\be
{i\cal M} = -i\lambda + {\cal O}(\lambda^2)
\label{mtreelevel}
\ee
Here we finally see the physical meaning of $\lambda$: it is just
(minus) the leading contribution to the ${\cal M}$-matrix, which encodes all the physical scattering data.

We also see that ${\cal M}$ differs in an important way from the four-point 
correlation function:
\be
\langle 0|T\Big(\phi(x_1) \cdots \phi(x_4) \int d^4y\, \phi^4(y)\Big)|0\rangle.
\ee
The latter, in momentum space, has four propagators and at leading order is equal to:
\be
-\lambda \prod^4_{i=1} {1 \over k^2_i - m^2 + i\epsilon} \, (2\pi)^4 
\delta^4(k_i + \cdots + k_r)
\ee
However ${\cal M}$ has no propagators for the external legs. It can be thought of as an ``amputated'' version of the four-point function, with the external legs ``chopped off''. 

The fact that the prescription for calculating the $\cS$-matrix requires omitting the external propagators resolves a puzzle we had encountered earlier. The external particles are physical and satisfy $k^2 = m^2$,
so, had they been present, the propagators $\frac{1}{k^2-m^2+i\epsilon}$ would have
diverged. Fortunately those propagators are absent and we can now safely place all external particles on-shell.

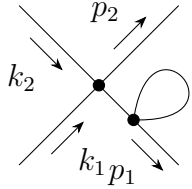
\begin{figure}[h!]
\begin{center}
\begin{tikzpicture}
\begin{feynman}
\vertex [dot] (c) {};
\vertex [above left=of c] (a);
\vertex [below left=of c] (b);
\vertex [above right=of c] (k);
\vertex [below right=of c] (l);
\vertex [dot, below right=of c, xshift=-6mm, yshift=6mm] (m) {};
\diagram* {
(a) -- [momentum'={[arrow shorten=0.1mm, arrow distance=2mm]\(k_2\)}] (c);
(m) -- [out=80, in=0, loop, min distance=1.4cm] (m)
(b) -- [momentum'={[arrow shorten=0.1mm, arrow distance=2mm]\(k_1\)}] (c);
(c) -- [momentum={[arrow shorten=0.1mm, arrow distance=2mm]\(p_2\)}] (k);
(c) -- [] (m) -- [momentum'={[arrow shorten=0.04mm, arrow distance=2mm]\(p_1\)}] (l);
};
\end{feynman}
\end{tikzpicture}
\caption{Diagram with both a self-interaction and an interaction between the two particles: the entire external leg must be amputated.}
\label{fourptamp}
\end{center}
\end{figure}

As a useful example, the diagram of Fig.\ref{fourptamp} should not be counted as a contribution to ${\cal M}$. In this diagram, from momentum conservation we see that the {\em internal} propagator between the two vertices carries a momentum $p_1$ which is on-shell. That would appear to give a divergent propagator. But in fact the loop is part of an external leg, and this entire external leg must be deleted. In other words we must amputate the full (and not free) external propagator.

To summarise, the prescription to compute the ${\cal T}$ or ${\cal M}$ matrix
is to keep only completely connected diagrams, amputate the full
external legs (including those diagrams where the external legs receive corrections), and only then, place the external particles on-shell. It follows that in the scalar theory we have been discussing, the diagrams that contribute to the $2\to 2$ scattering amplitude ${\cal M}$ are as shown in Fig.\ref{msecondord}:

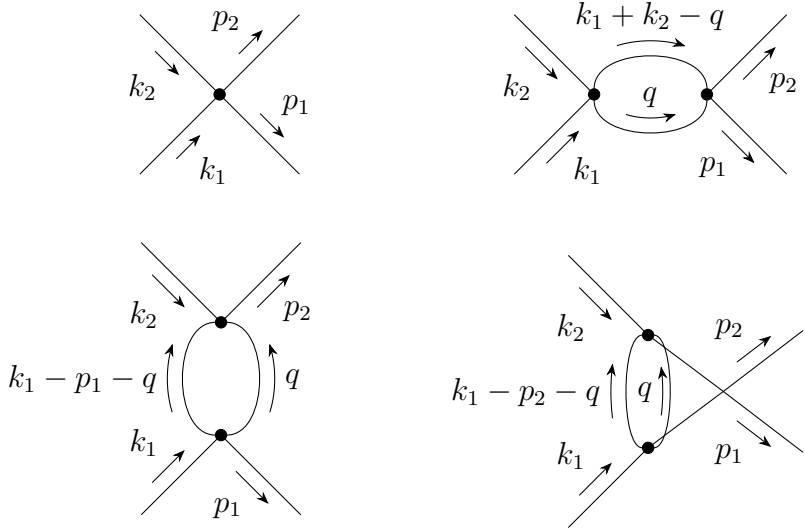
\begin{figure}[h!]
\begin{center}
\begin{tikzpicture}
\begin{feynman}
\vertex [dot] (c) {};
\vertex [above left=of c] (a);
\vertex [below left=of c] (b);
\vertex [above right=of c] (k);
\vertex [below right=of c] (l);
\diagram* {
(a) -- [momentum'={[arrow shorten=0.12mm, arrow distance=2mm]\(k_2\)}] (c);
(b) -- [momentum'={[arrow shorten=0.12mm, arrow distance=2mm]\(k_1\)}] (c);
(c) --  [momentum={[arrow shorten=0.12mm, arrow distance=2mm]\(p_2\)}] (k);
(c) -- [momentum={[arrow shorten=0.12mm, arrow distance=2mm]\(p_1\)}] (l);
};
\end{feynman}
\end{tikzpicture}
\hspace*{2cm}
\begin{tikzpicture}
\begin{feynman}
\vertex [dot] (c) {};
\vertex [above left=of c] (a);
\vertex [below left=of c] (b);
\vertex [dot, right=of c] (f) {};
\vertex [above right=of f] (k);
\vertex [below right=of f] (l);
\diagram* {
(a) -- [momentum'={[arrow shorten=0.1mm, arrow distance=2mm]\(k_2\)}] (c);
(b) -- [momentum'={[arrow shorten=0.1mm, arrow distance=2mm]\(k_1\)}] (c);
(f) -- [momentum'={[arrow shorten=0.1mm, arrow distance=2mm]\(p_2\)}] (k);
(f) -- [momentum'={[arrow shorten=0.1mm, arrow distance=2mm]\(p_1\)}] (l);
(c) -- [half left, looseness=1, momentum={[arrow shorten=0.11mm, arrow distance=2mm]\(k_1+k_2-q\)}] (f);
(c) -- [half right, looseness=1, momentum={[arrow shorten=0.11mm, arrow distance=2mm]\(q\)}] (f);
};
\end{feynman}
\end{tikzpicture}\\[6mm]
\hspace*{- 18mm}
\begin{tikzpicture}
\begin{feynman}
\vertex [dot] (c) {};
\vertex [above left=of c] (a);
\vertex [above right=of c] (b);
\vertex [dot, below=of c] (f) {};
\vertex [below left=of f] (k);
\vertex [below right=of f] (l);
\diagram* {
(a) -- [momentum'={[arrow shorten=0.1mm, arrow distance=2mm]\(k_2\)}] (c);
(c) -- [momentum'={[arrow shorten=0.1mm, arrow distance=2mm]\(p_2\)}] (b);
(k) -- [momentum={[arrow shorten=0.1mm, arrow distance=2mm]\(k_1\)}] (f);
(f) -- [momentum'={[arrow shorten=0.1mm, arrow distance=2mm]\(p_1\)}] (l);
(f) -- [half left, looseness=1, momentum={[arrow shorten=0.11mm, arrow distance=2mm]\(k_1-p_1-q\)}] (c);
(f) -- [half right, looseness=1, momentum'={[arrow shorten=0.11mm, arrow distance=2mm]\(q\)}] (c);
};
\end{feynman}
\end{tikzpicture}
\hspace*{1.3cm}
\begin{tikzpicture}
\begin{feynman}
\vertex [dot] (c) {};
\vertex [above left=of c] (a);
\vertex [above right=of c, xshift=1cm, yshift=-1cm] (b);
\vertex [dot, below=of c] (f) {};
\vertex [below left=of f] (k);
\vertex [below right=of f, xshift=1cm, yshift=1cm] (l);
\vertex (d) at (1,-0.75);
\diagram* {
(a) -- [momentum'={[arrow shorten=0.1mm, arrow distance=2mm]\(k_2\)}] (c);
(c) -- [] (d);
(d) -- [momentum'={[arrow shorten=0.1mm, arrow distance=2mm]\(p_1\)}] (l);
(k) -- [momentum={[arrow shorten=0.1mm, arrow distance=2mm]\(k_1\)}] (f);
(f) -- [] (d);
(d) -- [momentum={[arrow shorten=0.1mm, arrow distance=2mm]\(p_2\)}] (b);
(f) -- [half left, looseness=0.5, momentum={[arrow shorten=0.12mm, arrow distance=2mm]\(k_1-p_2-q\)}] (c);
(f) -- [half right, looseness=0.5, momentum={[arrow shorten=0.12mm, arrow distance=1mm]\(q\)}] (c);
};
\end{feynman}
\end{tikzpicture}
\caption{Diagrams contributing to the four-point scattering amplitude ${\cal M}$ upto second order. External propagators are drawn, but omitted from the calculation.}
\label{msecondord}
\end{center}
\end{figure}

To evaluate each such diagram, recall that we obtained the lowest-order result $i{\cal M}=-i\lambda$ from the fact that the interaction vertex carries a factor $-\frac{i}{4!}\lambda$ and there are $4!$ ways to construct the lowest-order diagram. For higher-order diagrams we simply continue assigning $-\frac{i}{4!}\lambda$ to every vertex, as well as $\frac{i}{q^2-m^2}$ for every internal line of 4-momentum $q_\mu$. Thereafter we carry out all possible contractions, integrate over all internal (loop) momenta, and identify the result with $i{\cal M}$.

The first diagram of Fig. \ref{msecondord} is trivial to compute and indeed we have already done it, the result being expressed in \eref{mtreelevel}. We would now like to compute higher-order corrections, embodied in the remaining three Feynman diagrams of this figure, each of which has one loop, as well as higher-loop diagrams. It is here that we will encounter the need for renormalisation. 

Before going on to compute loop diagrams, let us review some basics of spinor and vector fields. All that we have done so far can be extended to these fields in a relatively simple and natural way, with just a few complications to keep in mind.

\subsection{Vector fields}

A vector field is typically denoted $A_\mu(x)$. It is associated to a particle of spin 1 in
units of $\hbar$. The corresponding kinetic term in the Lagrangian is the Maxwell term:
\be
\begin{split}
{\cal L} &= - {1\over4} F_{\mu\nu} F^{\mu\nu}\\
 &= - {1\over4} (\partial_\mu A_\nu - \partial_\nu A_\mu) 
(\partial^\mu A^\nu - \partial^\nu A^\mu)  \\ 
&= - {1\over2} \partial_\mu A_\nu\partial^\mu A^\nu + {1\over2} 
\partial_\mu A_\nu \partial^\nu A^\mu
\end{split}
\label{gaugekin}
\ee
This Lagrangian is invariant under the gauge transformation:
\be
A_\mu \rightarrow A_\mu - \partial_\mu \alpha (x)
\label{freegt}
\ee
for an arbitrary function $\alpha (x)$. It describes a massless field. A mass term $m^2\,A_\mu A^\mu$ would violate gauge invariance.

Gauge invariance means that the space of gauge field configurations is
``degenerate''.  Any configuration $A_\mu (x)$ and another one $A_\mu
(x) - \partial_\mu \alpha (x)$ have the same value of the Lagrangian.  This is
not just a finite-parameter degeneracy but an infinite-parameter one,
parametrised by the space of functions $\alpha (x)$.
The physical interpretation is that the configurations
$A_\mu$, $A_\mu - \partial_\mu \alpha$ are {\em physically
equivalent}.  Thus the theory has {\em less} physical
content than it originally seemed to have. We could try to reduce it to
its physical degrees of freedom, but it can be shown that this would be at the cost of manifest Lorentz
invariance.

Physically observable quantities have to be gauge-invariant. We can compute gauge-dependent quantities, but they are to be considered as steps on the way to computing physical quantities. It turns out that off-shell amplitudes involving vector fields are gauge dependent. However, on-shell $\cS$-matrix elements, which we just introduced in the previous Section, are gauge invariant. 

Gauge invariance seems like a complicated and undesirable feature
because it introduces a redundancy in the field
configurations. However it is {\em necessary in order to reconcile Lorentz
invariance with unitarity} and is therefore considered a subtle and profound rather than undesirable feature.  

\subsection{Gauge fixing}

 Because of gauge invariance, the propagator of the field $A_\mu$ is not well-defined. The free propagator for any quantum field is obtained by inverting the operator appearing in the quadratic part of the action, but due to gauge invariance this operator is not invertible. For this reason we need to perform ``gauge fixing''. This is a procedure that explicitly imposes a supplementary condition (not included in Maxwell's equations) which breaks gauge invariance. Its role is to ensure that a single representative gets picked out of each collection. 
 
 Of course, it is essential to show at the very end that different gauge-fixing procedures do not lead to different physical systems, but rather to different mathematical descriptions of the same physical system. Calculations of physical (gauge-invariant) quantities should yield the same result, independent of the method of gauge-fixing.

\exercise{After integrating by parts, write down the action associated to the Maxwell Lagrangian of \eref{gaugekin} in the form:
$$
\int d^4x d^4y A^\mu(x){\cal D}_{\mu\nu}(x,y)A^\nu(y)
$$
where ${\cal D}$ is a differential operator in position space as well as a matrix in $\mu,\nu$ space. Show that ${\cal D}$  is not invertible.}

We now turn to the discussion of different methods of gauge-fixing, or equivalently different ``choices of gauge''. 

\subsubsection*{Coulomb gauge}

Coulomb gauge is a particularly intuitive gauge. Given any configuration $A_\mu (t,\vec x)$, let us single out the time component of the gauge field and define a gauge parameter:
\be
\alpha(t,\vec x) = \int^t A_0 (t',\vec x)\, dt'
\ee
If we perform a gauge transformation with this parameter we see that:
\be
A_0\to A_0 - \partial_0 \alpha = 0
\ee
Thus by a gauge transformation we can make $A_0$ disappear from the theory altogether. We
see that at least one component of the field is a ``gauge artifact'', rather than a physical field component that needs to be quantised.

All the gauge freedom has not yet been fixed. We can perform a further gauge transformation involving a parameter
$\alpha(\vec x)$ that is independent of time but otherwise arbitrary. After such a transformation, $A_0$ continues to vanish while $A_i$ changes:
\be
\begin{split}
A_0 &\rightarrow A_0 - \partial_0 \alpha(\vec x) = 0  \\
A_i &\rightarrow A_i - \partial_i \alpha(\vec x)  
\end{split}
\ee
We see that there remains a gauge
redundancy in the spatial components $A_i$.

To see how this can be fixed, consider the equations of motion: $\partial^\mu F_{\mu\nu} =
0$. If we choose the free index $\nu$ to be 0, we get:
\be
\partial^i F_{0i} = 0
\ee
Since we have set $A_0=0$, this implies
\be
\partial_0 (\partial^i\! A_i) = 0
\label{eomcoul}
\ee
so the spatial divergence of $A_i$ is time independent. Therefore if we set this to zero at a particular instant of time, it will remain zero forever. Accordingly we choose a  gauge parameter $\alpha(\vec x)$ satisfying $\del^i\del_i \alpha(\vec x)=\del^i A_i$. Performing a gauge transformation with this parameter sets $\del^iA_i=0$. Thus Coulomb gauge is defined by $A_0=0$ together with $\del^i A_i=0$.

Although this gauge can formally be reached (at least locally, though there can be subtle global issues) as above, 
implementing and working with it poses some challenges. The easy part is that we can set $A_0=0$ by hand, after
which the Maxwell equations simplify to:
\be
\square A_j = 0
\label{boxvec}
\ee
However the other constraint $\del^i\! A_i = 0$ involves a derivative on the field. Hence it must either be solved to eliminate one component (which involves an integration and thereby renders the theory nonlocal), or be retained and imposed by hand on physical quantities. In short, in Coulomb gauge, classical vector field configurations consist of solutions to \eref{boxvec} (i.e. three copies of the Klein-Gordon equation) subject to this latter constraint.

Since we eliminated $A_0$ using gauge transformations, it is clear that
there are no negative norm states. The three remaining fields $A_i$ can be
canonically quantised in terms of oscillators, all of which have
positive norm. The constraint $\del^i\!A_i=0$ reduces the physical Hilbert space built out of these oscillators to a subspace generated by only two independent oscillators. These are the two well-known physical polarisations of the photon. In this way, gauge invariance reduces four components of the vector field $A_\mu$ (required by special relativity) to two physical degrees of freedom (required by experiment) -- a miracle indeed!

\subsubsection*{Covariant gauges}
\label{lorgauge}

It is sometimes convenient to avoid breaking manifest Lorentz invariance, though there is a price to pay for that. To illustrate this point let us consider a method of gauge-fixing that respects Lorentz invariance. This is to impose the 
condition $\partial^\mu\! A_\mu (x) = 0$, called the Lorenz
gauge condition. 

In this gauge, the Lagrangian  becomes:
\be
{\cal L} = - {1\over2} (\partial_\mu A_\nu)^2 
= - {1\over2} (\partial_\mu A_0)^2 + {1\over2} \partial_\mu A_i 
\partial^\mu A_i 
\label{lorgaugecond}
\ee
So the $A_i$ behave like three massless Klein-Gordon fields. But in this gauge there is also $A_0$ to worry about. In fact, it now has a canonical
momentum (which it did not have before gauge-fixing). We see that:
\be
\pi^0 = \frac{\delta {\cal L}}{\delta \dot A_0}=-\dot  A_0
\ee
We will treat this field along with the three others as a valid
dynamical variable, and at the end, the constraint $\del^\mu\! A_\mu$ will be imposed. We will see in a moment that this removes {\em two} degrees of freedom, leaving us with two just as in Coulomb gauge.

The equations of motion arising from \eref{lorgaugecond} are: 
\be
\square A_\nu = 0 
\ee
As usual, we must check what residual gauge invariance is present
after fixing $\partial^\mu\! A_\mu = 0$. Sending $A_\mu \rightarrow
A_\mu - \partial_\mu \alpha$, we see that this preserves the gauge if
$\partial^\mu \partial_\mu \alpha = 0$. This is just the
equation of motion of a free massless scalar field.

The situation simplifies considerably in momentum space.
Let $\tilde A_\mu (k)$ be the Fourier transform of $A_\mu (x)$.  
The gauge condition becomes $k^\mu \tilde A_\mu (k) = 0$, and the 
residual gauge invariance becomes
\be
\tilde A_\mu (k) ~\to~ \tilde A_\mu (k) + ik_\mu \tilde \alpha (k)
\ee
where $\tilde\alpha$ is any function of $k$, subject to the condition
that $k^\mu k_\mu{\tilde \alpha}(k)=0$. This is satisfied if $k^2=0$, which is the case for an on-shell massless particle. Thus we see that the Lorenz gauge has two features:\\[2mm]
(i) a {\em condition}: $k^\mu \tilde A_\mu (k) ~\sim~ 0$,\\[2mm]
(ii) a {\em redundancy}: $\tilde A_\mu (k) \sim \tilde A_\mu (k) + 
ik_\mu {\tilde\alpha}(k)$,\\[2mm]
where the $\sim$ symbol means the configurations ${\tilde A}_\mu(k)$
and ${\tilde A}_\mu(k)+ik_\mu {\tilde \alpha}(k)$ are {\em
identified}\footnote{Note that ${\tilde A}_\mu$ is complex and satisfies ${\tilde A}^*_\mu(k)={\tilde A}_\mu(-k)$ due to the reality of $A_\mu(x)$.}. 

The above features guarantee that two  of the four possible
polarisations decouple from the theory.  To see this, 
consider the mode expansion of $A_\mu$:
\be
A_\mu (t,\vec x) = \int {d^3k \over (2\pi)^3} {1 \over \sqrt{2\omega_{\vec k}}}
\left(a_{\mu,\vec k}\, e^{-i{k\cdot x}} + a^\da_{\mu,\vec k}\, e^{i{k\cdot x}}
\right)
\ee
and quantise $a_{\mu,\vk},a^\da_{\mu,\vk}$ as usual by commutators.
Now consider the four states $a^\da_{\mu,\vec k}|0\rangle$. Naively these would all be one-particle states, but we need to impose the above gauge conditions on them. The gauge conditions imply that:
\be
\begin{split}
k^\mu a^\da_{\mu,\vec k}|0\rangle &\sim 0\nn\\
a^\da_{\mu,\vec k}|0\rangle &\sim a^\da_{\mu,\vec k}|0\rangle + i k_\mu\, 
\alpha_{\vec k}|0\rangle
\end{split}
\label{gaugecond}
\ee
where $\alpha_\vk$ is a mode of $\alpha$.

To find the physical states, consider the most general linear combination of the states $a^\da_{\mu,\vec k}|0\rangle$:
\be
\epsilon^\mu a^\da_{\mu,\vec k}|0\rangle
\label{genlin}
\ee
From the first condition in \eref{gaugecond} we see that one of these states, the one where $\epsilon^\mu$ is proportional to $k^\mu$, is unphysical. The second condition tells us that in addition, $\epsilon^\mu k_\mu=0$.

We can solve these conditions explicitly by choosing a convenient Lorentz frame. Since $k_\mu$ is a light-like vector, we can choose a frame in which $k_\mu = (k,0,0,k)$ for some number $k$. Raising the index using the Minkowski metric we have $k^\mu = 
(k,0,0,-k)$. Now the first condition gives:
\be
\left(a^\da_{0,\vec k}- a^\da_{3,\vec k}\right)|0\rangle =0
\ee
which means that
\be
a^\da_{0,\vec k}|0\rangle ~\sim~ a^\da_{3,\vec k}|0\rangle 
\ee
where again $\sim$ means we {\em identify} the two states.
The second condition gives: 
\be
\epsilon^0 + \epsilon^3 = 0
\ee
Therefore the states we are considering become
\be
\begin{split}
&\epsilon^0 a^\da_{0,\vec k}|0\rangle + \epsilon^1 a^\da_{1,\vec k}|0\rangle + 
\epsilon^2 a^\da_{2,\vec k}|0\rangle + \epsilon^3 a^\da_{3,\vec k}|0\rangle\\
&\qquad\qquad=\epsilon^0 \left(a^\da_{0,\vec k}|0\rangle - a^\da_{3,\vec k}|0\rangle\right) 
+ \epsilon^1 a^\da_{1,\vec k}|0\rangle + \epsilon^2 a^\da_{2,\vec k}|0\rangle 
 \\ 
&\qquad\qquad\sim \epsilon^1 a^\da_{1,\vec k}|0\rangle + \epsilon^2 a^\da_{2,\vec k}|0\rangle 
\end{split}
\ee
since $a^\da_{0,\vec k}|0\rangle - a^\da_{3,\vec k}|0\rangle \sim 0$.
Thus at the end we have two linearly independent states
\be
a^\da_{1,\vec k}|0\rangle, \  a^\da_{2,\vec k}|0\rangle
\label{twolin}
\ee
describing a transverse photon. We can equivalently say that an orthonormal basis for possible photon polarisations is given by the vectors:
\be
\epsilon^{(1)}_\mu = (0,1,0,0),\qquad\epsilon^{(2)}_\mu = (0,0,1,0)
\ee
in the frame where the photon momentum is proportional to $(1,0,0,1)$. In any other frame one just has to perform a Lorentz transformation to get the new polarisation vectors. The above vectors are defined by the following set of conditions:
\be
k\cdot \epsilon^{(i)}=0,\qquad \epsilon^{(i)}\nsim k,\qquad\epsilon^{(i)}\cdot\epsilon^{(j)}=-\delta^{ij}
\ee
These conditions are Lorentz-invariant and therefore hold in all frames.

We have proved the well-known fact that an electromagnetic wave oscillates transverse to its direction of propagation.
The special feature of the Lorenz gauge condition is that it preserves Lorentz invariance. Only in {\em implementing} it did we choose a Lorentz frame for $k_\mu$ and obtain a corresponding basis for the polarisation vectors. Note that when we convert Feynman diagrams to scattering matrix elements, we will have to include the polarisation vectors for each external vector field line. 

Instead of imposing the Lorenz gauge condition, it is more convenient to study a family of covariant gauges. These are obtained upon modifying the Maxwell Lagrangian by a term proportional to $(\del^\mu A_\mu)^2$, which breaks gauge invariance and depends on a real parameter $\xi$. The gauge-fixed Lagrangian is then:
\be
{\cal L}=-\frac14 F_{\mu\nu}F^{\mu\nu}-\frac{1}{2\xi}(\del^\mu A_\mu)^2
\label{gaugefix}
\ee
This is known as ``choosing the ``$R_\xi$ gauge''.
 
Since gauge invariance is broken, one can now invert the kinetic term and obtain a propagator. It is easy to establish that, in momentum space, this is:
\be
\langle 0|{\tilde A}_\mu(k){\tilde A}_\nu(0)|0\rangle \equiv i{\tilde \Pi}_{\mu\nu}(k)=-i\frac{\eta_{\mu\nu}-(1-\xi)\frac{k_\mu k_\nu}{k^2}}{k^2+i\epsilon}
\label{covprop}
\ee
In the limit $\xi\to 0$ one effectively implements the Lorenz gauge, since configurations not satisfying the gauge condition are completely suppressed. In this situation the propagator is:
\be
i{\tilde \Pi}_{\mu\nu}(k)=-i\frac{\eta_{\mu\nu}-\frac{k_\mu k_\nu}{k^2}}{k^2+i\epsilon}
\ee
which is easily seen to be transverse: $k^\mu {\tilde \Pi}_{\mu\nu}=0$. This is known as the Landau gauge propagator. 

For nonzero $\xi$, the gauge-fixed Lagrangian implements a weighted sum over different gauge conditions. It has the effect of breaking gauge invariance in a controlled way. A useful finite value is $\xi=1$, known as the Feynman gauge. Here the propagator simplifies greatly:
\be
i{\tilde \Pi}_{\mu\nu}(k)=-i\frac{\eta_{\mu\nu}}{k^2+i\epsilon}
\ee
As indicated above, one can use whichever gauge is convenient. Feynman gauge is convenient for its simplicity, while Landau gauge is convenient due to transversality of the propagator. 

\exercise{Derive the gauge-fixed propagator \eref{covprop}.}

\subsection{Spinor fields}

A spinor field is denoted $\psi_\alpha(x)$, though the spinor index is usually suppressed. The kinetic term includes a mass and is written:
\be
{\cal L} = \overline\psi(i\dsl - m)\psi
\label{diracac}
\ee
Since several indices are implicit, just for once we write the above expression explicitly:
\be
{\cal L} = \psi^\da_\alpha \gamma^0_{\alpha\beta} \left(i\gamma^\mu_{\beta\gamma} \partial_\mu - 
m \delta_{\beta\gamma}\right) \psi_\gamma
\ee
The free propagator for a spinor field is:
\be
\langle 0|T\left(\psi_\alpha(x) \overline\psi_\beta(y)\right)|0\rangle \equiv (S_F (x-y))_{\alpha\beta} = (i\gamma^\nu \partial_\nu + m)_{\alpha\beta} D_F(x-y). 
\ee
Using the momentum representation of $D_F(x-y)$, we have:
\be
S_F (x-y) = \int {d^4k \over (2\pi)^4}\, {i(\ksl +m) 
\over k^2 - m^2 + i\epsilon}\, e^{-ik\cdot(x-y)}
\ee
where we have used the notation $\ksl=\gamma^\mu k_\mu$.
It can be directly checked that this satisfies the desired equation and 
boundary conditions for a Feynman propagator.

Unlike vector fields, spinor fields do not introduce any essential new complication like gauge invariance. However since they are multi-component fields, we have to insert suitable external spinors when computing amplitudes with external spinor lines, as we will see shortly.

\subsection{Quantum Electrodynamics}

 We are now ready to couple vector and spinor fields, to make the theory of quantum electrodynamics.  The Lagrangian density is:
\be
{\cal L} = -{1\over4} F_{\mu\nu} F^{\mu\nu} + i\overline\psi \gamma^\mu 
(\partial_\mu+ieA_\mu) \psi - m \overline\psi \psi-\frac{1}{2\xi}(\del_\mu A^\mu)^2
\label{qedlag}
\ee
We have included the $R_\xi$ gauge-fixing term. Without this term, the rest of the Lagrangian is invariant under the gauge transformation: 
\be
A_\mu\to A_\mu -\frac{1}{e}\del_\mu \alpha(x), \quad \psi\to e^{i\alpha(x)}\psi
\ee
We now carry out perturbation theory in terms of the free fields $\psi,A_\mu$ treating the interaction term $e\overline\psi\gamma^\mu A_\mu \psi$ as a perturbation.

\subsection{Feynman rules}
\label{qedfeyn}

We start by formulating the Feynman rules for this theory. The
momentum-space propagators have already been derived:
\bea
\psi: &&\qquad \frac{i(\ksl+m)}{k^2-m^2+i\epsilon}\nn\\[2mm]
A_\mu: &&\qquad -i\,\frac{\eta_{\mu\nu}-(1-\xi)\frac{k_\mu k_\nu}{k^2}}{k^2+i\epsilon}\nn
\label{fermphotprop}
\eea
We need to put polarisation factors on the
external photon lines, as shown in Fig. \ref{extphot}. Similarly we need to put in external polarisation factors for spinors, as shown in Fig. \ref{extferm}.

\begin{figure}[H]
\begin{center}
\begin{tikzpicture}
\begin{feynman}
\vertex (a) {\(\mu\)};
\vertex [right=of a] (b);
\vertex [above right=of b] (c);
\vertex [below right=of b] (d);
\node [left=of a] {\(\overbracket[0.5pt][6pt]{\!\!A_\mu|\vp\,\rangle\!\!}~~=\)};
\diagram* {
(a) -- [boson, momentum=\(p\)] (b);
(b) -- [fermion] (c);
(d) -- [fermion] (b);
};
\node [right=of b] {\(=~~\epsilon_\mu(p)\)};
\end{feynman}
\end{tikzpicture}\\[4mm]
\begin{tikzpicture}
\begin{feynman}
\vertex (a) {\(\mu\)};
\vertex [left=of a] (b);
\vertex [above left=of b] (c);
\vertex [below left=of b] (d);
\node [left=of b] {\(\overbracket[0.5pt][6pt]{\!\!\langle \vp\,|A_\mu\!\!}~~=\)};
\diagram* {
(b) -- [boson, momentum=\(p\)] (a);
(b) -- [fermion] (c);
(d) -- [fermion] (b);
};
\node [right=of a] {\(=~~\epsilon^*_\mu(p)\)};
\end{feynman}
\end{tikzpicture}
\caption{Polarisation factors for external photon lines.}
\label{extphot}
\end{center}
\end{figure}

\begin{figure}[H]
\begin{center}
\begin{tikzpicture}
\begin{feynman}
\vertex (a) {};
\vertex [right=of a] (b);
\vertex [above right=of b] (c);
\vertex [below right=of b] (d);
\node [left=of a] {\(\overbracket[0.5pt][6pt]{\!\!\psi|\vp\,s\rangle\!\!}~{}_{\rm fermion}~~=~~~~~~\)};
\diagram* {
(a) -- [fermion, momentum=\(p\)] (b);
(b) -- [fermion] (d);
(b) -- [boson] (c);
};
\node [right=of b] {\(=~~u^s(p)\)};
\end{feynman}
\end{tikzpicture}\\[4mm]
\begin{tikzpicture}
\begin{feynman}
\vertex (a) {};
\vertex [left=of a] (b);
\vertex [above left=of b] (c);
\vertex [below left=of b] (d);
\node [left=of b] {\({}_{\rm fermion\,}\overbracket[0.5pt][6pt]{\!\!\langle \vp,s | {\overline\psi}\!\!}~~~=\)};
\diagram* {
(b) -- [fermion, momentum=\(p\)] (a);
(c) -- [fermion] (b);
(d) -- [boson] (b);
};
\node [right=of a] {\(=~~{\bar u}^s(p)\)};
\end{feynman}
\end{tikzpicture}
\caption{Polarisation factors for external fermion lines.}
\label{extferm}
\end{center}
\end{figure}

\begin{figure}[H]
\begin{center}
\begin{tikzpicture}
\begin{feynman}
\vertex (a) {};
\vertex [right=of a] (b);
\vertex [above right=of b] (c);
\vertex [below right=of b] (d);
\node [left=of a] {\(\overbracket[0.5pt][6pt]{\!\!{\overline\psi}|\vp\,s\rangle\!\!}~{}_{\rm anti-fermion}~~=~~~~~~\)};
\diagram* {
(b) -- [fermion, rmomentum=\(p\)] (a);
(d) -- [fermion] (b);
(b) -- [boson] (c);
};
\node [right=of b] {\(=~~{\bar v}^s(p)\)};
\end{feynman}
\end{tikzpicture}\\[4mm]
\begin{tikzpicture}
\begin{feynman}
\vertex (a) {};
\vertex [left=of a] (b);
\vertex [above left=of b] (c);
\vertex [below left=of b] (d);
\node [left=of b] {\({}_{\rm anti-fermion\,}\overbracket[0.5pt][6pt]{\!\!\langle \vp,s | {\psi}\!\!}~~~=\)};
\diagram* {
(a) -- [fermion, rmomentum=\(p\)] (b);
(b) -- [fermion] (c);
(d) -- [boson] (b);
};
\node [right=of a] {\(=~~v^s(p)\)};
\end{feynman}
\end{tikzpicture}
\caption{Polarisation factors for external anti-fermion lines.}
\label{extanti}
\end{center}
\end{figure}

Finally, the interaction vertex of QED is just the $\gamma$-matrix, depicted in Fig. \ref{intvert}.
\begin{figure}[H]
\begin{center}
\begin{tikzpicture}
\begin{feynman}
\vertex (a) {};
\vertex [dot, right=of a] (b) {};
\vertex [above right=of b] (c);
\vertex [below right=of b] (d);
\diagram* {
(a) -- [boson] (b);
(b) -- [fermion] (c);
(d) -- [fermion] (b);
};
\node [right=of b, xshift=1.8cm] {\(=~~-ie\gamma^\mu\)};
\end{feynman}
\end{tikzpicture}
\caption{The QED vertex}
\label{intvert}
\end{center}
\end{figure}
Depending on where this appears in a diagram, it can describe an
electron emitting a photon, or an electron absorbing a photon, or a
positron emitting a photon, or a positron absorbing a photon, or an
electron-positron pair annihilating into a photon, or a photon
dissociating into an electron-positron pair. 

If the photon momentum is chosen to be $(k,0,0,k)$, then transversality tells us that a basis for $\epsilon^r_\mu(k)$ is $(0,1,0,0)$ and $(0,0,1,0)$. These correspond to plane polarisation in the $x$ or $y$ direction. An alternative basis for $\epsilon^r_\mu(k)$ is $(0,1,i,0)$ and $(0,1,-i,0)$ which corresponds to circularly polarised light. 

\newpage

\section{Quantum field theory at one loop}
\label{qftoneloop}

\subsection{The logic of renormalisation}
\label{logicren}

Our goal is to compute corrections to $\cS$-matrix elements for quantum field theories. In the case of $2\to 2$ scattering in scalar field theory, to lowest order these are formally embodied in the last three diagrams of Fig. \ref{msecondord}. We will see in a moment that the loop integrals in these diagrams are divergent. But before we learn how to deal with the divergences, let us ask an important physical question.  How would we experimentally determine the parameter $\lambda$ in the Lagrangian? If we are working at tree level, the answer is easy. We determine the matrix element $\cM$ from experiment and then equate it to $\lambda$ using \eref{mtreelevel}. Until we do this, we cannot predict any experimental result starting from the theory. Once we have determined the value of $\lambda$ from {\em one} experiment, we can use field theory to predict the matrix element $\cM$ for any {\em other} experiment. This is when quantum field theory becomes predictive. 

The question now is, how do we determine $\lambda$ when we are working not at tree level but, say, at one-loop order in perturbation theory? The relation between $\cM$ and $\lambda$ will now be of the form:
\be
\cM=-\lambda + a\lambda^2
\label{Mcorrec}
\ee
for some computed quantity $a$, and this must be solved to determine $\lambda$ from the experimental measurement of $\cM$. But this means $\lambda$ can no longer be considered to be the strength of the interaction between two $\phi$-particles. Indeed, nature considers $\cM$ to be the coupling strength, while the parameter $\lambda$ in the Lagrangian has no direct physical meaning! The situation is complicated by two additional facts: (i) the quantity $a$ is divergent, (ii) the quantity $a$ depends non-trivially on external momenta. We will address both these issues shortly. But it is worth appreciating that even without either of these complications, the parameters in a Lagrangian are not directly equal to the result of a measurement, despite our careless habit of giving them names like ``charge'' and ``mass''. 

It is a little artificial to talk of experiments testing Scalar Field Theory (SFT), which is only a small sector of the entire Standard Model of Physics that describes self-couplings of the Higgs particle. Much more realistic would be to consider the theory of Quantum Electrodynamics (QED). Not only is this a larger sector of the Standard Model, but it is realistic to study this sector all by itself because we can consider processes (like electron-electron, electron-positron and electron-photon scattering) where the effects of both weak and strong interactions are negligible -- the first because they are weak, and the second because electrons do not participate. Unfortunately, due to the fact that vectors and spinors have multiple components that transform non-trivially among themselves under the the Lorentz group, QED is a lot more technically complicated than SFT. Accordingly, we will focus on both SFT and QED in what follows, but will prefer to talk about QED when comparison with experiments is invoked.

The Lagrangian density of QED for a single fermion species, which we usually take to be the electron/positron, has been written in \eref{qedlag}. Let us re-write it here (without the gauge-fixing term, for the moment):
\be
{\cal L}_{\rm QED}= -\frac14 F_{\mu\nu}F^{\mu\nu} +i\psibar\gamma^\mu\del_\mu\psi-m\psibar\psi-e\psibar\gamma^\mu\psi A_\mu
\ee
The analogous question to the one we asked above for SFT is: how do we determine the parameter $e$ (and also $m$) in this Lagrangian? These are commonly referred to as the electric charge and mass of the electron, respectively, but just as we saw in scalar field theory, here too this identification is not justified once we go beyond lowest-order perturbation theory.

As before, we select a matrix element $\cM$,  for example the one corresponding to $e^-e^-$ scattering, and measure it in an experiment\footnote{This is somewhat loose terminology because we really measure cross-sections rather than amplitudes.}. We also compute it in terms of the parameters $e$ and $m$ in the Lagrangian, and compare. This scattering proceeds by exchange of a virtual photon in the $t$ channel:

\begin{figure}[H]
\begin{center}
\begin{tikzpicture}
\begin{feynman}
\vertex (a) ;
\vertex [above=of a] (b);
\vertex [below=of a] (c);
\vertex [right=of a] (d);
\vertex [above=of d, yshift=-8mm] (e);
\vertex [below=of d, yshift=8mm] (f);
\vertex [right=of d] (g);
\vertex [above=of g] (h);
\vertex [below=of g] (i);
\diagram* {
(b) -- [fermion] (e) -- [fermion] (h);
(c) -- [fermion] (f) -- [fermion] (i);
(e) -- [boson, momentum=\(p\)] (f)
};
\end{feynman}
\end{tikzpicture}
\end{center}
\caption{$t$-channel exchange diagram.} 
\label{tchannel}
\end{figure}
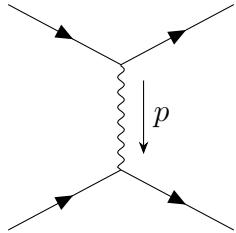

\noindent and one finds:
\be
\cM=e^2 f_1(\vp,m)+e^4 f_2(\vp,m)+\cdots
\ee
To gain a little insight, let us step back and recall how electromagnetism was understood in the days preceding QED. The Coulomb potential $V(r)$ between static particles (say, electrons) separated by a distance $r$ was measured to fall off like $\frac{1}{r}$. Hence the physical electric charge $e_{\rm phys}$ of the electron was defined via:
\be
V(x)=\frac{e_{\rm phys}^2}{4\pi r}
\ee
If we Fourier-transform this to momentum space, we find:
\be
\wV(\vcp)=\int d^3x~ e^{-i\vcp\cdot\vx}~\frac{e_{\rm phys}^2}{4\pi r}=\frac{e_{\rm phys}^2}{\vcp^{\,\,2}}
\label{coulf}
\ee
The question then is, what relation does $e_{\rm phys}$ bear to the parameter $e$ in the Lagrangian of \eref{qedlag}?

At tree level, the $t$-channel exchange diagram for $e^-e^-$ scattering tells us that:
\be
\cM\sim \frac{e^2}{p^2}\times \hbox{spinor factors}
\ee
where $p=p_3-p_1$ is the virtual momentum exchanged in the $t$-channel. If we write:
\be
p_1=(E_1,\vp_1), \qquad p_3=(E_3,\vp_3)
\ee
then in the limit of backward scattering we have $E_1=E_3$ and $\vp_1=-\vp_3$, so we have:
\be
\cM\sim \frac{e^2}{\vp^{\,\,2}}
\label{mqedtree}
\ee
This is precisely the Fourier transform of Coulomb's law upto an overall normalisation. Hence we can interpret the above equation as giving \eref{coulf}, with the identification $e_{\rm phys}=e$ to this order. However, as soon as we go to one-loop order, the matrix element for the same process turns out to be:
\be
\cM\sim \frac{e^2+a\,e^4}{\vp^{\,\,2}}
\ee
and we no longer have any reason to identify $e_{\rm phys}$ with $e$. The 	quantity $a$ is computed from a momentum integral (we will study these in great detail shortly) and turns out to be both momentum dependent and divergent. The divergence comes from the ultraviolet end of the momentum integral. 

Equating the RHS above to the one-loop corrected Coulomb potential:
\be
\wV(\vcp)_{\rm one-loop}= \frac{e^2+a\,e^4}{\vp^{\,\,2}}
\label{mqedloop}
\ee
we would conclude that:
\be
e_{\rm phys}^2=e^2+a\,e^4
\ee
with a divergent and momentum-dependent coefficient $a$. 

The situation looks rather hopeless! But we press on by supposing that a momentum cutoff $\Lambda$ is imposed on the integral. This renders $a$ finite but $\Lambda$-dependent. The original ultraviolet divergence will re-appear as soon as we try to take $\Lambda\to\infty$. And we must eventually take this limit, unless our theory has an in-built momentum cutoff for a physical reason (which it does not, in the case of QED or the Standard Model). However we hold off and perform some useful manipulations while keeping $\Lambda$ finite, and then take $\Lambda\to\infty$ at the very end.

With finite $\Lambda$, \eref{mqedloop} is replaced by:
\be
\wV(\vcp)_{\rm one-loop}=\frac{e^2+a(\vp,\Lambda)e^4}{\vp^{~2}\!}
\label{mqedcut}
\ee
If we use this to determine the physical value of the electric charge, we will get a variable answer depending on the momentum $\vp$ at which we are working. So let us pick a particular value of the exchanged momentum, say $\vp={\vec 0}$, and define:
\be
e_R^2=e^2+a({\vec 0},\Lambda)e^4
\ee
On the left hand side we have used the suffix $R$ for ``renormalised''. Now we invert the above equation to get:
\be
e^2=e_R^2-a({\vec 0},\Lambda)e_R^4
\label{ebare}
\ee
Recall that we were working in perturbation theory in $e$. But since at lowest order $e$ and $e_R$ are the same, the inverted relation is correct to the same order as the original one. This is why we did not keep higher-order terms in the above equation -- it would be inconsistent to do so.

Finally, inserting this into \eref{mqedcut} we get, to the order of perturbation theory in which we are working:
\be
\cM=\frac{e_R^2+\Big(a(\vp,\Lambda)-a({\vec 0},\Lambda)\Big)e_R^4}{\vp^{~2}}
\label{mrenorm}
\ee
This is a key result. Notice that there is a possibility of the divergences cancelling within the bracket. As an example, if 
\be
a\sim \log\frac{\vp^{~2}\!+m^2}{\Lambda^2}+{\cal O}\left(\frac{1}{\Lambda^2}\right)
\ee
 then we easily see that:
\be
\lim_{\Lambda\to\infty}\Big(a(\vp,\Lambda)-a({\vec 0},\Lambda)\Big)=\log\frac{\vp^{~2}+m^2}{m^2}
\ee
which is finite. The specific form for $a$ in the above example was simply suggested out of the blue, but we will see that such expressions arise quite generically at one loop for matrix elements in both SFT and QED.

To summarise, the matrix element $\cM$ has been calculated in terms of a finite renormalised coupling constant $e_R$, which in turn is determined by measuring $\cM$ at a particular value of the exchanged momentum, namely $\vp=0$. The equation \eref{mrenorm} is predictive, since it gives us the momentum-dependent value of the measurable matrix element $\cM$ in terms of a single measured parameter $e_R$. Similar considerations can be used to define a renormalised mass $m_R$, which would play a role in other matrix elements.

Notice how the parameter $e$ in the Lagrangian was simply a stepping stone to determine the dependence of $\cM$ on $e_R$. Thus it is incorrect to say that the original parameter $e$ is the electron charge $\sim 1.6\times 10^{-19}$ Coulomb, nor is the parameter $m$ equal to the electron mass $\sim 9.1\times 10^{-31}$ kg. 
The lesson we learn is that {\em parameters in the Lagrangian of a QFT are not physical quantities}. Moreover, since $e_R$ will be related to measured quantities, it must be cutoff-independent. Hence we interpret \eref{ebare} as saying that the parameter $e$ in the Lagrangian is cutoff-dependent. This will be elevated to a general philosophy: parameters in the Lagrangian are cutoff-dependent, while renormalised parameters are cutoff-independent. The latter are calibrated from experiment and determine physical quantities.

\subsection{Scalar field theory at one loop}

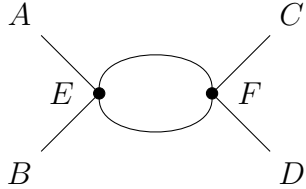
\begin{figure}[H]
\begin{center}
\begin{tikzpicture}
\begin{feynman}
\vertex [dot] (c) {};
\vertex [above left=of c] (a) {$A$};
\vertex [below left=of c] (b) {$B$};
\vertex [dot, right=of c] (f) {};
\vertex [above right=of f] (k) {$C$};
\vertex [below right=of f] (l) {$D$};
\node [left=of c, xshift=10mm] {$E$};
\node [right=of f, xshift=-10mm] {$F$};
\diagram* {
(a) -- [] (c);
(b) -- [] (c);
(f) -- [] (k);
(f) -- [] (l);
(c) -- [half left, looseness=1] (f);
(c) -- [half right, looseness=1] (f);
};
\end{feynman}
\end{tikzpicture}
\caption{A one-loop diagram with momenta suppressed and vertices labelled to facilitate counting.}
\label{symfact}
\end{center}
\end{figure}

Now we return to scalar field theory and actually calculate the diagrams that determine the $2\to 2$ $\cS$-matrix up to one loop. The diagrams in question are displayed in Figure \ref{msecondord}. The numerical factor for each diagram can be computed in the following way. Let us calculate it for the second diagram on the RHS, shown again in Fig. \ref{symfact}.

To create this, we have to bring down the interaction Hamiltonian twice from the exponential, which give a factor of $\half$ times the square of the coefficient of $\phi^4$. Next we contract points in all ways that lead to this diagram. Starting from the external point $A$, we must join it to one of the internal points $E,F$ (2 possible choices) and then pick one of the lines emanating from it (4 possible choices). Next the point $B$ has to attach to the same internal point, from which there are 3 ways corresponding to the remaining three lines coming out of it. The external point $C$ now has 4 choices and $D$ has 3 choices. Finally, $E$ and $F$ are left with two lines each coming out, which can be contracted in two distinct ways. 
Thus we have:
\be
\half\times \left(-i\frac{\lambda}{4!}\right)^2\times 2\times 4\times 3\times 4\times 3\times 2=-\frac{\lambda^2}{2}
\ee
The same factor will arise for the other two one-loop diagrams as well.

Next, let us define the integral $I(p)$ as:
\begin{equation}
I(p)=\int \frac{d^4 k}{(2\pi)^4}\left(\frac{1}{k^2-m^2+i\epsilon}\right)\left(\frac{1}{(p-k)^2-m^2+i\epsilon}\right)
\end{equation}
It is easily seen that the 4 point function, up to one-loop order, is: 
\begin{equation}
\begin{split}
i\mathcal{M}&=\left(-i\frac{\lambda}{4!}\right)\cdot4!+ \frac{1}{2}\left(-i\frac{\lambda}{4!}\right)^2 \cdot (4!)^2 \Big(i^2 I(p_1+p_2)+i^2 I(p_1-p_3)+i^2 I(p_1-p_4)\Big)\\&
=-i\lambda+\frac{\lambda^2}{2}\Big( I(p_1+p_2)+I(p_1-p_3)+I(p_1-p_4)\Big)
\end{split}
\label{Mmatrix}
\end{equation}
This has the form of \eref{Mcorrec}, as promised. The 4! and $(4!)^2$ factors in the numerator of different terms in the expression of $i\mathcal{M}$ come from symmetry factors. The extra $i^2$ factor before each integral comes because propagators are defined with an $i$ factor.

The job now  is to evaluate $I(p)$. It is convenient to  use the ``denominator trick'' which combines multiple denominators in integrands into a single one:
\begin{equation}
\frac{1}{AB}=\int_{0}^{1} \frac{dx}{\big[A+(B-A)x\big]^2}
\end{equation}
This can be proved easily by a change of variable, $y=A+(B-A)x$.
On using this identity, we find:
\begin{equation}
\begin{split}
\label{Ip}
I(p)&=\int_{0}^{1}dx\int\frac{d^4 k}{(2\pi)^4}\frac{1}{\big(k^2-m^2+x[(p-k)^2-k^2]~\big)^2}\\&
=\int_{0}^{1}dx\int\frac{d^4 k}{(2\pi)^4}\frac{1}{\big(\,(k-xp)^2-x^2 p^2-m^2+xp^2\big)^2}\\&
=\int_{0}^{1}dx\int\frac{d^4 k}{(2\pi)^4}\frac{1}{(k^2-\Delta)^2}
\end{split}
\end{equation}
where 
\be
\Delta\equiv m^2-p^2x(1-x)
\label{Deltadef}
\ee
Notice that $\Delta>0$ if $p^2< 4m^2$. Along the way, we made the shift $k^\mu\to k^\mu+xp^\mu$. This seems illegitimate: if the limits of integration are infinite then the answer is divergent, while if the integration is done with a momentum cutoff then the shift will also shift the cutoff. However, at the very end we are going to take the cutoff to infinity in a way that leaves behind a finite result, so we assume such a shift is permitted.

\begin{figure}[H]
	\begin{center}
		\resizebox{9cm}{!}{$
			\includegraphics{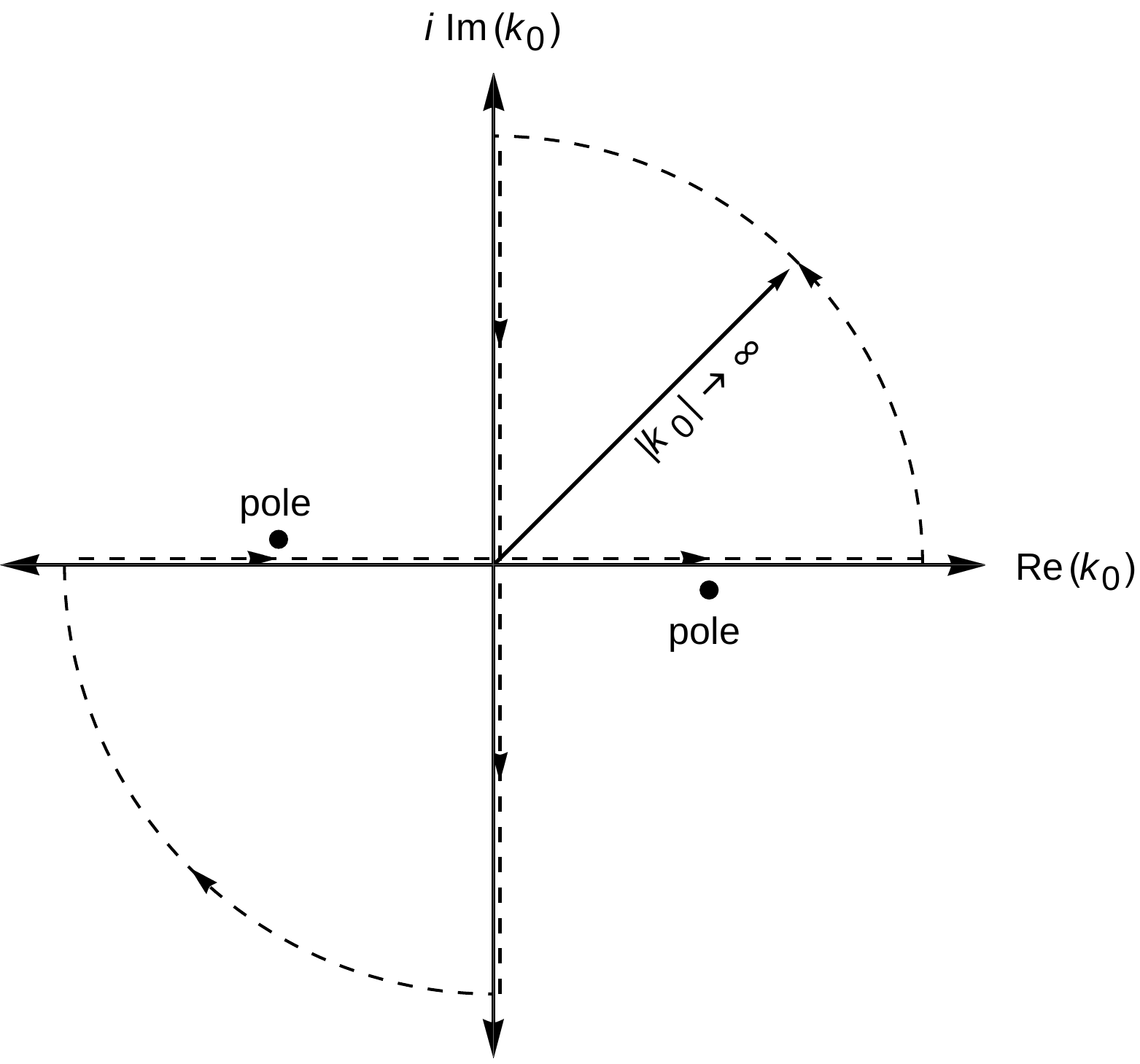}$}
		\caption{Wick rotation on complex $k_0$ plane.}
	\label{Wickrot}
	\end{center}
\end{figure}

The integrand in Eq.~\eqref{Ip} has two poles on the line of integration at $k_0=\pm\sqrt{|\vec{k}|^2+\Delta}$. So we follow  the so-called ``$i\epsilon$ prescription'' while defining $\Delta$, i.e. $\Delta\rightarrow\lim\limits_{\epsilon\rightarrow 0}(\Delta-i\epsilon)$. Then the poles no longer fall on the line of integration, but shift to $k_0=\pm\sqrt{|\vec{k}|^2+\Delta-i\epsilon}$. Now, we integrate along the closed contour shown as a dotted line in Fig. \ref{Wickrot}.

 As there is no pole inside the contour and the integrations along the curved lines are vanisingly small for $|k_0|\rightarrow \infty $ on the complex $k_0$ plane, the integral along the real $k_0$ axis becomes equal to the integral along the imaginary $k_0$ axis. This allows us to perform what is called a ``Wick rotation'', and evaluate the integral over imaginary values of $k_0$ ranging from $-i\infty$ to $i\infty$. Next we define a Euclidean momentum:
\be
k^\mu_E=(-ik_0,\vk)
\ee
whose zero component takes the normal range $-\infty$ to $\infty$. One sees that $k_E^2=-k_0^2+\vk^2=-k^2$. 

Since there is no pole on the imaginary $k_0$ axis, there is no need to use the $i\epsilon$ prescription after Wick rotation. Then \eref{Ip} becomes,
\begin{equation}
\label{Ip1}
I(p)=i\int_{0}^{1}dx\int\frac{d^4 k_E}{(2\pi)^4}\frac{1}{(k_E^2+\Delta)^2}
\end{equation}
Since the integral is rotationally symmetric, we can use:
\begin{equation}
\label{omegad}
d^4 k_E=k_E^3\, dk_E\, d\Omega_4\quad\textrm{and}\quad\int d\Omega_d=\frac{2\pi^{d/2}}{\Gamma(d/2)}
\end{equation}
where $\omega_d$ is the $d$-dimensional solid angle\footnote{Eq.~\eqref{omegad} can be easily obtained by writing the integral $\Big(\displaystyle\prod_{j=1}^{d}\displaystyle\int_{-\infty}^{\infty} dy_j\Big) \exp\Big[{-\sum_{j=1}^{d}y_j^2}\Big]$ in $d$-dimensional polar coordinates as $\displaystyle\int d\Omega_d \displaystyle\int_{0}^{\infty}r^{d-1}e^{-r^2}dr$. Next, we evaluate the integral in both forms and compare the two answers to extract $\int d\Omega_d$.}.

Now we should be all set to evaluate the integral, but as noted above, it is divergent. In fact it diverges logarithmically as $|k_E|\to\infty$. We have to introduce a prescription to regularise it, and we now turn to this.

\subsection{Regularisation of ultraviolet (UV) divergences}

\subsubsection*{(i) Momentum cut-off:}

In this regularisation, we put an upper bound (say $\Lambda$) on $|k_E|$ instead of integrating it all the way to infinity. At the end of the calculation we take the limit $\Lambda\rightarrow\infty$. Thus:
\begin{equation}
\begin{split}
\int\frac{d^4 k_E}{(2\pi)^4}\frac{1}{(k_E^2+\Delta)^2}
&=\frac{1}{(2\pi)^4}\int d\Omega_4\int_{0}^{\Lambda} \frac{k_E^3\, dk_E}{(k_E^2+\Delta)^2}\\[2mm]
&=\frac{1}{(2\pi)^4}\cdot (2\pi^2)\cdot \frac{1}{2} \Bigg[\ln\left(\frac{\Lambda^2+\Delta}{\Delta}\right)-\frac{\Lambda^2}{\Lambda^2+\Delta}\Bigg]
\end{split}
\end{equation}
Thus we find that, in the limit $\Lambda \gg \sqrt\Delta$, the divergent part can be written:
\begin{equation}
-\bigg(\frac{1}{16\pi^2}\bigg)\ln\bigg(\frac{\Delta}{\Lambda^2}\bigg)
\end{equation}
It follows that:
\be
\begin{split}
I(p)&=-\bigg(\frac{i}{16\pi^2}\bigg)\int_{0}^{1} dx \ln\bigg(\frac{\Delta}{\Lambda^2}\bigg)\\[2mm]
&=\frac{i}{8\pi^2}\ln\Lambda-\frac{i}{16\pi^2}\int_{0}^{1} dx \ln \Delta
\end{split}
\label{ILambda}
\ee
where $\Delta$ was defined in \eref{Deltadef}.

In principle we have achieved what we set out to do. With a cutoff, our momentum integral is finite and cutoff-dependent. Moreover, the cutoff dependence (for large $\Lambda$) only resides in the first term. The method is quite intuitive -- since the divergence arises from the ultraviolet region of large momentum, it makes sense to simply cut off that region. However, when applied to general diagrams, this method does not preserve Lorentz symmetry, and in gauge theories it creates difficulties in maintaining gauge invariance -- which is crucial to proving consistency of the renormalisation procedure to all orders. Hence we now turn to an alternative regularization method which is far less intuitive, but works much better in practice.

\subsection*{(ii) Dimensional regularisation:}

$I(p)$ as defined in \eref{Ip1} is divergent in dimension $d=4$, but is convergent in $d<4$. So, we perform the integral in $d$ dimensions and treat $d$ as a continuous variable close to four, i.e. $d=4-\epsilon$. To be sure, the idea of space-time dimension being non-integer is rather strange. However it is mathematically quite well-defined (as a result of several decades of work!) and we treat it as merely a tool to regularise divergent integrals. Moreover, as will emerge very clearly from our computations, it preserves both Lorentz invariance, manifested formally after Wick rotation as $SO(d)$ invariance, and also other symmetries like gauge invariance that play an essential role in the Standard Model. Of course, at the end of each calculation, we have to take the limit $\epsilon\rightarrow0$ to recover 4-dimensional physics.

To calculate the integral in $d$ dimensions, we use the following procedure. Replace the integration measure:
\be
\frac{d^4k}{(2\pi)^4}\to \frac{d^dk}{(2\pi)^d}
\ee
Hence the regulated integral of interest is:
\begin{equation}
I_d(p)=i\int_0^1 dx \int\frac{d^d k_E}{(2\pi)^d}\frac{1}{(k_E^2+\Delta)^2}
\ee
Next, using:
\be
d^d k_E=d\Omega_d\, k_E^{d-1}dk_E 
\ee
we find:
\be
I_d(p)=\frac{i}{(2\pi)^d}\int_0^1 dx \int d\Omega_d\int_{0}^{\infty} \frac{k_E^{d-1} }{(k_E^2+\Delta)^2}dk_E
\ee
At this point we use the identity:
\begin{equation}
\label{intformula}
\int_0^\infty \frac{y^a}{(y^2+\Delta)^b}\, dy=\Delta^{(\frac{a+1}{2}-b)}\, \frac{\Gamma(\frac{a+1}{2})\Gamma(b-\frac{a+1}{2})}{2\Gamma(b)}
\end{equation}
In turn, this formula follows from:
\be
\int_0^\infty \frac{z^{p-1}}{(1+z)^{p+q}}dz=B(p,q)=\frac{\Gamma(p)\Gamma(q)}{\Gamma(p+q)}
\label{eulerbeta}
\ee
valid when $\Re(p)>0,\Re(q)>0$. Here $B(p,q)$ is the Euler beta-function and the above integral is a standard one that can be found in tables of integrals. 

\exercise{Use \eref{eulerbeta} to derive \eref{intformula}.}

Returning to our calculation and using \eref{omegad} and \eref{intformula}, we have:
\be
\begin{split}
I_d(p)&=\frac{i}{(2\pi)^d}\cdot \frac{2\pi^{d/2}}{\Gamma(d/2)}\cdot\frac{\Gamma(d/2)\Gamma(2-d/2)}{2\Gamma(2)}\int_0^1 dx~\Delta^{(d/2-2)}\\&
=\frac{i}{(4\pi)^{d/2}}\cdot\Gamma(2-d/2)\cdot\int_0^1 dx~\Delta^{(d/2-2)}\\&
=i(4\pi)^{(\epsilon/2-2)}\cdot\Gamma(\epsilon/2)\cdot \int_0^1 dx~\Delta^{-\epsilon/2}
\end{split}
\label{Idp}
\end{equation}
where in the last step we used $d=4-\epsilon$.

Now let us expand this result about $\epsilon=0$, the desired limit. We use the following identities in \eref{Idp}:
\be
\begin{split}
\Gamma(z)&=\sfrac{1}{z}-\gamma_E+\mathcal{O}(z)\\
A^z&=1+z\ln A+\mathcal{O}(z^2)
\end{split}
\ee
where $\gamma_E\sim 0.577 $ is the Euler-Mascheroni constant, to get:
\begin{equation}
\label{Ipdim}
I_d(p)=\frac{i}{8\pi^2\epsilon}-\frac{i}{16\pi^2} 
\int_0^1 dx\, \ln \left(\frac{\Delta}{4\pi e^{-\gamma_E}}\right)+\mathcal{O}(\epsilon)
\end{equation}
where we recall that $\Delta=m^2-p^2x(1-x)$. Here we have dropped terms that vanish as $\epsilon\to 0$, just as we previously dropped terms that vanish as $\Lambda\to \infty$ in the cutoff scheme (henceforth we will also drop the $d$ subscript on $I_d$, for simplicity). 
Notice that in dimensional regularisation the UV divergence is expressed as a simple pole at $\epsilon=0$. Comparing this equation with \eref{ILambda} we notice that the two expressions are in fact quite similar. If we replace $1/\epsilon$ by $\ln\Lambda$ then the divergent first terms are identical. Also the momentum-dependent finite part is identical. This is reassuring, because ultimately different regularisations are supposed to be merely different ways of arriving at the same physical results (although with differing levels of convenience, particularly when more complicated integrals are involved). However we are still far from arriving at a physical result. All we have found up to now is a regulated integral that will diverge again if we try to remove the cutoff.

Before proceeding, let us note that there exists a dimension problem in \eref{Ipdim}. The second term above contains the logarithm of $\Big(\frac{\Delta}{4\pi e^{-\gamma_E}}\Big)$. This is a dimensionful quantity, since the mass dimension of $\Delta$ is 2. It is rather unusual, to say the least, to find quantities in physics that depend on logarithms of dimensionful quantities\footnote{The momentum cutoff does not have this problem since we find the logarithm of $\frac{\Delta}{\Lambda^2}$ which is indeed dimensionless.}. The solution to this problem, which we now discuss, leads to the understanding of a profound aspect of renormalisation. 

Let us first find the canonical dimension of various relevant quantities in $d=4-\epsilon$ dimensions. The action has to be dimensionless in units where $\hbar=1$. Imposing this requirement on $\int d^d x\,\del_\mu\phi\del^\mu\phi$ as well as $\lambda\int d^dx\, \phi^4$, we find the dimensions:
\be
[\phi]=\frac{d-2}{2}=1-\frac{\epsilon}{2},\qquad [\lambda]=4-d=\epsilon
\ee

\exercise{Verify the above formulae for the canonical dimensions of $\phi$ and $\lambda$.}

We see that $\lambda$ is dimensionless in four dimensions. But we are working away from four dimensions, at least temporarily, and to keep proper track of the dimensions in our formulae, it is best to keep $\lambda$ dimensionless even away from four dimensions. To this end, we replace $\lambda$ by $\lambda \mu^\epsilon$ where $\mu$ is an arbitrary constant with dimensions of mass: $[\mu]=1$. $\mu$ is called the ``renormalisation scale". Thus we re-do the above calculation after taking the interaction Hamiltonian to be:
\be
\mathcal{H}_I=\frac{\lambda\mu^\epsilon}{4!}\phi^4
\ee
Then \eref{Mmatrix} is replaced by:
\be
i\cM=\mu^\epsilon\bigg(-i\lambda+\frac{\mu^{\epsilon}\lambda^2}{2}\Big( I(p_1+p_2)+I(p_1-p_3)+I(p_1-p_4)\Big)\bigg)
\label{matrixel}
\ee
We have chosen to keep an overall factor $\mu^\epsilon$ outside the bracket, for reasons that will become apparent soon. Inside the bracket, at order $\lambda^2$ we have three terms, each of the form $\mu^\epsilon I(p)$ for different 4-momenta $p$. Now we expand $\mu^\epsilon=1+\epsilon\log\mu+\cdots$ and drop terms that vanish as $\epsilon\to 0$ to get:
\be
\begin{split}
\mu^{\epsilon}I(p)&= \frac{i}{8\pi^2\epsilon}-\frac{i}{16\pi^2}\int_0^1 dx\,\ln\Big(\frac{\Delta}{4\pi\mu^2 e^{-\gamma_E}}\Big)\\[2mm]
&=\frac{i}{8\pi^2\epsilon}-\frac{i}{16\pi^2}\int_0^1 dx\,\ln\Big(\frac{\Delta}{\widetilde{\mu}^2}\Big)
\end{split}
\label{lmu}
\ee
where in the second step we have defined $\widetilde{\mu}^2=4\pi\mu^2 e^{-\gamma_E}$. 

We see right away that the dimension problem has been solved, as the argument of the logarithm is now the ratio of $\Delta$ to the square of the renormalisation scale. This puts us in a position to carry out the renormalisation procedure, to which we will turn in the next subsection. It may seem that the introduction of the scale $\mu$ is just an artifice, but in fact is much more than that.

To complete the above calculation of the matrix element, we insert \eref{lmu} into \eref{matrixel} to find:
\be
i\mathcal{M}(s,t,u)=\mu^\epsilon\bigg[-i\lambda+\frac{\lambda^2}{2}\left(\frac{3i}{8\pi^2\epsilon}-\frac{i}{16\pi^2}F(s,t,u)\right)\bigg]
\label{Mmatrix.2}
\ee
where
\be F(s,t,u)=\displaystyle\int_0^1  dx\bigg[\ln\left(\frac{\Delta(s)}{\widetilde{\mu}^2}\right)+\ln\left(\frac{\Delta(t)}{\widetilde{\mu}^2}\right)+\ln\left(\frac{\Delta(u)}{\widetilde{\mu}^2}\right)\bigg]
\label{Fstu}
\ee
and the Mandelstam variables $s,t,u$ were defined in \eref{manddef}.

\subsection{Renormalised coupling and counterterms: a first look}

If we now take the difference between the matrix element for arbitrary 4-momenta $p_i$ (and corresponding Mandelstam variables $s,t,u$) and the same object for some chosen reference momenta $p_0^i$ (and its Mandelstam variables $s_0,t_0,u_0$) then we find: 
\begin{equation}
\begin{split}
\label{Mp-Mp0}
& \quad i\mathcal{M}(s,t,u)-i\mathcal{M}(s_0,t_0,u_0)=-\mu^\epsilon \left(\frac{i\lambda^2}{32\pi^2}\right)\big[F(s,t,u)-F(s_0,t_0,u_0)\big]\\&
=-\mu^\epsilon \left(\frac{i\lambda^2}{32\pi^2}\right)\displaystyle\int_0^1  dx\bigg[\ln\left(\frac{\Delta(s)}{\Delta(s_0)}\right)+\ln\left(\frac{\Delta(t)}{\Delta(t_0)}\right)+\ln\left(\frac{\Delta(u)}{\Delta(u_0)}\right)\bigg]
\end{split}
\end{equation}	
This is encouraging: the difference between the matrix elements at two different sets of momenta is finite as the cutoff is removed ($\epsilon\to 0$).

Next, we define the ``renormalised coupling constant" $\lambda_R$ via:
\be
i\mathcal{M}(s_0,t_0,u_0)=-i\mu^\epsilon\lambda_R
\label{scalarren}
\ee
Then, from \eref{Mmatrix.2}, we have
\begin{equation}
\lambda_R=\lambda-\frac{\lambda^2}{2}\Big[\frac{3}{8\pi^2\epsilon}-\frac{1}{16\pi^2}F(s_0,t_0,u_0)\Big]
\label{lambdaRexp}
\end{equation}
It follows that:
\begin{equation}
\label{lambdabare}
\lambda=\lambda_R+\frac{\lambda_R^2}{2}\Big[\frac{3}{8\pi^2\epsilon}-\frac{1}{16\pi^2}F(s_0,t_0,u_0)\Big]
\end{equation}
where again we must recall that we are working only to order $\lambda^2$, equivalently $\lambda_R^2$. 

Thus, finally we can write the matrix element in terms of the renormalised coupling:
\begin{equation}
\label{M1}
i\mathcal{M}(s,t,u)=-i\lambda_R-\frac{i\lambda_R^2}{32\pi^2}\big[F(s,t,u)-F(s_0,t_0,u_0)\big]
\end{equation}
This is a completely finite relation between physical quantities: the renormalised coupling $\lambda_R$ and the matrix element $\cM$. The original parameter $\lambda$ that appeared in the Lagrangian, which is denoted the ``bare coupling constant", no longer appears! We treat $\lambda_R$ as a finite quantity to be determined by experiment. This in turn means, from \eref{lambdabare}, that the bare coupling constant is not a finite quantity. It is merely the parameter in the original (``bare'') Lagrangian.

The above discussion contains the essence of the renormalisation programme. However it is convenient to re-formulate it in the language of ``counterterms''. For this, we rewrite the original Lagrangian in terms of $\lambda_R$ using \eref{lambdabare}. From now on, to avoid confusion, we denote the bare coupling constant as $\lambda_0$. Next we write the relation between bare and renormalised couplings as:
\be
\lambda_0=\lambda_R+\delta\lambda_R
\label{bareren}
\ee
From \eref{lambdabare} it is evident that, to one loop:
\be
\delta\lambda_R=\frac{\lambda_R^2}{2}\bigg(\frac{3}{8\pi^2\epsilon}-\frac{1}{16\pi^2}F(s_0,t_0,u_0)\bigg)
=\frac{\lambda_R^2}{32\pi^2}\bigg(\frac{6}{\epsilon}-F(s_0,t_0,u_0)\bigg)
\label{lambdaren}
\ee
Above, we noted that the bare Lagrangian $\cL$ in \eref{Lscalar} depends only on the bare coupling constant. However by simply inserting \eref{bareren} into $\cL$, we can re-express this same Lagrangian in terms of the renormalised coupling and an additional divergent quantity (called $\delta\lambda_R$ above):
\be
\cL=\half \del_\mu\phi\del^\mu\phi - \half m^2\phi^2 - \frac{\lambda_R}{4!}\phi^4 -\frac{\delta\lambda_R}{4!}\phi^4
\label{Lrpt}
\ee
We can write this as:
\be
\cL=\cL_R+\cL_{\rm counterterm}
\ee
where $\cL_R$ is the original Lagrangian but treated as a function of $\lambda_R$ instead of $\lambda$, and  the counterterm $\cL_{\rm counterterm}$ is just the last term in \eref{Lrpt}, proportional to $\delta\lambda_R$ which in turn is proportional to $\lambda_R^2$. This last term also contains all the cutoff dependence. Now we have a Lagrangian with two ``different'' couplings, one cutoff-independent and proportional to $\lambda_R$ and the other cutoff-dependent and proportional to $\lambda_R^2$. Thus if we carry out perturbation theory in $\lambda_R$ to order $\lambda_R^2$, we must use the first vertex either once or twice, but must use the second vertex only once. Then we will get both tree level and one-loop diagrams from the first vertex, of which the loop diagram will have a divergent part. But we also have a tree-level diagram from the second vertex which is already divergent. The two divergences will now cancel and we will be left with a finite result. This way of doing things is the basis of ``renormalised perturbation theory'', to which we will return in a subsequent section. 

Let us return now to \eref{lambdaRexp} relating the bare and renormalised couplings. This equation says that the renormalised coupling is equal to the bare coupling, plus a divergent correction proportion to the bare coupling squared, plus a finite part. In the above equation the finite part is proportional to $F(s_0,t_0,u_0)$. Now, while the divergent correction is uniquely defined in any specific regularisation scheme, the finite part is actually quite arbitrary. After all, we never specified the values of the momenta $p_0^i$ which determine $s_0,t_0,u_0$, so two different people carrying out the same procedure might choose different $p_0^i$  and arrive at different finite parts in the relation between their bare and renormalised couplings.

This leads us to think that we can generalise \eref{lambdaRexp} to:
\be
\lambda_R=\lambda_0-\frac{\lambda_0^2}{2}\Big[\frac{3}{8\pi^2\epsilon}-{\rm finite}\Big]
\label{finitepart}
\ee
where {\em the finite part is entirely up to us}! As long as the divergent part is chosen correctly, the UV divergence will cancel. Our original choice $\frac{1}{16\pi^2}F(s_0,t_0,u_0)$ for the finite part, as in \eref{lambdaRexp}, was dictated by our desire -- based on physics -- to compare the coupling at different scales of external momenta. But it may be more convenient to make a choice that simplifies the mathematics. The goal is to get a parameter $\lambda_R$ in terms of which ${\cal M}$ is a finite function, and this is ensured if the expression for $\lambda_R$ in terms of $\lambda_0$ contains the infinite part proportional to $\frac{1}{\epsilon}$ with the right coefficient. Different choices of the finite part are referred to as different {\em renormalisation schemes}. The scheme in which the finite part is proportional to $F(s_0,t_0,u_0)$ is called the {\em momentum subtraction} scheme.

Different schemes will give us different relationships between the measurable quantity ${\cal M}$ and the parameter $\lambda_R$. Thus even after renormalisation, $\lambda_R$ itself is not directly measurable. Different choices of the finite part in the counterterm will lead to different definitions of $\lambda_R$ and also different relationships between ${\cal M}$ and $\lambda_R$, with the measured value of ${\cal M}$ always being unaffected. 

At this stage it is very tempting to simply set the finite part in \eref{finitepart} to 0. What could be simpler? And indeed this turns out to be a very good scheme, in which:
\be
\lambda_R=\lambda_0-\frac{3\lambda_0^2}{16\pi^2\epsilon}
\label{lambdaRMS}
\ee
This is called the ``minimal subtraction'' or MS scheme, in which one only subtracts off the pole part of a divergent quantity. The matrix element in this scheme is:
\be
i\mathcal{M}(s,t,u)=-i\lambda_R-\frac{i\lambda_R^2}{32\pi^2}F(s,t,u)
\label{MSmat}
\ee
A better choice is to take the finite part in \eref{finitepart} to be $-3\ln(4\pi e^{-\gamma_E})$. This has a special feature: with this choice, the matrix element becomes:
\be
i\mathcal{M}(s,t,u)=-i\lambda_R-\frac{i\lambda_R^2}{32\pi^2}
\int_0^1  dx\bigg[\ln\left(\frac{\Delta(s)}{{\mu}^2}\right)+\ln\left(\frac{\Delta(t)}{{\mu}^2}\right)+\ln\left(\frac{\Delta(u)}{{\mu}^2}\right)\bigg]
\label{msamplitude}
\ee
We see that all dependence on the Euler constant $\gamma_E$ (residing in the $\widetilde{\mu}^2$ terms in \eref{Fstu}) has disappeared. This often turns out more convenient in practice, and is called the ``modified minimal subtraction'' or $\overline{\rm MS}$ scheme. In the MS and $\overline{\rm MS}$ schemes, our final answer does not depend on any arbitrary reference momenta, but only on physical quantities and on the renormalisation scale $\mu$. 

Due to the ambiguity in the choice of renormalisation scheme, there is a corresponding ambiguity in defining renormalised perturbation theory. The counterterm piece of $\cL$ necessarily contains a fixed divergent term,  but this can be accompanied by any finite terms of our choice. In the language of renormalised perturbation theory, it is this choice that specifies the renormalisation scheme.
 
It remains to clarify the physical role of $\mu$. An equation like \eref{msamplitude} may seem puzzling because it suggests that on varying a completely arbitrary parameter $\mu$ (which was introduced by us into the theory) a measurable matrix element takes different values. We will see in due course that this dependence is illusory. We will in the next subsection that the renormalised coupling $\lambda_R$ depends on $\mu$ and should really be written $\lambda_R(\mu)$. Thus the matrix element depends both directly and implicitly on $\mu$ and the two sources of $\mu$-dependence will ultimately cancel out.

\subsection{The $Z$-factor and the $\beta$-function}
\label{Zfactor}

Working with a toy model, namely scalar field theory with a $\lambda\phi^4$ interaction, we have already covered enough material to make a conceptual leap. We will define the coupling constant renormalisation factor that relates bare and renormalised couplings. Then we will implement the seemingly obvious fact that the bare coupling does not depend on the renormalisation scale $\mu$ that was introduced by hand in dimensional regularisation. The result will be an equation called the Renormalisation Group (RG) equation. This provides a deep insight into the very meaning of quantum field theory. It will also turn out to be of direct experimental relevance in the context of non-Abelian gauge theories.

To start, recall that in \eref{finitepart} we expressed the renormalised coupling $\lambda_R$ in terms of the bare coupling constant $\lambda_0$ in a cutoff-dependent way. For definiteness, let us work in the minimal subtraction scheme in which \eref{lambdaRMS} holds. This equation may be written:
\be
\lambda_0=Z_\lambda \lambda_R
\ee
where, to one loop:
\be
Z_\lambda=1+\frac{3}{16\pi^2\epsilon}\lambda_R
\label{Zlambda}
\ee
The multiplicative constant $Z_\lambda$ is called the ``renormalisation constant'' for the coupling $\lambda$. Such constants are generically divergent, just as in the above example (at least in four dimensions). To one-loop order, the divergent part consists of a simple pole in $\epsilon$ with a calculable coefficient, and then there is a finite part which depends on the renormalisation scheme. Knowledge of the $Z$-factor tells us how the theory is renormalised.

Now we focus on the role of the parameter $\mu$ with dimensions of mass, which was introduced in order to carry out dimensional regularisation. This was not a physical parameter of the theory and we described it at the outset as an artifice of the renormalisation procedure. It is time to demonstrate what this means. We start with the evident fact that the bare coupling  $\lambda_0$ was independent of $\mu$. However we must remember that what appeared in the Lagrangian was not $\lambda_0$, but rather the same quantity scaled by $\mu^\epsilon$. It is {\it this} object which should be independent of $\mu$. From this reasoning, we conclude that:
\be
\frac{\del}{\del\mu}\left(\mu^\epsilon Z_\lambda \lambda_R\right)=0
\ee
Carrying out the differentiation term by term and multiplying the result by $\mu\left(\mu^\epsilon Z_\lambda \lambda_R\right)^{-1}$, we find:
\be
\epsilon+\frac{1}{Z_\lambda}\mu\frac{\del Z_\lambda}{\del\mu}+\frac{1}{\lambda_R}\mu\frac{\del\lambda_R}{\del\mu}=0
\label{difflambda}
\ee
One must keep in mind that this is a delicate equation because $Z_\lambda$ also depends on $\epsilon$. So we cannot naively take $\epsilon\to 0$ just yet. Now besides its dependence on $\epsilon$, the only quantity on which $Z_\lambda$ depends is $\lambda_R$. As a result the last two terms are proportional to $\del\lambda_R/\del\mu$. Since $\epsilon$ is a finite quantity, it thus has to be the case that $\lambda_R$ depends on $\mu$!

If we now define the {\em $\beta$-function} of the theory:
\be
\beta(\lambda_R(\mu))\equiv \mu\frac{\del\lambda_R}{\del\mu}
\ee
then we can determine this function using \eref{difflambda}. At one-loop order, inserting \eref{Zlambda}, we have:
\be
\mu\frac{\del Z_\lambda}{\del\mu}=\frac{3}{16\pi^2\epsilon}\mu\frac{d\lambda_R}{d\mu}
\label{insertZ}
\ee
Inserting this in \eref{difflambda}, collecting terms and reorganising them (always bearing in mind that we are working only to order $\lambda_R^2$) we find the one-loop $\beta$ function of SFT:
\be
\beta(\lambda_R(\mu))
=-\epsilon\lambda_R +\frac{3}{16\pi^2}\lambda_R^2
\label{lambdarg}
\ee

\exercise{Carefully derive \eref{lambdarg} by combining Eqs.(\ref{Zlambda}), (\ref{difflambda}, (\ref{insertZ}).}

Notice that \eref{lambdarg} is finite as $\epsilon\to 0$ so we can take this limit at this stage. After we do so, we are left with a non-trivial $\beta$-function and moreover, the coefficient by which it is proportional to $\lambda_R^2$, namely $\frac{3}{16\pi^2}$, is inherited directly from the coefficient of the $\epsilon$-pole in \eref{Zlambda}. Thus, somewhat surprisingly we have obtained a {\em finite} equation that is determined entirely by the {\em divergent} behaviour of the theory!

The physical meaning of \eref{lambdarg} becomes apparent if we realise that different values of $\mu$ label different renormalisation schemes. Thus varying $\mu$ corresponds to considering a family of renormalised theories corresponding to the same bare theory. Suppose we start by renormalising the bare theory using some reference value of $\mu$, and then vary $\mu$ monotonically away from the original value. Then \eref{lambdarg} tells us that the value of $\lambda_R$ varies monotonically in the same direction. We can integrate \eref{lambdarg} (after taking $\epsilon\to 0$) to find:
\be
\lambda_R(\mu)=\frac{\lambda_R(\mu_0)}{1-\frac{3\lambda_R(\mu_0)}{16\pi^2}\log\frac{\mu}{\mu_0}}
\label{lambdasolv}
\ee
This equation clearly shows that as $\mu$ increases from some starting value $\mu_0$, the value of $\lambda_R$ keeps increasing. This is essentially due to the positive sign of the coefficient of $\lambda_R$ above.

Indeed, if we trust this equation even when $\mu$ is very large, we find that $\lambda_R$ appears to diverge at the value:
\be
\mu=\mu_0\exp\left(\frac{16\pi^2}{3\lambda_R(\mu_0)}\right)
\ee
There is a fallacy here, however. Much before $\lambda_R(\mu)$ can diverge, it becomes larger than 1. At such values, the entire notion of one-loop perturbation theory is invalid. Higher-loop corrections, all the way to infinite order, become important and we cannot possibly calculate them. Therefore we can only conclude that $\lambda_R(\mu)$ increases with increasing $\mu$ upto a point, and thereafter we do not know precisely what it does.

However, the conclusions we reach in the reverse direction are valid. As $\mu$ decreases, $\lambda_R(\mu)$ also decreases. The smaller the value of $\lambda_R(\mu)$, the smaller will be the corrections that are neglected when truncating the perturbative expansion to one loop. Accordingly, one-loop perturbation theory becomes more and more reliable. The lesson is that, within the $\mu$-dependent family of renormalised $\lambda\phi^4$ theories, it is best to take $\mu$ as small as possible to optimise perturbation theory. In a moment we will see what constrains such a choice.

Because large and small $\mu$ correspond, respectively, to the far ultraviolet and far infrared regimes of energy, we summarise the above observations by saying that $\lambda\phi^4$ theory is strongly coupled in the ultraviolet and weakly coupled in the infrared. Such a theory is called {\em infrared free}. We will soon see that quantum electrodynamics has the same behaviour. However, there are theories that can exhibit the opposite phenomenon, and important examples are provided by pure Yang-Mills theories and some of its matter-coupled generalisations. In such theories the $\beta$-functions have a negative sign, and as a result the analogue of \eref{lambdasolv} has a positive sign in the denominator. This causes the theory to be weakly coupled in the UV and strongly coupled in the IR -- exactly the reverse of what we found above.  Such theories are called {\em asymptotically free}.

From the discussion so far, it is not clear whether asymptotic/infrared freedom is simply a formal feature of the theory or has direct, observable physical consequences. To see that the latter is the case, let us return to the renormalised amplitude \eref{msamplitude}, which 
we now rewrite with a slightly improved notation:
\be
i\mathcal{M}(s,t,u)=-i\lambda_R(\mu)-\frac{i\lambda_R(\mu)^2}{32\pi^2}
\int_0^1  dx\bigg[\ln\left(\frac{\Delta(s)}{{\mu}^2}\right)+\ln\left(\frac{\Delta(t)}{{\mu}^2}\right)+\ln\left(\frac{\Delta(u)}{{\mu}^2}\right)\bigg]
\ee
The only change is that we have now explicitly acknowledged the dependence of $\lambda_R$ on $\mu$. 

This formula highlights the quandary about what is the ``best'' value of $\mu$ to choose. Remember that our single-minded goal is to make perturbation theory as good as possible, i.e. to minimise the contribution of unknown terms of higher order in $\lambda_R$ so that we can trust the terms we have calculated. So one may be tempted to choose $\mu$ to minimise the value of $\lambda_R$. But this alone does not guarantee a reliable perturbation theory! There is another potential problem: the size of the coefficients $\sim \log\frac{\Delta}{\mu^2}$. If these become too large due to an inappropriate choice of $\mu$, then they can destroy the validity of perturbation theory. Moreover it can be shown, by an analysis of Feynman diagrams, that correspondingly higher powers of these logarithms arise as we go to higher orders in perturbation theory.

We are led to consider four possible situations: (i) an infrared free theory, studied at low energy, (ii) an infrared free theory, studied at high energy, (iii) an asymptotically free theory, studied at high energy, (iv) an asymptotically free theory studied at low energy. Here, by ``energy'' we mean the typical scale of the external momenta in the process, which determines the quantities $\Delta$ above.

In case (i), the value of $\Delta$ is approximated by $m^2$ where $m$ is the mass of the relevant particle. So we can take $\mu\sim m$. If $\lambda_R(m)$ is now chosen to be small, we are done. Perturbation theory is good to one-loop order and higher-order terms are negligible. Also, over the entire range of energies from 0 to order $m$ we simply keep $\mu\sim m$ and there is no noticeable energy dependence in amplitudes. Indeed, the amplitude can be written:
\be
\mathcal{M}(s,t,u)\sim a\,\lambda_R(m)+ b\,\lambda_R(m)^2 + {\cal O}\big(\lambda_R(m)^3\big)
\ee
where $a,b$ are numbers of order 1. This is the case not just for scalar field theory but also for QED\footnote{However there are complications due to the exchange of photons, which are massless.}, and it accounts for the fact that scale dependence of the fine structure constant was not a fundamental issue in the early days of the subject. 

In case (ii), the mass $m$ is irrelevant and $\Delta$ is approximated by the value of $s,t$ or $u$ which is large. In that case, taking $\mu\sim m$ is no longer a good idea as it will result in large values of $\ln\left(\frac{\Delta}{\mu^2}\right)$. We should instead take $\mu\sim {\sqrt s}$. However, if $\lambda_R$ was small at $\mu\sim m$ then it will be larger for $\mu\sim \sqrt{s}$ as seen from \eref{lambdasolv}. This means that beyond some external energy, there is no choice of $\mu$ that will make perturbation theory reliable. There is just no way to make both the coupling and the coefficients that arise in $\cM$ small together for arbitrarily high energies.

\exercise{Imagine that $m=1$ MeV and $\lambda_R(m)$ is determined by a low-energy experiment to be $\frac{1}{100}$. Find the value of $\sqrt{s}$, in MeV, such that $\lambda_R(\sqrt{s})$ rises to $\frac{1}{10}$. Repeat to find $\sqrt{s}$ for which $\lambda_R(\sqrt{s})=1$.}

Now consider case (iii), an asymptotically free theory at high energy. The sign of the $\beta$-function for such a theory is opposite to what we calculated in scalar field theory, and this behaviour is exhibited in pure Yang-Mills theory, which will be discussed later on. Since we are considering high-energy processes, we would like to set $\mu\sim \sqrt{s}$ to minimise the log factors. Asymptotic freedom ensures that in the process, $\lambda(\sqrt{s})$ also decreases with increasing energy. In this situation, the analogue of ${\cal M}$ above can be written:
\be
\mathcal{M}(s,t,u)\sim a\,\lambda_R(\sqrt{s})+ b\,\lambda_R(\sqrt{s})^2 + {\cal O}\big(\lambda_R(\sqrt{s})^3\big)
\ee
If we plot the amplitude  as a function of energy, we see that the one-loop answer decreases {\em and} the ignored corrections become even more negligible. So we reliably conclude that the total  amplitude must decrease logarithmically with energy. This is an experimentally measurable (and measured) fact.

Finally consider case (iv), low energy physics in an asymptotically free theory. The fact that Yang-Mills theory is massless adds to the problem. It means that the effective $\Delta$ goes to zero as the energy decreases. But we cannot take $\mu$ arbitrarily small, as this would render the coupling constant large and perturbation theory would break down. It is generally agreed that low-energy processes in asymptotically free theories cannot be studied in perturbation theory.

\clearpage

\section{Quantum electrodynamics at one loop}
\label{qedoneloop}

\subsection{The divergent diagrams}

We now return to a theory with a sound physical interpretation, namely quantum electrodynamics. We have previously encountered the QED Lagrangian:
\begin{equation}
\mathcal{L}=-\frac{1}{4}F_{\mu\nu}F^{\mu\nu}+i\overline{\psi}(\slashed\partial+ie\slashed A)\psi-m\overline{\psi}\psi A^\mu-\frac{1}{2\xi}(\partial_\mu A^\mu)^2-J_\mu A^\mu
\end{equation}
We have already derived the Feynman rules. In particular the momentum-space fermion and photon propagators were summarised in \eref{fermphotprop}. In what follows we will work in Feynman gauge, $\xi=1$, where the photon propagator is particularly simple:
\be
-i\frac{\eta_{\mu\nu}}{p^2+i\epsilon}
\ee

This time, we have added a source term that couples the gauge field to an external electromagnetic current $J_\mu$. Note that since we are going to carry out dimensional regularisation, the parameter $e$ in the Lagrangian will be replaced by  $\mu^{\epsilon/2}e$ where, as before, $\mu$ is the renormalisation scale. This renders $e$ dimensionless in all space-time dimensions. 

In what follows, we examine all the one-loop divergent integrals that arise in QED. Among other things, this will enable us to make good on our promise in subsection \ref{logicren} of computing the one-loop corrected Coulomb potential in QED and showing that (in momentum space) it takes the form:
\be
\wV(\vp)=\frac{e^2+a(\vp,\Lambda)e^4}{\vp^{~2}\!}
\ee
with a dependence on the cutoff $\Lambda$, as predicted in \eref{mqedloop}. Of course, since we are going to work in the dimensional regularisation scheme, the cutoff is really $\epsilon$ rather than $\Lambda$.

There are three basic diagrams in QED that diverge at one loop.  

\begin{figure}[H]
\begin{tikzpicture}
\begin{feynman}
\vertex (a) ;
\vertex [right=of a] (b);
\vertex [right=of b] (c);
\vertex [right=of c] (d);
\diagram* {
(a) -- [boson, momentum=\(p\)] (b) -- [fermion, half left, momentum={[arrow shorten=0.1mm]\(k\)}] (c);
(c) -- [fermion, half left, momentum={[arrow shorten=0.1mm]\(k-p\)}] (b);
(c) -- [boson, momentum=\(p\)] (d)
};
\end{feynman}
\hspace*{5.5cm}
\begin{feynman}
\vertex (a) ;
\vertex [right=of a] (b);
\vertex [right=of b, xshift=8mm] (c);
\vertex [right=of c] (d);
\diagram* {
(a) -- [fermion, momentum=\(p\)] (b) -- [fermion, momentum={[arrow shorten=0.1mm]\(k\)}] (c);
(c) -- [fermion, momentum=\(p\)] (d);
(b) -- [boson, half left, momentum={[arrow shorten=0.13mm]\(p-k\)}] (c)
};
\end{feynman}
\hspace*{5.5cm}
\begin{feynman}
\vertex (g);
\vertex [above=of g] (a);
\vertex [below=of g] (d);
\vertex [below right=of a, yshift=5mm] (b);
\vertex [right=of g, xshift=16mm] (c);
\vertex [above right=of d, yshift=-5mm] (e);
\vertex [right=of c] (f);
\diagram* {
(a) -- [fermion, momentum=\(q_1\)] (b) -- [fermion, momentum={[arrow shorten=0.1mm]\(k\)}] (c);
(c) -- [fermion, momentum={[arrow shorten=0.1mm]\(k-p\)}] (e) -- [fermion, momentum=\(q_2\)] (d);
(b) -- [boson, momentum={[arrow shorten=0.1mm]\(q_1-k\)}] (e);
(c) -- [boson, momentum={[arrow shorten=0.1mm]\(p\,{=}\,q_1-q_2\)}] (f)
};
\end{feynman}
\end{tikzpicture}
\caption{The three basic divergent one-loop diagrams in QED: Vacuum polarisation, electron self-energy and vertex correction.} 
\label{oneloopQED}
\end{figure}
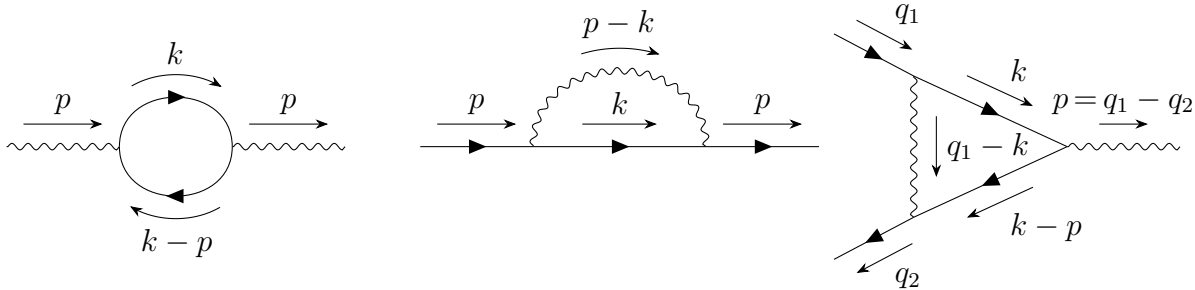

We will consider them one by one.  The first correction is the photon propagator correction, known as ``vacuum polarisation''. Inspection of the diagram shows that we have two powers of loop momenta in the denominator (one for each of the fermion propagators). Hence the naive UV behaviour of the loop integral takes the form: $\int\frac{d^4 k}{k^2}$ which is a quadratic divergence. However, on performing the calculation we will find that both the leading quadratic divergence and the subleading linear divergence cancel. This leaves a logarithmic divergence. It turns out that the underlying reason for such cancellations is gauge invariance. 

The second figure is the corrected electron propagator or ``self-energy correction''. In this case we have one fermion and one photon propagator, hence the naive behaviour is of the form $\int\frac{d^4 k}{k^3}$ which is a linear divergence. Again, the leading term cancels and we will find that the leading divergence is actually logarithmic. 

The third figure is, for obvious reasons, known as the ``vertex correction''. Here there are two fermion propagators and a photon propagator which (in Feynman gauge) is inverse quadratic in the momentum. Thus the total power of loop momenta in the denominator is 4, and the behaviour is $\sim \int \frac{d^4k}{k^4}$ which is again logarithmically divergent. 

Thus QED has the remarkable property that it has no divergence higher than logarithmic. Let us now carry out the actual computations. We will use the Feynman rules, derived in subsection \ref{qedfeyn}, to write down the diagrams to one-loop order. Note that the external lines will not be put on-shell since each of these divergent diagrams is to be thought of as {\em part} of a scattering amplitude rather than a physical scattering amplitude on its own. Then the external lines of these diagrams will become internal lines in some larger diagram. For this reason the external momenta in all these diagrams are kept completely arbitrary during the calculation. 

Recall that we denoted the bare coupling constant of scalar field theory as $\lambda_0$ in the previous chapter. In a similar spirit, we henceforth denote the bare electric charge by $e_0$, to make explicit the distinction with the renormalised charge $e_R$. Since the mass is another parameter that will eventually be renormalised in a similar way, we also denote the bare mass by $m_0$. At this stage we are also going to drop the explicit $i\epsilon$ term in all denominator factors, for ease of notation and to avoid possible (though unlikely) confusion with $\epsilon=4-d$. 

\subsection{Vacuum polarisation}
\label{vacpolsec}

Defining the corrected photon propagator as $iG^{\mu\nu}$, the Feynman rules tell us that to one-loop order:
\begin{equation}
\begin{split}
iG^{\mu\nu}(p)&=\frac{-i\eta^{\mu\nu}}{p^2}+\frac{-i\eta^{\mu\rho}}{p^2}i\Pi_{2\rho\lambda}(p)\frac{-i\eta^{\lambda\nu}}{p^2}\\&
=\frac{-i\eta^{\mu\nu}}{p^2}+\frac{-i\Pi^{\mu\nu}_2(p)}{(p^2)^2}
\end{split}
\end{equation}
where we have used $\Pi^{\mu\nu}$ to denote the contribution of just the loop, without the external propagators. 
We easily see that:
\begin{equation}
i\Pi_2^{\mu\nu}(p)=-\frac{(-ie_0)^2}{2}\cdot 2\cdot\int \frac{d^4 k}{(2\pi)^4}\frac{\tr[\gamma^\mu(\slashed k-\slashed p +m_0)\gamma^\nu (\slashed k+m_0)]}{(k^2-m_0^2)({(p-k)^2-m_0^2})}
\end{equation}
Here, the minus sign at the beginning, as well as the trace, are associated to the fermion loop. The factor  $\frac{(-ie_0)^2}{2}$ comes from $[\exp(-i\mathcal{H}_I)]$ as usual, while the next factor of 2 is the symmetry factor. This arises because there are two ways for the external photons to be connected to two vertices. Once this connection is chosen, the fermions in the two vertices contract with each other in a unique way and there are no more factors\footnote{This is because the fermions are complex. Their propagators therefore carry arrows depicting charge propagation, and these arrows have to join consistently due to charge conservation.}.  On evaluating the trace and performing the loop integral, we get:
\begin{equation}
\Pi_2^{\mu\nu}(p)=e_0^2(-p^2 \eta^{\mu\nu}+p^\mu p^\nu)\Pi_2(p^2)
\label{PIredef}
\end{equation}
where:
\begin{equation}
\Pi_2(p^2)=\frac{1}{6\pi^2\epsilon}-\frac{1}{2\pi^2}\int_{0}^{1}dx \textrm{ } x(1-x) \ln \bigg(\frac{\Delta}{\widetilde{\mu}^2}\bigg)+\mathcal{O}(\epsilon)
\label{pitwo}
\end{equation}
with $\Delta=m_0^2-p^2 x(1-x)$. It follows that:
\begin{equation}
iG^{\mu\nu}(p)=\frac{-i\eta^{\mu\nu}\big[1-e_0^2 \Pi_2(p^2)\big]}{p^2}+(p^\mu p^\nu\textrm{term})
\label{vacpol}
\end{equation}
where we have separated off the term proportional to $p^\mu p^\nu$ at the end. The reason is that such terms are gauge artifacts and we can ignore them. The physical content of the one-loop-corrected  photon propagator is contained in the piece proportional to $\eta^{\mu\nu}$. 

Recall that in subsection \ref{logicren} we defined electric charge by considering a particular matrix element $\cM$ for $e^--e^-$ scattering that involved photon exchange in the $t$-channel. Now that we have computed the one-loop correction to the photon propagator, we can imagine inserting it into this matrix element to find the one-loop correction to the electric charge. Choosing the $t$-channel 4-momentum to have vanishing time component, we find from \eref{vacpol} that the Coulomb potential energy in momentum space $\wV(\vp)$ gets corrected to:
\begin{equation}
\label{vtil}
\widetilde{V}(\vp)=\frac{e_0^2}{\vp^{\,2}}\Big(1-e_0^2 \Pi_2(-\vp^{\,2})\Big)
\end{equation}
This confirms the expectation of \eref{mqedloop}. We see that the effective electric charge is modified due to vacuum polarisation, leading to a correction to Coulomb's law at large momenta (or short distances) proportional to $
e_0^4$.

Now, we choose a reference momentum $p_0$ and define the renormalised electric charge $e_R$ such that $e_R^2=e_0^2-e_0^4 \Pi_2(p_0^2)$ (recall that $e_0$ is the bare electric charge). This can be inverted to give:
\be
e_0^2=e_R^2\Big(1+e_R^2\Pi_2(p_0^2)\Big)
\ee
Using \eref{pitwo} and taking a square root on both sides, this can be written:
\be
e_0=e_R\Big(1+\frac{e_R^2}{12\pi^2\epsilon}+\hbox{finite}\Big)
\label{erenorm}
\ee
We will extract an important consequence from this in the following subsection.

A  convenient choice (but by no means the only one) in the present case is to take $p_0^\mu=0$, which sets $\Delta(p_0)=m_0^2$. Then, up to order $e_R^4$, Eq.~\eqref{vtil} can be written as:
\begin{equation}
	\vp^{\,2}\widetilde{V}(\vp)=e_R^2-e_R^4\Big(\Pi_2(-\vp^{\,2})-\Pi_2(0)\Big)=e_R^2+\frac{e_R^4}{2\pi^2}\int_{0}^{1}dx \textrm{ } x(1-x) \ln \bigg(\frac{\Delta(p)}{m_R^2}\bigg)
\end{equation}
In the above expression we have replaced $m_0$ by $m_R$ since the difference is higher-order in perturbation theory.
To find the Coulomb potential we assume $p^2\ll m_R^2$. Then:
\begin{equation}
\ln\bigg(\frac{\Delta(p)}{m_R^2}\bigg)=\ln\bigg(1+\frac{{\vp}^{\,2} x(1-x)}{m_R^2}\bigg)\approx \bigg(\frac{{\vp}^{\,2} x(1-x)}{m_R^2}\bigg)
\end{equation}
from which it follows that:
\begin{equation}
\widetilde{V}(p)=\frac{e_R^2}{\vp^{\,2}}+\frac{e_R^4}{60\pi^2 m_R^2}
\end{equation}
Fourier transforming back to position space, we see that:
\be
V(x)=\frac{e_R^2}{4\pi|\vec{x}|}+\frac{e_R^4}{60\pi^2 m_R^2}\delta^3(\vec{x})
\ee
The extra term is called the ``Uehling potential energy". This term is physically measurable and contributes about $2\%$ to the Lamb shift. 

It may seem more natural to define the renormalised electric charge using the vertex correction, since the diagram directly represents the interaction of a photon and an electron. In fact this can be done, but the gauge symmetry of QED ensures that various different ways of determining the renormalised electric charge give the same answer. We will see how this works when we study Ward identities. 

\subsection{Electron self-energy}

Next we would like to compute the one-loop correction to the electron propagator, given by the second diagram in Figure \ref{oneloopQED}:
\begin{equation}
iG(p)=\frac{i}{\slashed p-m_0}+\frac{i}{\slashed p-m_0}i\Sigma_2(p)\frac{i}{\slashed p-m_0}
\label{Gzero}
\end{equation}
where:
\begin{equation}
\label{sigma2}
\begin{split}
\hspace{11mm}
i\Sigma_2(p)&=(-ie_0)^2 \int\frac{d^4 k}{(2\pi)^4}\gamma^\mu\frac{i(\slashed k+m_0)}{k^2-m_0^2}\gamma^\nu \frac{-i\eta_{\mu\nu}}{(k-p)^2}\\
&= 2e^2\int\frac{d^4 k}{(2\pi)^4}\cdot\frac{\slashed k-2m_0}{(k^2-m_0^2)(k-p)^2}\\&
=2e^2\int_0^1 dx \int\frac{d^4 k}{(2\pi)^4}\cdot \frac{\slashed k+x\slashed p-2m_0}{(k^2-\widetilde{\Delta})^2}
\end{split}
\end{equation}
Here the subscript ``2'' on $\Sigma(p)$ indicates that it is of second-order in the coupling. Note that in the first step we used $\gamma^\mu(\slashed k+m_0)\gamma_\mu=-2\slashed k+4m_0$, and in the second step we defined:
\be
\widetilde{\Delta}=(1-x)(m_0^2-p^2x)
\ee

\exercise{Verify the identity $\gamma^\mu(\slashed k+m_0)\gamma_\mu=-2\slashed k+4m_0$.}

The momentum integral can now be done, leading to:
\be
\begin{split}
i\Sigma_2(p)&=\frac{ie_0^2}{8\pi^2}\int_0^1 dx\,(\slashed px-2m)\Bigg(\frac{2}{\epsilon}-\ln\Big(\frac{\widetilde{\Delta}}{\widetilde{\mu}^2}\Big)+\mathcal{O}(\epsilon)\Bigg)\\
&= \frac{e_0^2}{8\pi^2\epsilon}\big(\slashed p - 4m_0\big)+\frac{e^2}{8\pi^2}\int_0^1 dx \big(2m_0-\slashed p x\big)\ln\frac{\widetilde\Delta}{{\widetilde\mu}^2}
\label{sigmatwocomp}
\end{split}
\ee
It follows that the divergent part of $\Sigma_2(p)$ is:
\be
\Big[\Sigma_2(p)\Big]_{div.}=\frac{e_0^2}{8\pi^2 \epsilon}\big(\slashed p-4m_0\big)
\label{divSigma}
\ee
We can interpret the term proportional to $\frac{m_0}{\epsilon}$ as a divergent contribution to the mass, which can then be dealt with by performing mass renormalisation -- just as we performed charge renormalisation above. However, the divergent term proportional to $\slashed p$ seems quite puzzling. What are we supposed to renormalise to account for this? Certainly we cannot renormalise the value of momenta, which are physical numbers. The correct answer is that such a term contributes to renormalisation of the {\em field}, in this case the fermion field $\psi$. This process is known as field renormalisation or (historically) wave-function renormalisation. 

Thus we must allow for the possibility that the electric charge parameter $e$, the mass $m$ and the fields $\psi,A_\mu$ appearing in the original QED Lagrangian are {\em all} ``bare'' objects. Accordingly we endow all of them with the subscript ``0'' and the bare QED Lagrangian \eref{qedlag} (including the gauge-fixing term) now reads:
\be
{\cal L} = -{1\over4} F_{0\,\mu\nu} F_0^{\mu\nu} + i\overline\psi_0 \gamma^\mu 
(\partial_\mu-ieA_{0\,\mu}) \psi_0 - m_0 \overline\psi_0 \psi_0-\frac{1}{2\xi_0}(\del_\mu A_0^\mu)^2
\label{qedlagbare}
\ee
where, as the notation suggests, $F_{0\,\mu\nu}\equiv \del_\mu A_{0\,\nu}-\del_\nu A_{0\,\mu}$.

With the above results it is, in principle, straightforward to complete the one-loop renormalisation process by expressing all the bare parameters and fields: $e_0,m_0,\psi_0, A_{0\,\mu}$ in terms of the corresponding renormalised parameters and fields: $e_R,m_R,\psi_R, A_{R\,\mu}$. Physical quantities computed with the bare Lagrangian, and then expressed in terms of renormalised parameters, will be finite functions of the latter. However instead of following the same path as before, we make a small but significant change of emphasis. This leads to ``renormalised perturbation theory'', to which a brief introduction was given in the previous Section. This will ultimately provide the most practical and modern way to understand the predictions of perturbative QFT. After introducing this concept in the following section, we return to the study of electron propagator corrections on which we embarked above. 

\clearpage

\section{Renormalised perturbation theory}

\subsection{The setup and the electron propagator revisited}

We now re-state what we have been doing in slightly different language. So far, we have expressed the bare Lagrangian in terms of bare parameters (charges, masses and fields), then performed loop calculations with these parameters, and finally expressed the answers in terms of renormalised parameters. Henceforth, instead, we express the bare Lagrangian directly in terms of renormalised parameters (charges, masses, fields) plus {\it counterterms} that are cutoff-dependent. Now we carry out perturbation theory using not just the normal vertices, but also new vertices contained in the counterterm Lagrangian. The role of the latter is precisely to cancel all divergences in loop diagrams that involve the former. This turns out to be much more convenient, as we will see. 

We start by proposing a set of relations between bare and renormalised quantities:
\be
\begin{split}
e_0&=Z_e\, e_R\\
m_0&=Z_m\, m_R\\
\psi_0&=\sqrt{Z_2}\,\psi_R\\
A^\mu_0&=\sqrt{Z_3}\,A^\mu_R\\
\xi_0&=Z_\xi \,\xi_R
\end{split}
\label{renparam}
\ee
where the various $Z$'s are renormalisation factors that are cutoff-dependent and diverge in the limit that the cutoff is removed. Of these, we have already computed $Z_1$ which can be read off from \eref{erenorm}.

For future convenience we also define $Z_1=Z_eZ_2\sqrt{Z_3}$. This will be the factor that renormalises the coupling of the vector potential $A_\mu$ to the current $\psibar\gamma^\mu\psi$ and is therefore of historical importance, though it is determined in terms of other renormalisation constants. Henceforth we will denote the renormalisation constants as a set $Z_i$ where $i$ takes the values $1,2,3,e,m,\xi$. We are already familiar with the fact that parameters like $e,m$ have to be renormalised, and renormalising $\xi$ may not seem too surprising since it is also a parameter in the gauge-fixed theory. However we are shortly going to see that, as a consequence of \eref{divSigma}, renormalisation of the {\em fields} $\psi,A_\mu$ is also necessary. 

Let us start by writing each of the renormalisation constants as:
\be
Z_i=1+\delta_i
\label{deltadef}
\ee
This split reflects the fact that at tree level the value of each $Z_i$ is just 1 since there is no renormalisation at this order. The $\delta_i$ are contributions from loop corrections and therefore start at one-loop order, which means they are at least of ${\cal O}(e_R^2)$. In these lecture notes all explicit calculations will only be done to one-loop order, which will greatly simplify our manipulations. 

Inserting \eref{renparam} into \eref{qedlagbare} and using the definition of $Z_1$ given above, we find:
\be
\label{lagz}
\begin{split}
\mathcal{L}=&-\frac{1}{4} Z_3 F_{R\mu\nu}F_R^{\mu\nu}+Z_2\overline{\psi}_R(i\slashed \partial-Z_m m_R^{})\psi_R^{}\quad\quad\quad\quad\\&
-\mu^{\epsilon/2}Z_1 e_R^{}\overline{\psi}_R\gamma_\mu\psi_R^{} A_R^\mu-\frac{Z_3}{Z_\xi}\frac{1}{2\xi_R} (\partial_\mu A_R^\mu)^2\\
&=-\frac{1}{4} F_{R\mu\nu}F_R^{\mu\nu}+\overline{\psi}_R(i\slashed \partial-m_R^{})\psi_R^{}-\mu^{\epsilon/2}e_R^{}\overline{\psi}_R\gamma_\mu\psi_R^{} A_R^\mu-\frac{1}{2\xi} (\partial_\mu A_R^\mu)^2\\&
+\Big[-\frac{\delta_3}{4}  F_{R\mu\nu}F_R^{\mu\nu}+i\delta_2 \overline{\psi}_R\slashed \partial\psi_R^{}-(\delta_m+\delta_2)m_R^{}\overline{\psi}_R\psi_R^{}-\delta_1 e_R^{}\overline{\psi}_R\gamma_\mu\psi_R^{} A_R^\mu+...\Big]
\end{split}
\ee
where, recalling that $Z_1\equiv Z_e Z_2\sqrt{Z_3}$ and working only up to one-loop level, we have $\delta_1=\delta_e+\delta_2+\frac{1}{2}\delta_3$. Notice that we have replaced $e_0$ by $\mu^{\epsilon/2} e_0=\mu^{\epsilon/2}Z_e e_R$. The reason, as explained earlier, is that $e_0$ should be kept dimensionless away from $d=4$. The expression inside square brackets above  is called the ``counterterm Lagrangian". Being linear in the various $\delta_i$, it starts at one-loop order. A technical point to note is that, precisely because we are working only up to one-loop order, we can equally well allow the coupling constants and masses in the counterterm Lagrangian to be $(e_0,m_0)$ or $(e_R,m_R)$ as we please, and similarly for the fields. This is because within the counterterm Lagrangian the difference between these expressions gives us terms of higher order (at least two-loop order) in the coupling, and we have decided to neglect those here.

Recalling \eref{erenorm} we see that we have already determined the infinite part of $\delta_3$:
\be
\delta_e=\frac{e_R^2}{12\pi^2\epsilon}+\textrm{finite}
\label{deltae}
\ee
To fix the other renormalisation constants, we return to the electron propagator. The two-point Green's function for the electron field in QED, discussed in the previous section, is:
\be
G_0(p)=\int d^4x~ e^{ip\cdot(x-y)} \bra{0} T\big(\psi_0(x)\overline\psi_0(y)\big)\ket{0}
\label{G0}
\ee
where we have introduced the subscript $0$ on $G$ because it involves unrenormalised fields. 
To one-loop order, we calculated it to be:
\be
iG_0(p)=\frac{i}{\slashed p-m_0}+\frac{i}{\slashed p-m_0}i\Sigma_2(p)\frac{i}{\slashed p-m_0}
\ee
This is simply a restatement of \eref{Gzero} in the notation of renormalised perturbation theory.

Next we define the renormalised two-point function:
\be
G_R(p)=\int d^4x~ e^{ip\cdot(x-y)} \bra{0} T\big(\psi_R(x)\overline\psi_R(y)\big)\ket{0}
\label{GR}
\ee
In view of Eqs.(\ref{renparam}),(\ref{deltadef}),(\ref{G0}), we can write:
\be
\begin{split}
iG_R(p) &= \frac{1}{Z_2} iG_0(p)\\
&= \frac{1}{1+\delta_2}\Bigg(\frac{i}{\slashed p-m_R-\delta_m m_R}+\frac{i}{\slashed p-m_R}i\Sigma_2(p)\frac{i}{\slashed p-m_R}\Bigg)
\end{split}
\ee
Note that in the second term inside the brackets, we just replaced $m_0$ by $m_R$ and ignored the extra $\delta_m$ term. As explained earlier, this kind of replacement is legitimate because we are working to one loop and the second term, being proportional to $\Sigma_2$, is already of order $e_R^2$. For the same reason, the $\delta_2$ in the expansion of the prefactor $\frac{1}{1+\delta_2}=1-\delta_2+\cdots$ need not be multiplied with the second term. We thus get:
\be
\begin{split}
iG_R(p) &\simeq \frac{1}{1+\delta_2}\frac{i}{\slashed p-m_R-\delta_m m_R}+\frac{i}{\slashed p-m_R}i\Sigma_2(p)\frac{i}{\slashed p-m_R}\\
&\simeq \frac{i}{\slashed p-m_R^{}}+\frac{i}{\slashed p-m_R^{}}\Big(i\delta_2\slashed p-i(\delta_2+\delta_m)m_R^{}+i\Sigma_2(p)\Big)\frac{i}{\slashed p-m_R^{}}
\end{split}
\end{equation}
Recall that $\Sigma_2(p)$ was computed in \eref{sigmatwocomp}.

\exercise{Carefully derive the second line of the above equation from the first line, always dropping any term of order beyond $e_R^2$. Verify that all factors and signs are correct.}

It is now clear that to remove one loop divergences from $G_R(p)$, we need:
\begin{equation}
 \delta_2\slashed p-(\delta_2+\delta_m)m_R^{}+\Sigma_2(p)=\textrm{finite}
\label{finiteSigma}
\end{equation}
Using Eq.~\eqref{divSigma} this becomes:
\be
\Big(\delta_2+\frac{e_R^2}{8\pi^2 \epsilon}\Big)\slashed p-\Big(\delta_2+\delta_m+\frac{e_R^2}{2\pi^2 \epsilon}\Big)m_R^{}=\textrm{finite}
\label{deltm}
\ee
which in turn implies that:
\be
\delta_2=-\frac{e_R^2}{8\pi^2 \epsilon}+\textrm{finite}, \quad\quad\delta_m=-\frac{3e_R^2}{8\pi^2 \epsilon}+\textrm{finite}
\label{deltavals}
\ee
In particular,  our calculation has established that $\delta_2$ has a divergent part, confirming that the field renormalisation constant $Z_2$ for the electron -- parametrised here in terms of $\delta_2$ -- is really necessary.

\subsection{Renormalisation schemes}

We now return to a crucial feature of perturbative renormalisation that was briefly discussed around \eref{MSmat} in the context of scalar field theory. The infinite parts of the renormalisation constants are uniquely determined to any given loop order, and we have computed some of them to one loop in the steps leading to \eref{deltavals}. However the finite parts of these constants are entirely up to us. The choice of a finite part is called a ``renormalisation scheme'' or ``renormalisation prescription''. At first glance this freedom is puzzling, since physical quantities will then depend on our arbitrary choice of the finite parts. The resolution is that when we work in perturbation theory, there is always some contribution to the exact answer that we have neglected -- namely the higher-order corrections. However, precisely what part of the exact answer is neglected by doing perturbation theory to some finite order, depends on the expansion parameter $e_R$.  Thus different subtraction schemes will return one-loop answers for physical quantities that differ in accuracy as compared to the exact answer. This point is extremely subtle and difficult to grasp at first, but will become more familiar as we go on.

As we observed in Subsection \ref{Zfactor}, there are two prominent subtraction schemes that are commonly used in the context of dimensional regularisation. One is ``minimal subtraction'', known for short as MS. In this scheme one subtracts only the infinite part, for example the terms proportional to $\frac{1}{\epsilon}$ in \eref{deltavals}. This scheme turns out to be particularly effective, compared with other schemes, when working with non-Abelian gauge theory -- as we will do soon. There is a modified version of the MS scheme called ``modified minimal subtraction'' and conventionally denoted $\overline{\rm MS}$. In this scheme, the finite part is chosen to remove $\ln(4\pi e^{-\gamma_E})$ from the physical answer. This is equivalent to saying that it replaces $\ln{\widetilde\mu}^2$ factors in expressions like \eref{sigmatwocomp} by corresponding factors of $\mu^2$. That is convenient because it means the mass scale $\mu$ that was initially chosen to define dimensional regularisation appears unchanged in the logarithms in the final answer\footnote{Recall that $\widetilde\mu$ was defined just after \eref{lmu}.}. The $\overline{\rm MS}$ scheme is extremely natural and popular. But in what follows, we will sometimes use renormalisation prescriptions that are neither of the above, and will indicate their virtues. The choice of ``best'' prescription to use depends significantly on the physics of the problem at hand, and there is no single optimal choice. 
 
\exercise{From Eqs.(\ref{deltae}),(\ref{deltavals}) we see that for QED in the MS scheme, $\delta_e=\frac{e_R^2}{12\pi^2\epsilon}$, $\delta_2=-\frac{e_R^2}{8\pi^2\epsilon}$ and $\delta_m=-\frac{3e_R^2}{8\pi^2\epsilon}$. Find the values of $\delta_e,\delta_2,\delta_m$ in the $\overline{\rm MS}$ scheme.}
 
\subsection{1PI diagrams and pole mass}
\label{1PI}

In this section we examine an issue regarding the physical masses of elementary particles that arises when a theory is renormalised. The mass $m$ of a particle is, of course, experimentally measurable. In a renormalised theory, one might guess that the role of the physical mass $m$ is played by the renormalised mass $m_R$. But things are not so simple, as we now show.

In S-matrix theory there is a precise definition of the observed physical mass. Consider scalar particles first. Notice that the tree-level propagator in momentum space has a pole when $p^2=m^2$ and moreover the residue at this pole is $i$. Since at tree-level there is no renormalisation, this parameter $m$ where the pole occurs is precisely the physical mass. Once we incorporate loop effects, the location of the pole will shift. S-matrix theory then tells us that the physical mass, denoted the ``pole mass'' $m_p$, is the location of the pole in the full loop-corrected propagator. Further, there is a condition that the residue at this pole must continue to be $i$. As we will soon see explicitly, the value of $m_p$ as a function of $m_R$ will be renormalisation-scheme-dependent. Hence depending on the scheme, we may or may not have $m_p=m_R$. However $m_p$ will always be the physical mass. 

For fermions, the conditions are implemented by requiring:
\be
\lim_{\slashed p\to m_p} G_R(p)\sim \frac{i}{\slashed p - m_p}
\ee
and this is what we will consider below. The notation may seem a little confusing since $\slashed p$ is a matrix while $m_p$ is a number, however we can always rationalise the denominator by multiplying above and below by $\slashed p+m_p$ and then the denominator is the same as that for a scalar particle. 

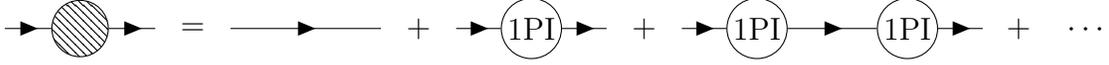
\begin{figure}[H]
\begin{center}
\begin{tikzpicture}
\begin{feynman}
\vertex[blob] (m) at (0,0) {};
\vertex (a) at (-1,0);
\vertex (b) at (1,0);
\diagram* {
(a) -- [fermion] (m) -- [fermion] (b);
};
\draw (1.5,0) node {$=$};
\end{feynman}
\begin{feynman}
\vertex (a) at (2,0);
\vertex (b) at (4,0);
\diagram* {
(a) -- [fermion] (b);
};
\draw (4.5,0) node {$+$};
\end{feynman}
\begin{feynman}
\vertex[circle, draw=black, inner sep=0pt, minimum size=8mm] (m) at (6,0) {1PI};
\vertex (a) at (5,0);
\vertex (b) at (7,0);
\diagram* {
(a) -- [fermion] (m) -- [fermion] (b);
};
\draw (7.5,0) node {$+$};
\end{feynman}
\begin{feynman}
\vertex[circle, draw=black, inner sep=0pt, minimum size=8mm] (m) at (9,0) {1PI};
\vertex (a) at (8,0);
\vertex[circle, draw=black, inner sep=0pt, minimum size=8mm] (n) at (11,0) {1PI};
\vertex (c) at (12,0);
\diagram* {
(a) -- [fermion] (m) -- [fermion] (n) -- [fermion] (c);
};
\draw (13,0) node {$+\quad \cdots$};
\end{feynman}
\end{tikzpicture}
\end{center}
\caption{Fermion propagator obtained by summing all 1PI contributions.}
\label{1PI.expansion}
\end{figure}

In order to look for the pole mass, we must find the correction to the bare mass in the {\em denominator} of the propagator. To do this, there are some preliminary steps we must carry out. Let us define a class of Feynman diagrams called ``one-particle irreducible'', or 1PI for short. These are the diagrams which cannot be cut into two separate diagrams by slicing just one line. It is clear that any 1PI diagram must contain at least one loop. Now let us focus on the electron propagator. In Figure \ref{1PI.expansion}, we depict it as a combination of 1PI segments. Note that each 1PI contribution on the RHS of the figure contains contributions summed to all loop orders. 

We see that it is built up by stringing together a chain of 1PI contributions, each connected to the next one by free propagators. This allows a nice re-organisation of the perturbation series. Without calculating any individual 1PI diagram, we we will express the full propagator, as a function of the 1PI diagram, in the form of an infinite geometric series. Then we will sum up this series and use the result to understand where the pole of the interacting propagator is located. 

\subsubsection{Electron propagator revisited}

Henceforth we define $i\Sigma(p)$ as the sum of all 1PI diagrams for the fermion propagator. Notice that $\Sigma_2(p)$, which was defined earlier, is simply the second-order (equivalently, one-loop) contribution to $\Sigma(p)$, so the notation is consistent. But unlike $\Sigma_2$, the full $\Sigma$ contains different powers of $e$ (i.e. $e^2$ for one-loop diagrams, $e^4$ for two-loop diagrams and so on)\footnote{It is important to keep in mind that, just for these manipulations, we are not making any one-loop approximation but keeping all orders in perturbation theory. When we revert to the one-loop approximation, this will be explicitly noted.}. 

From Figure \ref{1PI.expansion}, the expression for  the bare propagator in terms of $\Sigma$ is seen to be:
\be
iG_0(p)=\frac{i}{\slashed p-m_0}\Bigg(1+\big(i\Sigma\big)\frac{i}{\slashed p-m_0}+\big(i\Sigma\big)\frac{i}{\slashed p-m_0}\big(i\Sigma\big)\frac{i}{\slashed p-m_0}+.....\Bigg)
\label{Gseries}
\ee
This geometric series can be summed up to find:
\be
iG_0(p)=\frac{i}{\slashed p-m_0+\Sigma(p)}
\label{Gsummed}
\ee

\exercise{Consider two matrices $A,B$ where $B$ is proportional to a small parameter, and show that:
$$
(A+B)^{-1}=A^{-1}-A^{-1}B A^{-1}+A^{-1}B A^{-1}B A^{-1}+\cdots
$$
Hence derive \eref{Gseries} from \eref{Gsummed}.}

Now, from Eq.~\eqref{GR}, we know that:
\be
\begin{split}
iG_R(p)&=\frac{1}{Z_2}iG_0(p)\\
&= \frac{1}{1+\delta_2}\left(\frac{i}{\slashed p -m_0+\Sigma(p)}\right)
\end{split}
\ee
Multiplying out the denominators and using $m_0=m_R(1+\delta_m)$ we find:
\be
\begin{split}
iG_R(p)&=\frac{i}{\slashed p-m_R+\delta_2 \slashed p-(\delta_2+\delta_m) m_R+(1+\delta_2)\Sigma(p)-\delta_2\delta_m m_R}\\
&=\frac{i}{\slashed p-m_R^{}+\Sigma_R(p)}
\label{GRsummed}
\end{split}
\end{equation} 
where  $\Sigma_R(p)$ is defined as 
\begin{equation}
\label{sigmar}
	\Sigma_R(p)=\delta_2 \slashed p-(\delta_2+\delta_m) m_R^{}+(1+\delta_2)\Sigma_(p)-\delta_2\delta_m m_R
\end{equation}
At this point we specialise to one-loop order by neglecting terms of order $e_R^4$ or higher. Thereupon $\Sigma_R(p)$ becomes equal to the expression already found in \eref{finiteSigma}:
\be
\Sigma_{2,R}(p)= \delta_2\slashed p-(\delta_2+\delta_m)m_R^{}+\Sigma_{2}(p)
\label{SigtwoR}
\ee
The virtue of summing the geometric series is that $\Sigma_{2,R}(p)$ now appears in the denominator of the propagator, which makes it possible for us to locate the pole correctly. 

As noted above, the value of $\slashed p$ at which the renormalised propagator $iG_R(p)$ has a simple pole with residue $i$ is called the pole mass. Eq.~\eqref{GRsummed} tells us that $G_R(p)$ does not diverge at $\slashed p=m_R$. Instead, it  has a simple pole at $\slashed p=m_R-\Sigma_R(p)$. Hence the pole mass $m_p$ is defined via the equation: 
\begin{equation}
\label{mp}
m_p=m_R-\Sigma_R(p)\Big|_{\slashed p=m_p}
\end{equation}
The fact that the residue at the pole is $i$ additionally implies:
\be
\label{dsigma}
\lim\limits_{\slashed p\rightarrow m_p} \frac{i(\slashed p- m_p)}{\slashed p-m_R^{}+\Sigma_R(p)}=i
\ee
Expanding the denominator around $\slashed p=m_p$, it is easily seen that this is equivalent to:
\be
\lim\limits_{\slashed p\rightarrow m_p} \frac{1}{1+\frac{d}{d\slashed p}\big[\Sigma_R(p)\big]}=1
\ee
which in turn implies:
\be
\frac{d\Sigma_R(p)}{d\slashed p}\bigg|_{\slashed p=m_p}=0
\label{rescond}
\ee
Thus we have two conditions, \eref{mp} and \eref{rescond}. The first one defines $m_p$ and the second sets the residue of the propagator equal to $i$. 

From \eref{mp} we can see what it takes to have $m_p=m_R$. For this, we must choose the subtraction scheme so that:
\be
\Sigma_R(p)\Big|_{\slashed p=m_p}=0
\label{Sigzero}
\ee
The renormalisation scheme in which this holds is called ``on-shell subtraction''. This differs from both MS and $\overline{\rm MS}$ schemes. 


\begin{figure}[H]
\begin{eqnarray*}
	\feynmandiagram[baseline=(a5.base)]{a1--[]--a2--[]--a3[crossed dot]--[]--a4--[]--a5,a2--[fermion]a3--[fermion]a4};
	\quad \quad &\implies& i\big[\delta_2\slashed p-(\delta_m+\delta_2)m_R^{}\big]\\[3mm]
	\feynmandiagram[baseline=(a5.base)]{a1--[]--a2--[]--a3[crossed dot]--[]--a4--[]--a5,a2--[photon]a3--[photon,]a4};
	\quad \quad&\implies& -i\delta_3 p^2 \eta^{\mu\nu}+\textrm{gauge fixing term}\\[3mm]
	\feynmandiagram[baseline=(b3.base)]{a1[]--[]--a2--[]--a3,b1--[]--b2[crossed dot]--[]--b3,c1--[]--c2--[]--c3,b3--[photon]b2, c1--[fermion]b2--[fermion]a1};
	\quad\quad &\implies& -ie_R\delta_1\gamma^\mu
\end{eqnarray*}
\caption{Feynman rules for counterterms in QED.}
\end{figure}


Let us implement this up to one-loop order. Then $\Sigma_{2,R}(p)$ is given by \eref{SigtwoR}. The quantity $\Sigma_2(p)$ on the RHS of this equation is just the one-loop 1PI contribution $\Sigma_2(p)$ that was calculated in \eref{sigmatwocomp} except that in the framework of renormalised perturbation theory, its arguments $e_0,m_0$ should be replaced by $e_R,m_R$:
\begin{equation}
\label{tsigma2}
\Sigma_2(p)=\frac{e_R^2}{8\pi^2}\int_0^1 dx\,(\slashed px-2m_R^{})\bigg[\frac{2}{\epsilon}-\ln\frac{(1-x)(m_R^2-p^2 x)}{\widetilde{\mu}^2}+\mathcal{O}(\epsilon)\bigg]
\end{equation}

Now let us take $\slashed p\to m_R$ and also require $m_R=m_p$. This will determine the finite parts of $\delta_j$ (see \eref{deltavals}). With these choices, \eref{tsigma2} becomes:
\begin{equation}
\begin{split}
\Sigma_2(p)\Big|_{\slashed p=m_p}&=\frac{e_R^2}{8\pi^2}\,m_p\int_0^1 dx\,(x-2)\bigg(\frac{2}{\epsilon}+\ln\Big(\frac{\widetilde{\mu}^2}{m_p^2}\Big)-2\ln(1-x)\bigg)\\[2mm]
&=\frac{e_R^2}{8\pi^2}\,m_p\bigg(- \frac{3}{\epsilon}-\frac{3}{2}\ln\Big(\frac{\widetilde{\mu}^2}{m_p^2}\Big)-\frac{5}{2}\bigg)
\end{split}
\end{equation}
It then follows from \eref{SigtwoR} that to one loop,
\begin{equation}
\Sigma_{2,R}(p)\Big|_{\slashed p=m_p}=-\delta_m m_p+\Sigma_2(p)\Big|_{\slashed p=m_p}
\end{equation}
Now \eref{Sigzero} tells us that the LHS of the above equation vanishes, hence:
\begin{equation}
\begin{split}
\delta_m&=\frac{1}{m_p}\Sigma_2(p)\Big|_{\slashed p=m_p}\\[2mm]
&=\frac{e_R^2}{8\pi^2}\bigg( -\frac{3}{\epsilon}-\frac{3}{2}\ln\Big(\frac{\widetilde{\mu}^2}{m_p^2}\Big)-\frac{5}{2}\bigg)
\end{split}
\end{equation}
The infinite part of $\delta_m$ agrees with what we had established in \eref{deltavals}. But now we have also determined the finite part.

Notice that we did not yet determine $\delta_2$. It is easy to understand why this happened. The location of the pole in the propagator provides information about mass renormalisation, while it is the residue at that pole that fixes the normalisation of the field (if we re-scale the field even by a finite factor, the two-point function will be scaled). This is precisely the remaining condition, \eref{rescond}, that sets the residue at the pole to be $i$. Let us now use this to determine $\delta_2$.

Differentiating \eref{tsigma2} and using $\slashed p^2=p^2$, we find: \begin{equation}
\frac{d\Sigma_2(p)}{d\slashed p}\bigg|_{\slashed p=m_p}=\frac{e_R^2}{8\pi^2}
\int_0^1 dx\,x\,\bigg[\frac{2}{\epsilon}-
\frac{2(2-x)}{1-x}-\ln\Big(\frac{m_p^2(1-x)^2}{\widetilde{\mu}^2}\Big)\bigg]
\end{equation}
Unfortunately, the integral: 
\be
\displaystyle\int_0^1 dx \,\frac{2x(2-x)}{1-x}
\ee
diverges at $x=1$. To regulate this divergence, we redefine the photon propagator:
\be
\frac{-i\eta^{\mu\nu}}{p^2}\quad\to\quad\frac{-i\eta^{\mu\nu}}{p^2-m_\gamma^2}
\label{ircutoff}
\ee 
where a photon mass $m_\gamma$ has been introduced as a regulator. On making this change, we find:
\begin{equation}
\Sigma_2(p,m_\gamma)=\frac{e_R^2}{8\pi^2}\int_0^1\! dx\,(\slashed px-2m_R^{})\Bigg[\frac{2}{\epsilon}-\ln\bigg(\frac{(1-x)(m_R^2-p^2 x)+xm_\gamma^2}{\widetilde{\mu}^2}\bigg)\Bigg]
\end{equation}
From this it follows that:
\begin{equation}
\frac{d\Sigma_2(p,m_\gamma)}{d\slashed p}\bigg|_{\slashed p=m_p}=\frac{e_R^2}{8\pi^2}\int_0^1\!\!\! dx\,x\,\Bigg[\frac{2}{\epsilon}\,-\,\frac{2(x-2)(x-1)m_p^2}{x m_\gamma^2 \,+\,(x-1)^2m_p^2}-\ln\bigg(\frac{(x-1)^2 m_p^2+xm_\gamma^2}{\widetilde{\mu}^2}\bigg)\Bigg]
\end{equation}
Notice that the integrand no longer diverges near $x=1$, due to the $m_\gamma$ term. Hence we may perform the integral. This gives a rather complicated answer but we only need to keep the terms that survive as 
$m_\gamma\rightarrow0$. With a little effort, we end up with:
\begin{equation}
\frac{d\Sigma_2(p)}{d\slashed p}\bigg|_{\slashed p=m_p}=\frac{e_R^2}{8\pi^2}\bigg[\frac{1}{\epsilon}+\frac{1}{2}\ln\Big(\frac{\widetilde{\mu}^2}{m_p^2}\Big)+\frac{5}{2}+\lim\limits_{m_\gamma\rightarrow0}\ln\Big(\frac{m_\gamma^2}{m_p^2}\Big)\bigg]
\end{equation}
Now using \eref{SigtwoR} and \eref{rescond} we get:
\begin{equation}
\frac{d\Sigma_{2,R}(p)}{d\slashed p}\bigg|_{\slashed p=m_p}=\delta_2+\frac{d\Sigma_2(p)}{d\slashed p}\bigg|_{\slashed p=m_p}=0
\end{equation}
This determines the remaining renormalisation constant:
\be
\delta_2=\frac{e_R^2}{8\pi^2}\bigg[-\frac{1}{\epsilon}-\frac{1}{2}\ln\Big(\frac{\widetilde{\mu}^2}{m_p^2}\Big)-\frac{5}{2}-\lim\limits_{m_\gamma\rightarrow0}\ln\Big(\frac{m_\gamma^2}{m_p^2}\Big)\bigg]
\label{delta.2.onshell}
\ee
Again, the divergent part (the pole in $\epsilon$) agrees with what we predicted in \eref{deltavals}, and on-shell subtraction has determined the finite part as well. However this time there is a new kind of divergence that arises when we take $m_\gamma\to 0$. Divergences of this kind, which arise when fields are massless, are called ``infrared (IR) divergences''. These are very different from the UV divergences that are the main subject of these notes. They are not taken care of by renormalisation, but rather arise from the physics of massless or very light particles. In the present case, it is a sign that on-shell renormalisation schemes do not work well in the presence of massless particles. We would not have encountered this IR divergence if we had used the MS or $\overline{\rm MS}$ schemes.

We conclude this subsection with a few comments about the advantages and disadvantages of on-shell subtraction. 
The advantages are as follows. In on-shell subtraction, $\delta_2$ and $\delta_m$ have terms that depend logarithmically on the renormalisation scale $\mu$. It is easy to verify that, when inserted into physical quantities like $G_R(p)$, these terms completely remove the dependence on $\mu$ in those answers and replace it by a dependence on the pole mass $m_p$. This means we no longer have to worry about having physical quantities depend (or even appear to depend) on an arbitrary renormalisation scale. Second, on-shell subtraction is not special to dimensional regularisation -- in any regularisation scheme, we can examine the pole in the propagator and require it to equal the renormalised mass.  The disadvantages are that the finite part one wants to subtract is rather complicated, and also there are sometimes IR divergences that require introducing a photon mass as a regulator.

It may seem from the above that the advantages of on-shell subtraction outweigh the disadvantages, but in fact the reverse is the case. The apparent dependence of physical quantities on $\mu$ in the MS scheme is a bonus, since it allows a simple and transparent derivation of renormalisation group (RG) equations -- we have seen a preliminary set of results in this direction for scalar field theory and will see more shortly. Moreover, in the MS and $\overline{\rm MS}$ schemes, the kind of IR divergences we encountered above do not occur. Finally, these latter schemes are helpful in preserving gauge invariance and performing multi-loop-order renormalisation. 
 
We conclude that dimensional regularisation, the trick of continuing the number of spacetime dimensions and subtracting just the pole parts (or an additional fixed factor depending on the Euler constant $\gamma_E$) is, for most purposes, the simplest and most effective way to renormalise continuum QFT and especially gauge theories. It is somewhat surprising that the most artificial scheme from the physical point of view, which requires continuing to an unphysical number of dimensions, is so successful! A number of essential developments in the renormalisation of gauge theories have relied on this approach while other schemes like momentum cutoff, Pauli-Villars regularisation etc are less often used. 

\subsubsection{Photon propagator revisited}

The one-loop corrected photon propagator has been discussed in Section \ref{vacpolsec}. We will now revisit what we did there, but in the context of renormalised perturbation theory. In particular this will determine the remaining renormalisation constant $\delta_3$. 

Recalling \eref{vacpol}, in renormalised perturbation theory the analogous equation is:
\begin{equation}
\label{grmn}
iG_R^{\mu\nu}=\frac{-i\eta^{\mu\nu}}{p^2}\Big[1- e_R^2\Pi_2(p^2)-\delta_3\Big]
\end{equation}
Again we have retained only the $\eta^{\mu\nu}$ part of the loop-correction term, and also defined the 1PI contribution $\Pi^{\mu\nu}(p)$ in terms of a scalar quantity $\Pi(p)$ just as was done in \eref{PIredef}. The new features in renormalised perturbation theory are that (i) $e_R$ appears in place of $e_0$, (ii) the counterterm $\delta_3$, associated to the renormalisation of the photon field, appears.

\exercise{Starting from the Lagrangian of renormalised perturbation theory, \eref{lagz}, derive the $\delta_3$ term in \eref{grmn}.}

Just as we did for the electron propagator above, we now make a 1PI decomposition of the photon propagator:

\begin{figure}[H]
\begin{center}
\begin{tikzpicture}
\begin{feynman}
\vertex[blob] (m) at (0,0) {};
\vertex (a) at (-1,0);
\vertex (b) at (1,0);
\diagram* {
(a) -- [boson] (m) -- [boson] (b);
};
\draw (1.5,0) node {$=$};
\end{feynman}
\begin{feynman}
\vertex (a) at (2,0);
\vertex (b) at (4,0);
\diagram* {
(a) -- [boson] (b);
};
\draw (4.5,0) node {$+$};
\end{feynman}
\begin{feynman}
\vertex[circle, draw=black, inner sep=0pt, minimum size=8mm] (m) at (6,0) {1PI};
\vertex (a) at (5,0);
\vertex (b) at (7,0);
\diagram* {
(a) -- [boson] (m) -- [boson] (b);
};
\draw (7.5,0) node {$+$};
\end{feynman}
\begin{feynman}
\vertex[circle, draw=black, inner sep=0pt, minimum size=8mm] (m) at (9,0) {1PI};
\vertex (a) at (8,0);
\vertex[circle, draw=black, inner sep=0pt, minimum size=8mm] (n) at (11,0) {1PI};
\vertex (c) at (12,0);
\diagram* {
(a) -- [boson] (m) -- [boson] (n) -- [boson] (c);
};
\draw (13,0) node {$+\quad \cdots$};
\end{feynman}
\end{tikzpicture}
\end{center}
\caption{Photon propagator obtained by summing all 1PI contributions.}
\label{1PI.expansion.photon}
\end{figure}

The 1PI contribution to one loop is:
\begin{equation}
\Pi_2(p^2)=\frac{1}{6\pi^2\epsilon}-\frac{1}{2\pi^2}\int_{0}^{1}dx \textrm{ } x(1-x) \ln \bigg\{\frac{m_R^2-p^2 x(1-x)}{\widetilde{\mu}^2}\bigg\}+\mathcal{O}(\epsilon)
\end{equation}
Now we can find the pole mass for the photon. As before, this is done by summing up  the contribution of all 1PI diagrams to the full photon propagator, $i\Pi^{\mu\nu}(p)$.
 
Gauge invariance requires $p_\mu\Pi^{\mu\nu}=0$ (this is an example of a Ward identity, which we will discuss in more detail later on), hence the photon propagator must be of the form:
\begin{equation}
i\Pi^{\mu\nu}(p)=i(-\eta^{\mu\nu}p^2+p^\mu p^\nu)\Pi(p^2)
\label{propform}
\end{equation} 
where $\Pi(p^2)$ is regular at $p^2=0$. Focusing as before only on the $\eta^{\mu\nu}$ part of the free propagator, the summed photon two-point function is:
\begin{equation}
\begin{split}
iG^{\mu\nu}_2(p)&=\frac{-i\eta^{\mu\nu}}{p^2}+\frac{-i\eta^{\mu\rho}}{p^2}i\Pi_{\rho\lambda}(p)\frac{-i\eta^{\lambda\nu}}{p^2}+...\\
&=\frac{-i\eta^{\mu\nu}}{p^2}\left(1-\Pi(p^2)+\big(\Pi(p^2)\big)^2+\cdots \right)\\
&=\frac{-i\eta^{\mu\nu}}{p^2(1+ \Pi(p^2))}
\end{split}
\end{equation}
where in the second step we used \eref{propform} and in the last step we summed the geometrical series.
This shows that $G^{\mu\nu}_2(p)$ still has a singularity at $p^2=0$, even after summing all 1PI diagrams. Thus the pole mass of the photon, which was originally zero, remains zero upon incorporating loop corrections! Ultimately this is due to gauge invariance, which allowed us to extract an overall tensor factor of order $p^2$ in \eref{propform}.

To one-loop order we have $\Pi(p^2)=e_R^2\Pi_2(p^2)+\delta_3$. In the on-shell subtraction scheme we determine $\delta_3$ by requiring $\Pi(p^2)=0$ on-shell, i.e. at $p^2=0$. This gives:
\begin{equation}
\delta_3+e_R^2\Pi_2(p^2)\bigg|_{p=0}=0
\ee
From \eref{pitwo} we then find:
\be
\delta_3=-\frac{e_R^2}{6\pi^2\epsilon}
-\frac{e_R^2}{12\pi^2}\ln\frac{\widetilde{\mu}^2}{m_R^2}
\ee
On the other hand in the MS scheme the renormalisation constant $\delta_3$ is just the pole part, namely:
\be
\delta_3=-\frac{e_R^2}{6\pi^2\epsilon}
\end{equation}
while in the $\overline{\rm MS}$ scheme it is:
\be
\delta_3=-\frac{e_R^2}{6\pi^2\epsilon}
-\frac{e_R^2}{12\pi^2}\ln\Big(4\pi e^{-\gamma_E}\Big)
\ee

\subsection{Relations among renormalisation constants}

To summarise our calculations of renormalisation constants so far, we have computed the values of $\delta_2,\delta_3,\delta_e,\delta_m$ in various schemes. The last remaining renormalisation constant (not counting the one for the gauge-fixing parameter $\xi$) is $\delta_1$, which is not really an independent constant but is equal to $\delta_e+\delta_2+\half\delta_3$ (see the comment below \eref{lagz}). Adding this into the list, the results in the MS scheme can be summarised as:
\be
\delta_1=-\frac{e_R^2}{8\pi^2\epsilon},\quad \delta_2=-\frac{e_R^2}{8\pi^2\epsilon},\quad \delta_3=-\frac{e_R^2}{6\pi^2\epsilon},\quad \delta_e=\frac{e_R^2}{12\pi^2\epsilon},\quad \delta_m=-\frac{3e_R^2}{8\pi^2\epsilon}
\label{alldeltavals}
\ee
We notice an apparently coincidental feature that, to this order and in this scheme, $\delta_1=\delta_2$. In fact this is not a coincidence but a consequence of a famous relation, $Z_1=Z_2$, which is exact to all orders in perturbation theory. This relation follows from gauge invariance and can be proved using the Ward identities. Instead of providing a detailed derivation of this fact, we will present a shorter and somewhat intuitive argument. 

Recall that $Z_1$ renormalises the fermion-photon interaction:
\begin{equation}
\label{z1z2z3}
-e_0\overline{\psi}_0\gamma_\mu\psi_0 A_0^\mu= -Z_1 e_R^{}\overline{\psi}_R\gamma_\mu\psi_R^{} A_R^\mu
\end{equation}
while $Z_2$ renormalises the fermion kinetic term:
\begin{equation}
\label{fermionkin}
i\overline{\psi}_0 \slashed \partial\psi_0=
i Z_2\overline{\psi}_R \slashed \partial\psi_R
\end{equation}
Now define $\hat{A}_0^\mu=e_0A_0^\mu$. In terms of this scaled vector field, the bare fermion kinetic term and fermion-photon interaction term can be written together as:
\be
\overline{\psi}_0(i\slashed\partial- \hat{\!\!\slashed{A}}_0)\psi_0
\label{barescaled}
\ee
In these variables, the gauge transformation can be written in a form that is completely independent of coupling constants:
\be
\delta \hat{A}_0^\mu=\partial^\mu \alpha,\qquad  \delta\psi=i\alpha\psi
\ee
Rewriting \eref{barescaled} in terms of renormalized fields, we have:
\begin{equation}
\overline{\psi}_0\big(i\slashed\partial- \hat{\!\!\slashed{A}}_0\big)\psi_0= Z_2\overline{\psi}_R\Big(i\slashed\partial-\frac{Z_1}{Z_2}~ \hat{\!\!\slashed{A}}_R\Big)\psi_R
\label{renormgauge}
\end{equation}
The gauge transformation above, being independent of coupling constants, should be a symmetry of the classical as well as the quantum theory. Clearly it leaves invariant the LHS of \eref{renormgauge}, which describes the classical action. Therefore it should continue, without any changes, to be a symmetry of the RHS of the same equation. This will only be the case if $Z_1=Z_2$.

One consequence of this crucial relation is that $e_0 A^\mu_0=e_R^{} A^\mu_R$. Another consequence, following from \eref{z1z2z3}, is that $Z_e=\frac{1}{\sqrt{Z_3}}$ or equivalently $\delta_e=-\frac{1}{2}\delta_3$, which is manifestly true in \eref{alldeltavals} but again, is actually true to all orders. Finally, note that the ratio of electric charges of the up-type quark and the electron is experimentally determined to be a fixed fractional number, $-\frac{2}{3}$. That this statement is an exact one, unaffected by renormalisation, is again a consequence of $Z_1=Z_2$.

\subsection{Vertex correction}

We now turn to the last divergent diagram at one loop in QED, namely the three-point vertex -- the last diagram in Figure \ref{oneloopQED}. The calculation will not present any significant new conceptual features, but we will go through it both for completeness and because it provides a verification of the relation $Z_1=Z_2$.

We take the fermion momenta to be $q_1$ (incoming) and $q_2$ (outgoing). The outgoing photon momentum $p^\mu$ is clearly equal to $q^\mu_1-q^\mu_2$. In order to focus only on the one-particle irreducible three-point interaction, we amputate external propagators. Thereafter, for the present discussion, we take $q_1,q_2$ to be on-shell, but $p$ is allowed to remain off-shell. Denote the resulting object by $-ie_R\Gamma^\mu$. Clearly $-ie_R\Gamma^\mu$ is given by computing $\langle A^\mu(y)\psi(x_1)\psibar(x_2)\rangle$ and then amputating the external legs. At tree level we have:
\be
\Gamma^\mu=\gamma^\mu
\ee
The corresponding matrix element is defined as:
\be
i\cM^\mu =-ie\,\ubar(q_2)\Gamma^\mu u(q_1)
\ee

Using Lorentz invariance and the on-shell conditions $\slashed{q_1}u(q_1)=mu(q_1)$ and $\slashed{q_2} {\bar u}(q_2)=m {\bar u}(q_2)$, we can parametrise $\Gamma^\mu$ as:
\be
\Gamma^\mu(q_1,q_2)=A(p^2)\gamma^\mu + B(p^2)(q_1+q_2)^\mu + C(p^2)(q_1-q_2)^\mu
\label{Gammaparam}
\ee
where the coefficients $A,B,C$ can depend on $p^2$ as indicated, and will of course depend on the mass and coupling constant as well. 

The single one-loop diagram contributing to $\Gamma^\mu$ is called the vertex correction. We denote this correction by $i\cM_2^\mu$. It is the contribution that arises by bringing down one interaction term $\int A_\nu(x)\psibar(x)\gamma^\nu\psi(x)$ from the interaction Hamiltonian and contracting it to get:
\be
\begin{split}
&\langle \left(\int d^4x\, A_\nu(x)\psibar(x)\gamma^\nu\psi(x)\right)A^\mu(y)\psi(x_1)\psibar(x_2)\rangle\\
&=
\int d^4 x \,\langle A_\nu(x)A^\mu(y)\rangle \langle \psibar(x)\gamma^\nu\psi(x)\psi(x_1)\psibar(x_2)\rangle\\
&=
\int d^4 x \,\langle A_\nu(x)A^\mu(y)\rangle \langle j^\mu(x)\psi(x_1)\psibar(x_2)\rangle
\end{split}
\ee
Here we have denoted the bilinear $\psibar\gamma^\mu \psi$ by $j^\mu$, which classically is the conserved current of the charged fermion. The second factor in the last line is precisely what we want to compute. After transforming to momentum space, we may write:
\be
i\cM_2^\mu(p,q_1,q_2)=\int d^4x\,d^4x_1\,d^4x_2\,e^{ip\cdot x}e^{iq_1\cdot x_1}e^{-iq_2\cdot x_2}
\langle j^\mu(x)\psi(x_1)\psibar(x_2)\rangle
\ee

Before computing this, let us see what happens if we contract the above equation with $p_\mu$. On the RHS we find the Fourier transform of the three-point function:
\be
\langle \del_\mu j^\mu(x)\psi(x_1)\psibar(x_2)\rangle
\ee
Because $j^\mu(x)$ is a conserved current, one might think that this should vanish. In fact, the above statement is true only as long as $x_1,x_2$ are well-separated from $x$. If we lift this restriction then there are extra terms arising from the possible coincidence of $x_1$ or $x_2$ with $x$. Indeed, the correct identity turns out to be:
\be
\langle \del_\mu j^\mu(x)\psi(x_1)\psibar(x_2)\rangle=-\delta^4(x-x_1)\langle\psi(x_1)\psibar(x_2)\rangle
+\delta^4(x-x_2)\langle\psi(x_1)\psibar(x_2)\rangle
\label{deljcorr}
\ee
A simple derivation may be found in Section 14.8.1 of \cite{Schwartz:2013pla} and the reader is urged to look it up. It is also shown there that, if we sandwich the above expression between on-shell spinors, then the ``contact terms'' on the RHS vanish. Thus we have sketched a proof that, at least to one-loop order, 
the vertex satisfies: 
\be
p_\mu \ubar(q_2)\Gamma^\mu u(q_1)=0
\label{divGamma}
\ee
This is known as a Ward identity. It is true only under the condition, which we imposed above, that $q_1,q_2$ are on-shell. Also though we only need this one-loop version of the Ward identity in what follows, it can be shown that the above identity holds to all orders in perturbation theory and even non-perturbatively.

We are now in a position to explain the result, promised several times in the preceding sections, that $\delta_1=\delta_2$ follows from gauge invariance. 
For this, let us sandwich \eref{deljcorr} between {\em off-shell} spinors. Thus we are no longer taking $q_1,q_2$ be on-shell momenta. In this case the contact terms no longer vanish. It is straightforward to show that in this case, the analogue of the Ward identity is the more general result: 
\be
p_\mu \ubar(q_2)\Gamma^\mu u(q_1)= i\left(G^{-1}(q_1)-G^{-1}(q_2)\right)
\label{WTid}
\ee
where $G(q)$ is the full electron propagator with (off-shell) momentum $q$.This is called the Ward-Takahashi identity. The RHS above clearly vanishes if both $q_1$ and $q_2$ are on-shell, and in this limit the Ward-Takahashi identity reduces to the Ward identity.

\exercise{Starting from \eref{deljcorr}, verify \eref{WTid}. Note that this also verifies \eref{divGamma} when the fermion lines are on-shell.}

The above identity allows us to relate renormalisation constants. Consider $\Gamma^\mu$ in the limit that the photon momentum vanishes. We have:
\be
\Gamma^\mu(p\to 0)=\frac{1}{Z_1}\gamma^\mu
\ee
On the other hand, we know that:
\be
G(q)=\frac{iZ_2}{\slashed{q}-m}
\ee
Inserting these into \eref{WTid} we see that:
\be
\frac{1}{Z_1}\slashed{p}=\frac{1}{Z_2}\left(\slashed{q_1}-\slashed{q_2}\right)
\ee
Since $p=q_1-q_2$, it follows that $Z_1=Z_2$. 

Let us now ask what conditions the gauge-invariance equation \eref{divGamma} imposes on the coefficients $A,B,C$ in \eref{Gammaparam}. Recalling that $p^\mu=q_1^\mu-q_2^\mu$, we have:
\be
p_\mu\,\ubar(q_2)\gamma^\mu u(q_1)=\ubar(q_2)(\slashed{q_1}-\slashed{q_2}) u(q_1)
=0
\ee 
by the on-shell conditions, and similarly:
\be
p_\mu (q_1+q_2)^\mu=q_1^2-q_2^2=0
\ee
Hence the requirement in \eref{divGamma} places no conditions on the coefficients $A$ and $B$. However since $p\cdot(q_1-q_2)\ne 0$, we must have $C(p^2)=0$.

Thus $\Gamma^\mu$ is determined in terms of two scalar functions $A(p^2),B(p^2)$. By convention a slightly different parametrisation of $\Gamma^\mu$ is preferred. For this we use the Gordon identity:
\be
\ubar(q_2)\gamma^\mu u(q_1)=\frac{1}{2m}\ubar(q_2)\Big( (q_1+q_2)^\mu+i\Sigma^{\mu\nu}p_\nu\Big)u(q_1)
\ee
where $\Sigma^{\mu\nu}\equiv \frac{i}{2}[\gamma^\mu,\gamma^\nu]$.

\exercise{Derive the Gordon identity, which holds when the spinors $u(q_i)$ are on-shell.}

As a consequence of this identity and the fact that $C=0$, we can choose $\Gamma^\mu$ to be proportional to the two 4-vectors $\gamma^\mu, \Sigma^{\mu\nu}p_\nu$. Thus we have:
\be
\Gamma^\mu =\gamma^\mu F_1(p^2)  +\frac{i\Sigma^{\mu\nu}}{2m}p_\nu F_2(p^2)
\label{GammaFparam}
\ee
where the relations are always assumed to hold after sandwiching between on-shell spinors. It only remains to compute the functions $F_1,F_2$.

\exercise{Find the precise relation between $F_1,F_2$ in \eref{GammaFparam} and $A,B$ in \eref{Gammaparam}.}

In renormalised perturbation theory, we have:
\be
-ie_R\Gamma^\mu=-ie_R\gamma^\mu+i\cM_2^\mu -ie_R\delta_1\gamma^\mu
\label{Gammaform}
\ee
Once we perform a one-loop computation to determine $\cM_2$, we will be able to read off the value of $\delta_1$.

At tree level, $F_1=1$ and $F_2=0$, and both receive corrections at one loop. Indeed from the above equation we see that the $F_2$ term, if any, resides entirely in $\cM_2$ which is a one-loop contribution. Moreover the counterterm proportional to $\delta_1$ in \eref{Gammaform} contributes only to $F_1$. Hence we will be in trouble if $F_2$ diverges -- there would be no available counterterm to cancel it! Fortunately, the internal consistency of quantum electrodynamics is so powerful that $F_2$ turns out to be finite. 

Following the usual Feynman rules for QED we easily write down the one-loop integral:
\be
i\cM_2^\mu = (-ie_R)^3\int \frac{d^4k}{(2\pi)^4}\ubar(q_2)\gamma^\nu\frac{\slashed p +\slashed k +m_R}{(p+k)^2-m_R^2}
\gamma^\mu \frac{\slashed k +m_R}{k^2-m_R^2}\gamma_\nu u(q_1)\frac{1}{(k-q_1)^2}
\ee
Let us first verify that the contribution to $F_2$ is finite. From \eref{GammaFparam} we see that the $F_2$ contribution arises from the term linear in $p$ in the above equation. Once we pick this, the rest of the numerator is at most linear in the loop momentum $k$ while the denominator goes like $\sim k^6$. Including four powers of $k$ from the integration measure, we see that the numerator scales at most like $\sim k^5$ and hence the loop integral is indeed convergent. This term is physically important as it contributes to the anomalous magnetic moment of the electron. A straightforward calculation shows that:
\be
F_2(p^2=0)=\frac{e^2}{8\pi^2}=\frac{\alpha}{2\pi}
\ee
This is a famous result, first derived in 1948 by J. Schwinger.

Since our primary focus here is on renormalisation, we now move on to the computation of $F_1$. 
For this, we use an identity that allows us to combine three denominators:
\be
\frac{1}{ABC}=2\int_0^1 dx \int_0^1 dy \int_0^1 dz~ \delta(x+y+z-1)\frac{1}{(Ax+By+Cz)^3}
\ee
Using this one can show that:
\be
i\cM_2^\mu = -e_R f(p^2)\gamma^\mu + \cdots
\label{mtwoexp}
\ee
where we have only shown the term that contributes to $F_1$, and:
\be
f(p^2)=-2ie_R^2\int \frac{d^4k}{(2\pi)^4}\int_0^1 dx\,dy\,dz~\delta(x+y+z-1)\frac{k^2-2(1-x)(1-y)p^2-2(1-4z+z^2)m_R^2}{(k^2-\Delta)^3}
\label{fpsq}
\ee
Here,
\be
\Delta=m_R^2(1-z)^2 - xy\,p^2
\ee
In addition to the usual UV divergence of this integral (both numerator and denominator grow as $\sim k^6$), there is also an IR divergence. The former is handled via dimensional regularisation as usual, while for the latter we give the photon a small but finite mass $m_\gamma$ just as we did in \eref{ircutoff} while evaluating the electron propagator. The evaluation of this momentum integral is technically more involved than previous ones that we have dealt with, so we only quote the answer: 
\be
f(p^2)=\frac{e_R^2}{8\pi^2}\Bigg(\frac{1}{\epsilon}-\half +\int_0^1 dx\,dy\,dz\left[\frac{p^2(1-x)(1-y)+m_R^2(1-4z+z^2)}{\Delta+zm_\gamma^2}+\ln\frac{{\tilde \mu}^2}{\Delta+zm_\gamma^2}\right]\Bigg)
\ee
The detailed steps can be found in [Schwarz]. We may now insert the above expression as well as \eref{mtwoexp} into \eref{Gammaform} to get:
\be
-ie_R\Gamma^\mu=-ie_R\gamma^\mu\Big(1+f(p^2)+\delta_1\Big)
\ee
ignoring terms not proportional to $\gamma^\mu$ which, as we have seen, are finite.

At this stage, we can choose to carry out on-shell renormalisation. The on-shell condition is $p^2=0$ so all we need is to set $\delta_1=-f(0)$. From \eref{fpsq} we easily find:
\be
\delta_1=-\frac{e_R^2}{8\pi^2}\left(\frac{1}{\epsilon}+\half\ln\frac{\tilde\mu^2}{m_R^2}+\frac52+\ln\frac{m_\gamma^2}{m_R^2}\right)
\label{delta.1.onshell}
\ee
Comparing with \eref{delta.2.onshell}, we see that to one-loop order, the prediction $\delta_1=\delta_2$ has been verified.

While we have found $\delta_1=\delta_2$ in the on-shell renormalisation scheme, it is easy by inspection to find the corresponding quantities in the MS or $\overline{\rm MS}$ schemes and verify that they are equal to each other. For example, in MS we have:
\be
\delta_1=\delta_2=-\frac{e_R^2}{8\pi^2\epsilon}
\label{somedeltas}
\ee
Since we have reached the end of our computations of renormalisation constants in QED, let us for completeness list the MS values of the remaining constants that we computed at one-loop order:
\be
\label{otherdeltas}
\delta_m=-\frac{3e_R^2}{8\pi^2 \epsilon},\qquad \delta_e=-\frac{1}{2}\delta_3=\frac{e_R^2}{12\pi^2 \epsilon}
\ee

\subsection{Generalities about renormalisation}

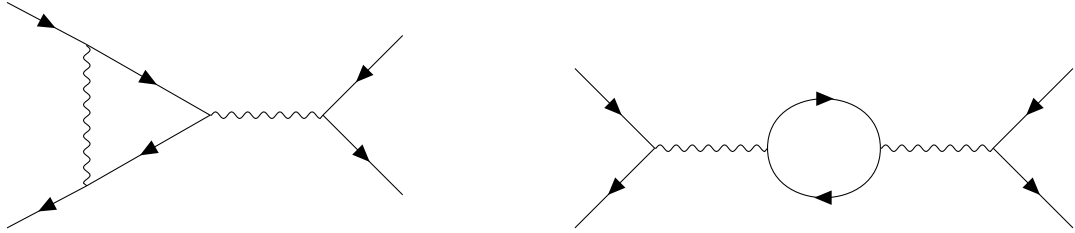
\begin{figure}[H]
\begin{center}
\begin{tikzpicture}
\begin{feynman}
\vertex (g);
\vertex [above=of g] (a);
\vertex [below=of g] (d);
\vertex [below right=of a, yshift=5mm] (b);
\vertex [right=of g, xshift=12mm] (c);
\vertex [above right=of d, yshift=-5mm] (e);
\vertex [right=of c] (f);
\vertex [above right=of f] (h);
\vertex [below right=of f] (i);
\diagram* {
(a) -- [fermion] (b) -- [fermion] (c);
(c) -- [fermion] (e) -- [fermion] (d);
(b) -- [boson] (e);
(c) -- [boson] (f);
(h) -- [fermion] (f) -- [fermion] (i)
};
\end{feynman}
\end{tikzpicture}
\hspace*{20mm}
\begin{tikzpicture}
\begin{feynman}
\vertex (c);
\vertex [above left=of c] (a);
\vertex [below left=of c] (b);
\vertex [right=of c] (f);
\vertex [right=of f] (g);
\vertex [right=of g] (h);
\vertex [above right=of h] (k);
\vertex [below right=of h] (l);
\diagram* {
(a) -- [fermion] (c) -- [fermion] (b);
(k) -- [fermion] (h) -- [fermion] (l);
(c) -- [boson] (f)
(g) -- [boson] (h)
(f) -- [fermion, half left] (g);
(g) -- [fermion, half left] (f);
};
\end{feynman}
\end{tikzpicture}
\caption{Some additional one-loop-divergent diagrams.} 
\label{divdiag}
\end{center}
\end{figure}

We have now computed the one-loop divergences for three basic Green's functions: the photon propagator (``vacuum polarisation''), the electron propagator (``electron self-energy'') and the electron-photon vertex (``vertex correction'') and showed how these are renormalised. We will not discuss higher loops in these notes. But even at one loop, it would seem that there is much left to be done. Certainly there are many more Feynman diagrams, such as the ones shown in Figure \ref{divdiag}, that are divergent at one loop.

The good news is that {\em all such diagrams are finite to a given loop order if the theory has been renormalised to render the three basic Green's functions finite.} To appreciate this result, recall that in Section \ref{1PI} we defined the notion of one-particle irreducible, or 1PI, diagrams. Clearly all Feynman diagrams can be obtained by attaching the external legs of 1PI diagrams with free propagators. This process does not introduce any new loop diagrams, hence there are no new divergences. Thus it is sufficient to render all 1PI diagrams finite by renormalisation, because this in turn will ensure that all diagrams will be finite.

Turning now to the 1PI diagrams, we may define their {\em superficial degree of divergence} $D$ as the total number of powers of loop momenta (including the integration measure) in the integral. A simple way to compute this is to scale all loop momenta by a common factor: $k_\mu^I\to \lambda k_\mu^I$, then the diagram will be scaled by a factor $\lambda^D$ and this tells us the value of $D$. Clearly the diagram is divergent if $D\ge 0$ (when $D=0$ it is said to be logarithmically divergent). However for $D<0$ the diagram may be divergent or finite. For example it could have some loop integrals that diverge and contribute a total power of $\lambda^{d_1}$ with $d_1\ge 0$, while other loop integrals converge and contribute a total power of $\lambda^{d_2}$ with $d_2<0$. The overall superficial degree of divergence of the diagram will then be  $D=d_1-d_2$ and this can very well be negative. Thus it would appear that we need to worry about both classes of diagrams, those with $D\ge 0$ and those with $D<0$. 

Before addressing what to do next, let us outline a simple proof that $D$ depends only on the {\em process} being considered (defined by the external lines) and not by the individual diagram contributing to it. This enormously simplifies the task of computing $D$. The proof goes as follows. In QED in 3+1 space-time dimensions, each loop momentum measure $d^4k$ gives us 4 powers of momentum. Each photon propagator behaves for large $k$ like $k^{-2}$, while each fermion propagator behaves as $k^{-1}$. Thus if we have $\ell$ loops in the diagram, $I_f$ internal fermion propagators and $I_\gamma$ internal photon propagators, then the superficial degree of divergence $D$ of the graph is:
\be
D=4\ell-2I_\gamma-I_f
\ee
Now the value of $\ell$ can in turn be determined in terms of the number of internal propagators and the number of vertices $V$. Each internal propagator gives rise to a momentum integral when it is Fourier transformed to momentum space. Thus there is a total of $I_f+I_\gamma$ momentum integrals to start with. However each vertex removes one of them by imposing momentum conservation. Therefore it seems we should subtract $V$. But that is too much, because overall momentum conservation is already imposed. Thus vertices contribute only $V-1$ delta-functions, and we have:
\be
\ell=I_f+I_\gamma-(V-1)
\ee
Now let $E_f,E_\gamma$ be the number of external fermion and photon propagators respectively. Then we
can find relations between the total number of (internal as well as external) propagators of a given type, and the number of vertices. For example, each vertex has exactly one photon propagator. If this is internal it is shared between two vertices, while if it is external then it joins onto exactly one vertex. Thus we find:
\be
V=2I_\gamma+E_\gamma
\ee
Applying the same logic to fermion vertices, but noting that there are two fermion propagators at each vertex, we find:
\be
2V=2I_f+E_f
\ee
Between the four equations above, we can eliminate $\ell,I_f,I_\gamma$ and $V$ to get:
\be
D=4-E_\gamma-\sfrac32 E_f
\ee
As promised, this depends only on the process being considered, not the loop order to which we are working or any specific diagram in that order. For example we see that the amplitude for electron-positron annihilation ($E_f=4,E_\gamma=0$) has $D=-2$. In fact the only diagrams with $D\ge 0$ are the photon propagator with $D=2$, the electron propagator with $D=1$, the QED vertex ($E_\gamma=1,E_f=2$) with $D=1$ and the photon four-point function with $D=0$. 

Now, a profound result called the BPHZ theorem assures us that we do not need to worry about diagrams with $D<0$. The theorem says we need to introduce counterterms {\em only for 1PI diagrams with $D\ge 0$}. Once we have done this, all diagrams with $D<0$, even if divergent in the bare theory, will automatically be rendered finite by the counterterms. For QED, an immediate consequence is that the electron propagator correction, photon propagator correction and vertex correction are the only diagrams for which we need counterterms (the photon four-point function does not feature due to gauge invariance, see below). Thus our calculations in this chapter are sufficient to renormalise all QED amplitudes at one-loop. But in fact the BPHZ theorem is true to all orders in perturbation theory, and it follows that QED is renormalisable to all orders. This result was arrived at after decades of hard work by many researchers, and is a key reason why quantum field theory is taken seriously at all.

An added wrinkle due to gauge invariance is that the true degree of divergence is often less than that counted by the method given above. In fact we have seen that the photon propagator and electron propagator are only logarithmically divergent at one loop, corresponding to $D=0$, rather than having $D=2,1$ respectively as given by ``naive'' power counting. Similarly it turns out that the photon four-point function is finite due to gauge invariance. While these properties are easy to show explicitly at one loop, they persist at all loops and can be proved using Ward-Takahashi identities. In the presence of gauge invariance, the most powerful method to prove all-orders renormalisability is based on BRST invariance, which we will encounter towards the end of these notes.

\newpage

\section{Renormalization group equations}

\subsection{The beta function}
	
In the process of renormalisation we were forced to introduce a mass scale. This fact is most evident in dimensional regularisation, where the parameter $\mu$ was brought in to make formulae dimensionally consistent. In some schemes this scale dependence can be hidden -- for example in the on-shell scheme we picked an external momentum to set the scale, while in other schemes we renormalised at zero momentum. The point is that in principle there is always a scale introduced by the renormalisation procedure. The question then is, in what way do the predictions of the theory depend on the renormalisation scale?

The answer is extraordinarily subtle and beautiful. The theory as a whole is independent of the choice of the scale. However, the consequences of perturbation theory are indeed scale-dependent. To get the most reliable results to a given order of perturbation theory, one has to choose a scale that minimises the magnitude of the higher-loop corrections that we have ignored. This is often possible, and has become a routine procedure when theoretical calculations in high-energy physics are to be compared with experiment.

 Because the scale is manifest in dimensional regularisation, we will exclusively consider that procedure from now on, together with the minimal subtraction (MS) scheme. Recall that in this scheme the Lagrangian of QED is given by \eref{lagz} where all the $\delta_i$ are pure simple poles in $\epsilon$. Focusing on charge renormalisation, we have:
\be
e_0=\mu^\frac{\epsilon}{2}Z_e e_R, \qquad Z_e=1+\delta_e,\qquad \delta_e=-\frac{e_R^2}{12\pi^2\epsilon}
\ee
As we have seen before, the bare parameter $e_0$ must not depend on the renormalisation scale $\mu$. This leads to the $\beta$-function equation, which is derived as follows:
 \be
 \label{eze}
 \frac{de_0}{d\mu}=\frac{d}{d\mu}(\mu^{\epsilon/2} Z_e e_R^{})=0
 \ee
 Expanding the RHS and rearranging, we find that:
 \be
 \mu\frac{d e_R^{}}{d\mu}=-\frac{\epsilon e_R^{}}{2}-\frac{e_R^{}}{Z_e}\left(\mu\frac{dZ_e}{d\mu}\right)
 \ee
 Now to one loop order,
 \begin{equation}
\mu\frac{dZ_e}{d\mu}=\mu\frac{d\delta_e}{d\mu}=\frac{e_R^{}}{6\pi^2\epsilon}\Big(\mu\frac{de_R^{}}{d\mu}\Big)
\end{equation}
Inserting this back into \eref{eze} and keeping only terms up to order $e_R^4$ on the RHS, we find:
\begin{equation}
\label{beta1}
\mu\frac{de_R^{}}{d\mu}=-\frac{\epsilon e_R^{}}{2}+\frac{e_R^3}{12\pi^2}+\mathcal{O}(e_R^5)
\end{equation}
Taking $\epsilon\rightarrow 0$ and defining the {\em beta-function}:
\be
\beta(e_R(\mu))\equiv \mu\frac{de_R^{}}{d\mu}
\ee 
we finally get, for QED at one loop:
\begin{equation}
\label{beta2}
\beta(e_R)=\frac{e_R^3}{12\pi^2}
\end{equation}
This equation is easily integrated to find the relation between $e_R$ at two different scales $\mu_0$ and $\mu$:
\begin{equation}
\label{ersq}
e_R^2(\mu)=\frac{e_R^2(\mu_0)}{1-\frac{e_R^2(\mu_0)}{6\pi^2}\ln\big(\frac{\mu}{\mu_0}\big)}
\end{equation}
It should be evident that this result is closely analogous, both in the derivation and the outcome, to the one derived in \eref{lambdasolv} for scalar field theory. 

\begin{figure}[H]
\begin{center}
\includegraphics[scale=0.5]{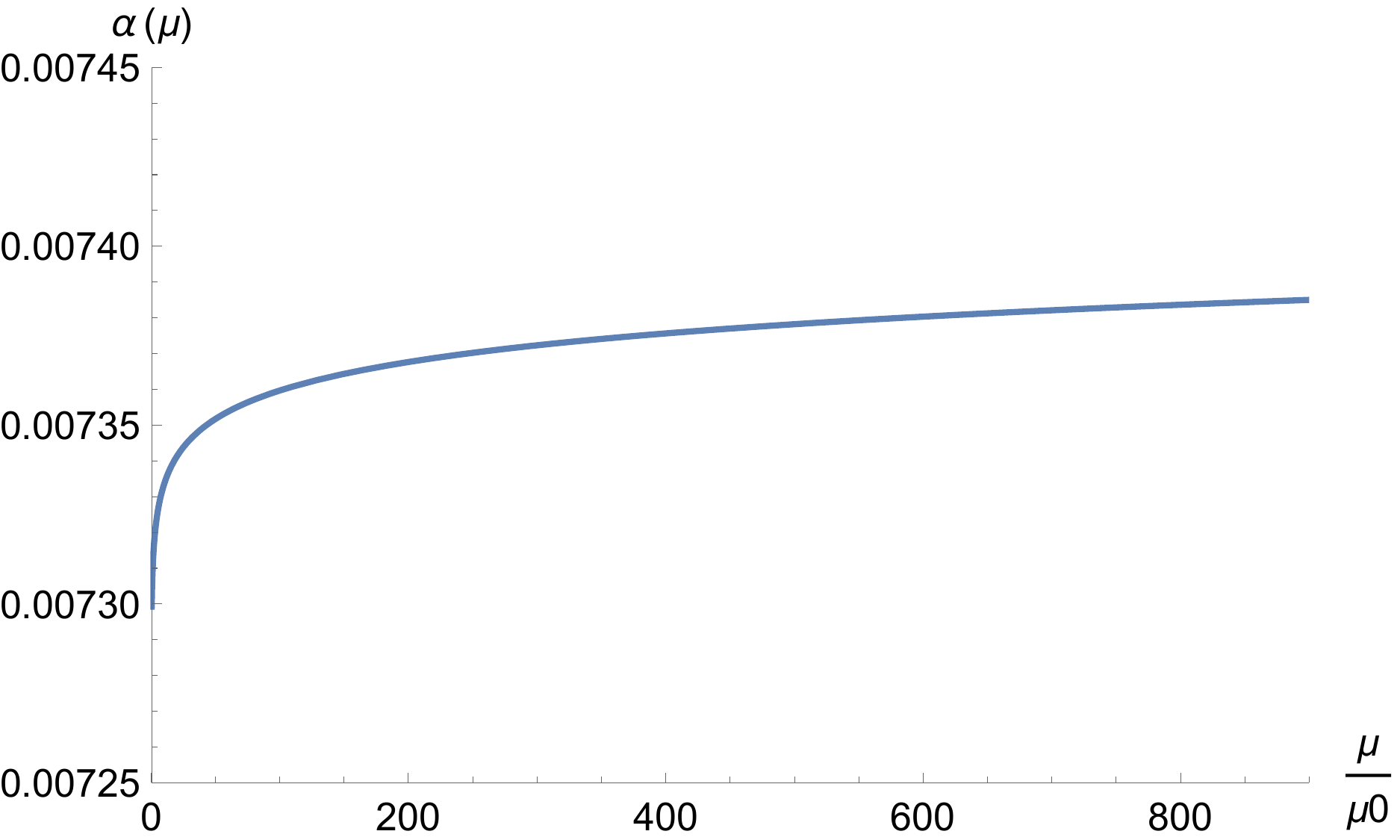}
\caption{$\alpha(\mu)$ vs $\mu$ for one-flavour QED.}
\end{center}
\end{figure}

Let us try to see what this means physically for electrodynamics. Suppose an experiment is carried out to measure $e_R$ at a scale $\mu_0=0.5$ MeV. This is a relatively low energy, at which we define the usually quoted value of the fine structure constant $\alpha\sim \frac{1}{137}$. Thus we have:
\be
e_R^2(\mu_0=0.5{\rm ~MeV})=4\pi\alpha\sim 0.0917
\ee	
Now as we increase $\mu$ above $\mu_0$, the denominator on the RHS of \eref{ersq} decreases, leading to a (very slow) increase in $e_R(\mu)$. Suppose $\mu=100$ GeV, our formula predicts that the renormalised charge has grown to $e_R^2\sim 0.0935$. This is a significant, measurable difference from the low-energy value, indicating that the fine structure constant has grown from $\alpha\sim 1/137$ to $\alpha\sim 1/134$. Experiment confirms the growth of $\alpha$ but the measured result is different -- one finds that $\alpha\sim 1/127$ at around 100 GeV. The reason is that our calculation was done assuming the photon couples only to the electron, while in nature it couples to all charged leptons and quarks.  If we carry out the correct computation in the full Standard Model, its result is experimentally confirmed to a high degree of accuracy. 

The unmistakable conclusion from this analysis is that coupling constants are not constant! They {\em effectively} vary with the energy scale at which they are computed, and measured. 

We have no particular reason to think that Quantum Electrodynamics is valid at arbitrarily high energy scales. It could well be subsumed into a larger theory (unified or otherwise) of which it merely describes the low-energy behaviour in the electromagnetic sector. Without making any physical assumptions about this question, let us ask for what value of $\mu$ the denominator of the RHS of \eref{ersq} becomes vanishingly small and the renormalised coupling based on a one-loop calculation diverges. This value, known as the ``Landau pole'', can be treated as a characteristic label of the theory. Thus, one sets:
\be
\frac{e_R^2(\mu_0)}{6\pi^2}\ln\frac{\mu}{\mu_0}=1
\label{setmu}
\ee
This is solved by $\mu=\Lambda_{\rm QED}$ where:
\be
\Lambda_{\rm QED}=\mu_0\,e^{\frac{6\pi^2}{e_R^2(\mu_0)}}\sim 10^{283}{\rm ~MeV}
\ee
We can write $e_R^2(\mu)$ in terms of $\Lambda_{\rm QED}$ as follows:
\be
e_R^2(\mu)=\frac{6\pi^2}{\ln\big(\frac{\Lambda_{QED}}{\mu}\big)}
\label{eR.lamqed}
\ee
This is a fascinating formula, despite the unphysicality of the scale $\Lambda_{\rm QED}$. There is simply no dependence on the dimensionless quantity $e_R$, at any scale, on the RHS! It has been replaced with a {\em dimensional} quantity, $\Lambda_{\rm QED}$, which characterises the theory, and the above formula gives the value of the coupling constant at any arbitrary energy scale $\mu$ in terms of the ratio of that scale to the scale associated with the theory itself. The disappearance of $e_R$ in favour of $\Lambda_{QED}$ is known as ``dimensional transmutation''. 

As already emphasised in the context of scalar field theory (see Subsection \ref{Zfactor}), one is certainly not claiming that the electric charge ever grows to infinity in the UV. Long before one reaches the value of $\mu$ that satisfies \eref{setmu}, the electric charge becomes greater than 1 and thereafter perturbation theory becomes invalid (and one-loop perturbation theory, even more so!). One should simply treat \eref{setmu} as a definition of a scale $\Lambda_{\rm QED}$ that then permits a change of variables leading to \eref{eR.lamqed}.

The corresponding formulae for non-Abelian gauge theories turn out to be of utmost importance in understanding their dynamics. Let us anticipate what will happen in this case. For a generic field theory with a renormalised coupling $e_R$, we can parametrise its $\beta$-function as:
\be
\beta(e_R)=a_1 e_R^3+a_2 e_R^5+\cdots
\ee
It can be shown that each $a_n$ is calculable via loop diagrams up to $n$ loops. We calculated the one-loop contribution $a_1$ for QED and found it to have the positive value $\frac{1}{12\pi^2}$. What if in some theory we found $a_1<0?$. The coupling constant would then run as:
\be
e_R^2(\mu)=\frac{e_R^2(\mu_0)}{1+2|a_1| e_R^2(\mu_0)\ln\big(\frac{\mu}{\mu_0}\big)}
\ee
The negative sign of $a_1$ has a profound consequence. In this case the coupling constant {\em decreases} with increasing energy. As a result, perturbation theory becomes better and better (instead of getting worse as in QED). Moreover, the characteristic energy scale of such a theory, defined by vanishing of the denominator on the RHS,  is:
\be
\Lambda = \mu_0\,e^{-\frac{1}{2|a_1|e_R^2(\mu_0)}}
\ee
In the context of quantum chromodynamics, this scale is known as $\Lambda_{\rm QCD}$. Again it marks the scale at which perturbation theory breaks down, but there are two crucial differences from the Abelian case. One is that due to the negative sign of $a_1$, perturbation theory is valid at energies {\em above} rather than below the scale $\Lambda_{\rm QCD}$. The second is that the scale itself turns out to lie in an experimentally accessible range of energies. We will return to this point after calculating $a_1$. 

In terms of this scale the running coupling can finally be written:
\be
e_R^2(\mu)=\frac{1}{2|a_1|\ln\big(\frac{\mu}{\Lambda_{\rm QCD}}\big)}
\ee
Note the inverted ratio between $\Lambda$ and $\mu$ relative to \eref{eR.lamqed}. This nicely captures the decrease of the coupling with an increase in energy scale, and tells us that as the scale $\mu$ {\em decreases} down to $\Lambda$, the coupling constant becomes large and perturbation theory breaks down. 

Thus if $a_1<0$, we can reliably conclude that the corresponding theory is weakly coupled at high energies. It also appears to be strongly coupled at low energies, but this conclusion is tempered by fact that  close to $\mu=\Lambda_{\rm QCD}$ we simply cannot trust perturbation theory. Such theories are said to be {\em asymptotically free}. This concludes our brief preview of the beta-function for non-Abelian gauge theories, which will be discussed in more detail in subsequent chapters.

\subsection{Anomalous dimensions}

So far we have only considered the flow of coupling constants with energy scale. Let us now apply the same logic to study the flow of the renormalised mass.
What are the physical consequences of the bare mass being independent of the renormalisation scale $\mu$? This time the starting point is:
\be
m_0=Z_m m_r,\qquad Z_m=1+\delta_m, \qquad \delta_m=-\frac{3e_R^2}{8\pi^2\epsilon}
\ee
Now, we have:
\be
\frac{dm_0}{d\mu}=\frac{d}{d\mu}(Z_m m_R^{})=0
\label{diffm}
\ee
which, from the above equations, implies that:
\be
\begin{split}
\frac{1}{m_R^{}}\frac{d m_R^{}}{d\mu}&=-\frac{1}{Z_m}\frac{d Z_m}{d\mu}\\
&=-\frac{1}{1+\delta_m}\frac{d \delta_m}{d\mu}\\
&=\Big(\frac{3e_R^{}}{4\pi^2\epsilon}+\mathcal{O}(e_R^{3})\Big)\frac{d e_R^{}}{d\mu}
\label{mderiv}
\end{split}
\ee

\exercise{Show that $m_0$ has dimensions of mass in any space-time dimension. This is why we did not need to insert a power of $\mu$ along with it, as we did for the electric charge.}

We now define the {\em anomalous dimension} $\gamma_m$ associated to the fermion mass by:
\be
\gamma_m(e_R,m_R)\equiv\frac{\mu}{m_R^{}}\frac{\partial m_R^{}}{\partial\mu}=\frac{\partial \log m_R}{\partial \log\mu}
\label{anomdim}
\ee
In the present example of QED we have:
\begin{equation}
\begin{split}
\gamma_m(e_R,m_R) &=\frac{3e_R^{}}{4\pi^2\epsilon}\,\beta(e_R)\\[1mm]
&=-\frac{3 e_R^2}{8\pi^2}
\end{split}
\end{equation}
In the first step we used \eref{mderiv} and in the next step we used \eref{beta1}. Notice that at the end we got a finite contribution from the first term in $\beta$ which vanishes as $\epsilon\to 0$\footnote{The general rule is that this term, the first term in \eref{beta1}, can be discarded by taking $\epsilon\to 0$ only at the very end of a calculation. In the case of anomalous dimensions, it is precisely this term that gives the one-loop result.}. 

Since we already know $e_R(\mu)$ from the $\beta$-function equation, \eref{anomdim} can be solved to find the ``running mass'' $m_R(\mu)$ as a function of the renormalisation scale.

\exercise{Verify the above computations.\\[-5mm]

{\em Exercise:} Use the available information to calculate $m_R(\mu)$. 
}

One also has anomalous dimensions associated to fields. For example in QED, given that the fields are renormalised as:
\be
\psi_0=\sqrt{Z_2}\,\psi_R,\qquad A^\mu_0=\sqrt{Z_3}\,A^\mu_R
\ee
we may define the anomalous dimensions:
\be
\gamma_2\equiv \frac{\mu}{Z_2}\frac{\del Z_2}{\del\mu},\qquad
\gamma_3\equiv \frac{\mu}{Z_3}\frac{\del Z_3}{\del\mu}
\ee
The physical significance of these anomalous dimensions is the following. In the bare Lagrangian, viewed at the classical level, the mass parameter and the fields have fixed scale dimensions, for example $[m]=1$, $[A^\mu]=1$ and $[\psi]=\frac32$. Naively we might have expected that these scale dimensions would hold in the quantum theory as well. However we have seen that even a dimensionless coupling constant runs slowly with a change of renormalisation scale. Anomalous dimensions tell us, in a similar way, that the scaling dimensions of the mass and the fields of the theory depart from their classical value, the corrections being small in perturbation theory and slowly (logarithmically) varying with $\mu$. 

\subsection{The Callan-Symanzik equation}

Having carried out the renormalisation programme and observed how couplings and masses run with energy scale, it is time to study amplitudes. We start with a bare Green's function:
\begin{equation}
G_0\big(\{x_i\},\{y_j\},\{z_j\}\big)
\equiv\bra{0}T \Big(A_{0}^{\mu_1^{}}(x_1^{}) \cdots A_{0}^{\mu_n^{}}(x_n^{})\,\psi_0^{}(y_1^{})\overline{\psi}_0(z_1^{}) \cdots \psi_0^{}(y_p^{})\overline{\psi}_0(z_p^{})\Big)\ket{0}
\end{equation}
Using the relation between bare and renormalised fields, we can rewrite this as:
\be
G_0\big(\{x_i\},\{y_j\},\{z_j\}\big)
=Z_3^{n/2}Z_2^p\, \widetilde{G}_R\big(\{x_j\},\{y_j\},\{z_j\}\big) 
\ee
Taking the Fourier transform with respect to all positions on both sides and dropping the common momentum-conserving $\delta$-functions, we have an analogous relation between the bare and renormalised momentum-space Green's functions:
\begin{equation}
{\tilde G}_0\big(\{p_j\}\big)=Z_3^{n/2}Z_2^p\, {\tilde G}_R\big(\{p_j\},e_R^{},m_R^{},\mu\big)
\end{equation}
Now it is clear that the bare Green's function is independent of the renormalisation scale $\mu$. Thus we have:
\be
\mu\frac{d G_0}{d\mu}=0
\ee
This means there must be a cancellation between the many dependencies of $G_0$ on $\mu$, via $Z_2,Z_3,e_R,m_R$ which all depend on $\mu$, as well as its direct dependence. Expanding the RHS we find:
\be
 \Big(\mu\frac{\partial}{\partial\mu}+\beta\frac{\partial}{\partial e_R^{}}+m_R^{}\gamma_m\frac{\partial}{\partial m_R^{}}+\frac{n}{2}\gamma_3+p\gamma_2\Big)G_R=0
\ee
where $\gamma_m\gamma_2,\gamma_3$ are the anomalous dimensions defined in the previous subsection.

The above equation is called the ``Callan Symanzik equation". It tells us how the value of the Green's function varies with a change in renormalisation scale, in terms of the variation of the coupling constants, masses and fields in terms of which the Green's function is expressed. In turn, this gives us the scale-dependence of scattering cross-sections which are physically measurable.

\newpage

\section{Non-abelian gauge theory}
\label{nagtoneloop}

\subsection{The Yang-Mills Lagrangian}

We now turn to the discussion of non-Abelian gauge theories, also known as Yang-Mills theories. These are Lagrangian field theories with a multiplet of vector fields $A_\mu^a$ where the index $a$ runs over all the fields in the multiplet, coupled to suitable multiplets of scalar or spinor fields. 

The original idea behind this class of theories was to generalise the gauge invariance of QED to a non-Abelian, or non-commuting, context. In QED,  the combination of two successive gauge transformations is another transformation by the sum of the parameters:
\be
\begin{split}
\psi~&\to~ e^{i\alpha(x)}\psi~ \to ~e^{i\beta(x)}e^{i\alpha(x)}\psi=e^{i(\alpha(x)+\beta(x))}\psi\\
A_\mu~&\to~A_\mu-\frac{1}{e} \del_\mu\alpha~\to~A_\mu-\frac{1}{e}\del_\mu\alpha-\frac{1}{e}\del_\mu\beta=A_\mu-\frac{1}{e}\del_\mu\big(\alpha(x)+\beta(x)\big)
\end{split}
\ee
Clearly the result is the same if we perform the transformation in the reverse order. 

In non-Abelian theories, instead, we choose multiplets of fields that transform by a {\em matrix-valued} gauge parameter. Because matrix multiplication is non-commutative, two successive gauge transformations performed in different orders lead to different results. It is a remarkable quirk of history that Yang and Mills started from this aesthetic principle, with the hope of imposing local gauge symmetry on isospin multiplets. As it turns out, their results did not teach us anything about isospin (which is only an approximate global symmetry). Instead the theories they proposed have emerged, unexpectedly and over several decades, as accurate descriptions of the fundamental weak and strong nuclear interactions in nature. 

Let us describe Yang-Mills theories based on the Lie algebra SU(N). The adjoint representation of this group has $N^2-1$ generators, so we take a family $A_\mu^a$ of vector fields with $a=1,2,\cdots,N^2-1$. We then seek a Lagrangian density for these fields that is invariant under the infinitesimal transformations\footnote{If we think of Abelian theory as the special case when the structure constants vanish and the gauge group is U(1), the sign in the gauge transformation here is opposite to what we had chosen in the Abelian case. The conventions in these notes follow those of \cite{Peskin:1995ev}.}:
\begin{equation}
\delta A^a_{\mu}= \frac{1}{g} (D_{\mu} \alpha)^a=\frac{1}{g}\Big(\partial_{\mu} \alpha^a(x) + g f^{abc}A^b_{\mu} \alpha^c(x)\Big),
\label{Agauge}
\end{equation}
where $\alpha^a(x)$ are a set of gauge parameters that are arbitrary functions of $x^\mu$, $f^{abc}$ are the structure constants of SU(N) and $g$ will be identified with a coupling constant in the Lagrangian. Without the second term, these look just like the Abelian gauge transformations of QED, with an independent parameter for each vector field. However the second term mixes the vector field component $A_\mu^a$ with all the others. 

In the Abelian (QED) case the gauge transformation of the vector field was inhomogeneous, while the Maxwell field strength $F_{\mu\nu}=\del_\mu A_\nu - \del_\nu A_\mu$ was gauge invariant and could be used as a building block for a gauge-invariant Lagrangian. In the non-Abelian case the gauge transformation again has an inhomogeneous term. Now a key insight of Yang and Mills was that a generalisation of the Maxwell field strength, defined by:
\be
F_{\mu\nu}^a=\del_\mu A_\nu^a -\del_\nu A_\mu^a +gf^{abc}A_\mu^bA_\nu^c
\label{fmndef}
\ee
transforms homogeneously under the non-Abelian gauge transformation:
\be
\delta F_{\mu\nu}^a = f^{abc}F_{\mu\nu}^b \alpha^c
\label{fmnvar}
\ee
While $F$ is not actually gauge  invariant, the path from here to a gauge-invariant Lagrangian is straightforward. We simply write:
\begin{equation}
\mathcal{L}=-\frac{1}{4}F^a_{\mu \nu}F^{\mu \nu a} 
\label{pureYM}
\end{equation} 
where a sum over $a$ is implied. 
This is easily verified to be invariant under the above non-Abelian gauge transformation. At finite $g$ it is an interacting theory, which may be seen by inserting \eref{fmndef} in \eref{pureYM} and expanding out all the terms. One finds terms with three and four powers of $A_\mu^a$, which are interaction terms. In the limit  $g\to 0$
it reduces to several copies of the free Maxwell Lagrangian.

The above theory (called ``pure Yang-Mills theory'') can be generalised in many different ways to incorporate scalars and fermions. We will first generalise it with the addition of a fermion multiplet $\psi_i$, $i=1,2,\cdots,N$ which transforms under the gauge transformation as:
\be
\delta\psi_i=i\alpha_a(x)T^a_{ij}\psi_j
\label{psigauge}
\ee 
where $T^a_{ij}$ are Hermitian matrices defining some representation of the Lie algebra SU(N). The finite transformation from which this originates is:
\be
\psi_i\to \left(e^{i\alpha^a(x)T^a}\right)_{ij}\psi_j
\ee
and this corresponds, as promised in the introduction of this chapter, to multiplication of the collection of spinor fields $\psi_i$ by a matrix in a representation of the group SU(N).

A gauge-invariant Lagrangian density for the fermion and gauge-field system together is then:
\begin{equation}
\mathcal{L}=-\frac{1}{4}F^a_{\mu \nu}F^{\mu \nu a} +i \bar{\psi_i}\big(\delta_{ij}  \slashed \partial - i g \slashed A^a T^a_{ij} \big)\psi_j -m \bar{\psi_i}\psi_i,
\label{YMfermions}
\end{equation} 

We said above that the parameter $\alpha^a(x)$ is an arbitrary function of $x^\mu$. However this was an over-simplification. We have earlier claimed that gauge invariance refers to a redundancy in the description of the theory. However a gauge transformation that is non-trivial at infinity cannot be a redundancy. For example, if a vector field has certain boundary conditions specified at infinity, then those conditions will not be preserved by such a transformation. Hence the true redundancy is generated by gauge parameters that fall off at infinity. Such transformations do not generate a conserved Noether current. It is only this class of $\alpha^a(x)$ that should be called gauge transformations. 

Separately one may consider the case of constant $\alpha^a$, which clearly do not fall off anywhere. The resulting transformations:
\be
\delta A^a_{\mu}=  f^{abc}A^b_{\mu} \alpha^c,\qquad
\delta\psi_i=i\alpha^a T^a_{ij}\psi_j
\label{globalsym}
\ee
are not redundancies but genuine symmetries of the Lagrangian, and we call these global symmetries. Unlike gauge symmetries, they give rise to a conserved current via Noether's theorem. Using the standard procedure, one finds the conserved current for the symmetry \eref{globalsym} of \eref{YMfermions} to be:
\begin{equation}
J^a_{\mu}=-\bar{\psi_i}\gamma_{\mu}T^a_{ij}\psi_j + f^{abc} A^{\nu b} F^c_{\mu \nu},
\label{Jcons}
\end{equation}
We can verify that, using the equations of motion, this current satisfies the conservation equation:
\begin{equation}
\partial^\mu J^a_{\mu}=0.
\end{equation}

Now we encounter an intriguing fact. Although conserved, this current is not gauge covariant. A covariant current would transform as:
\be
\delta J_\mu^a=f^{abc} A^b_\mu \alpha^c
\label{covcurr}
\ee
but this is not how $J_\mu^a$ transforms when we vary $\psi_i$ and $A_\mu^a$, on which it depends, following Eqs.(\ref{Agauge}),(\ref{psigauge}). The reason is that $A_\mu^a$ (and not just its field strength) appears explicitly in $J_\mu^a$, and the inhomogeneous part of the transformation of this $A_\mu^a$ factor leads to a result different from \eref{covcurr}.

\exercise{Calculate $\delta J^a_\mu$ for the above current under a gauge transformation and verify that it does not agree with \eref{covcurr}.}

A consequence of the above is that if we define a charge:
\be
Q^a\equiv \int d^3x~J_0^a(x)
\ee
then $Q^a$ is conserved, but it is not gauge covariant.

Now there is a different current that we can define by picking just the first term of \eref{Jcons}:
\be
j_\mu^a=-\bar{\psi_i}\gamma_{\mu}T^a_{ij}\psi_j
\ee
This current is gauge covariant. However it is not conserved: $\del^\mu j_\mu^a\ne 0$. Instead it satisfies a ``covariant conservation law'':
\be
D^\mu j_\mu^a\equiv \del^\mu j_\mu^a +gf^{abc}\alpha^b j_\mu^c=0
\ee
where $D^\mu$ is a covariant derivative. However, it is the ordinary (not covariant) divergence which tells us whether a current is really conserved and here that is not the case. Hence the associated charge:
\be
q^a\equiv \int d^3x~ j_0(x)
\ee
though gauge covariant, is also not conserved. It is fascinating that in a non-Abelian gauge theory we do not have any charge that is simultaneously gauge covariant and conserved. 

Note that the issue is specifically with covariance of the charge under {\em local} gauge transformations. As far as the global transformations \eref{globalsym} are concerned, we have a legitimate conserved current $J_\mu^a$ and a conserved charge $Q^a$ and they have all the properties we would expect under the global symmetry. But there is a problem because of the presence of local gauge invariance in the theory. As we have observed earlier, a local gauge invariance is a redundancy of the description and in its presence, only gauge-invariant quantities can be physical observables. If $Q^a$ had been covariant, we could have constructed an invariant out of it -- for example by taking bilinears -- but since it is not, we cannot find any gauge invariant conserved charge. 

For completeness as well as future use, let us mention here how to couple scalar fields, rather than fermions, to a non-Abelian gauge field. The procedure is quite similar to that for fermions. We introduce scalar multiplets $\phi_i$, $i=1,2,\cdots,N$ which transform under gauge transformations as:
\be
\delta\phi_i=i\alpha_a(x)T^a_{ij}\phi_j
\label{phigauge}
\ee 
where $T^a_{ij}$ are again matrices defining some suitable representation of the Lie algebra SU(N). As before, this can be considered the infinitesimal limit of the finite transformation:
\be
\phi_i\to \left(e^{i\alpha^a(x)T^a}\right)_{ij}\phi_j
\ee
The appropriate gauge-invariant Lagrangian density for the coupled scalar and gauge-field system is then:
\begin{equation}
\mathcal{L}=-\frac{1}{4}F^a_{\mu \nu}F^{\mu \nu a} +\half (D_\mu\phi)_i(D^\mu\phi)_i-V(\phi_i)
\label{YMscalars}
\end{equation} 
where:
\be
(D_\mu\phi)_i\equiv \left(\del_\mu \delta_{ij}-igA_\mu^aT^a_{ij}\right)\phi_j
\label{scalarcov}
\ee
and the potential $V(\phi_i)$ is chosen to be separately gauge-invariant\footnote{The above normalisation is conventional for real representations. For complex representations, the kinetic term for $\phi$ is replaced by $(D_\mu\phi)^{\dagger\,i}(D^\mu\phi)^i$.}.

\exercise{Compute the conserved current arising from the global version of the gauge symmetry in the case of scalars, and verify that it is not gauge-invariant -- just as in the case of fermions.}

\subsection{Quantising Yang-Mills theory: Faddeev-Popov ghosts}

As with QED, gauge invariance in Yang-Mills theory needs to be fixed using a gauge-fixing term. However, due to the nonlinearity of the gauge transformation in Yang-Mills theory, there is a residual complication involving a certain determinant, that has to be taken care of after fixing the gauge. We will see that this forces us to add some new fields, known as ``Faddeev-Popov ghosts'', in the Lagrangian. These propagate only inside loops. 

The considerations below will be accessible only to readers familiar with the path integral approach to quantum field theory. Unfortunately this topic could not be reviewed in the present notes. 

With this warning, let us go back to Abelian gauge theory  and try to better understand the $R_\xi$ gauge-fixing term $-\frac{1}{2\xi}\left(\del_\mu A^\mu\right)^2$ that was introduced in \eref{gaugefix}. Classically this changes the equations of motion in a way that breaks gauge invariance. In the context of the path integral the implications are somewhat more subtle. Let us start with the gauge-invariant path integral (without a gauge-fixing term):
\be
\int [dA]~e^{iS[A]}
\ee
In this context, gauge invariance gives rise to a problem of overcounting which we can see as follows. Given any particular configuration $A_\mu(x)$, the path integral measure $[dA]$ includes an integration over the ``gauge orbit'' of this configuration -- the family of all configurations related to the original one by gauge transformations. By gauge invariance the action is the same for all configurations on a given orbit. Hence the integration over the gauge orbit contributes an infinite ``volume'' and it becomes a problem to treat this. Ideally one would like, instead, to pick a single element from every gauge orbit and integrate only over such elements.

Now by adding a gauge-fixing term in the Lagrangian, one is distinguishing different points in a gauge orbit and one may heuristically think this lifts the enormous degeneracy. With gauge fixing, the original path integral changes as follows: 
\be
\int [dA]~e^{iS[A]}\qquad \to \qquad \int [dA]~e^{iS[A]-\frac{i}{2\xi}\int d^4x \left(\del_\mu A^\mu\right)^2}
\label{gaugefixterm}
\ee
But it is not clear why this replacement is justified. Can we be sure that we did not simply change the original theory to a new one? We will now show that the above replacement is correct for QED. In the process, we will make the above considerations more precise and will also identify a crucial difference between Abelian and non-Abelian gauge theories. Then we  will learn how to deal with the latter case.

For the Abelian case, we pick a particular gauge potential $A_\mu(x)$ and define\footnote{Hopefully the context makes it clear that the $\sim $ does not mean Fourier transform here.}:
\be
\tA_\mu(x)=A_\mu(x)-\frac{1}{e}\del_\mu \alpha(x)
\label{abeliangt}
\ee
This is a family of configurations $\tA_\mu(x)$ as $\alpha(x)$ ranges over all functions, and is called the gauge orbit of $A_\mu$. Clearly if $A_\mu(x)$ gives rise to some field strength $F_{\mu\nu}(x)$, then all the $\tA_\mu(x)$ give rise to exactly the same field strength. Thus along any gauge orbit, the action takes the same value. Now because we can choose any function $\alpha(x)$, there are infinitely many configurations belonging to the orbit. As  a result, one gets an infinite contribution corresponding to the ``volume'' (in field space) of the orbit. Without going into the formal mathematics, it is worth noting that the volume of function space, i.e. the ``number'' of distinct functions on a space, is a much ``larger'' infinity than, say, the volume of some ordinary space like the real  line.  

To avoid this problem, one must pick precisely one representative configuration from every gauge orbit. For example, we may require $\del_\mu A^\mu$ to be equal to a fixed function $\omega(x)$. This will certainly cut down the gauge orbits, hopefully reducing each one to a single configuration from each orbit\footnote{Whether it actually does so or not depends on details of the gauge-fixing condition and the boundary conditions on the fields.}. To implement this, we would like to insert a rule into the path integral that says ``require each configuration $A_\mu(x)$ to satisfy a condition that picks one representative from each gauge orbit''. A naive solution is to insert a $\delta$-function(al) into the path integral:
\be
\int [dA]e^{iS[A]}\qquad \to \qquad \int [dA]e^{iS[A]}~\delta\left(\del_\mu A^\mu -\omega(x)\right)
\label{naivegaugefix}
\ee
This is an infinite-dimensional $\delta$-function that equates $\del_\mu A^\mu$ with $\omega$ at every point of space-time! However, it can be dealt with, at least formally. The real problem is that our $\delta$-function fixes not $A_\mu$ itself, but a {\em function} of $A_\mu$ -- in this example, its divergence. 

To see why this is a problem, consider the following exercise in ordinary multi-variable integration. Suppose $x_i$ is a set of $N$ real variables, each one ranging over the full real line, and suppose $g_j(x_i)$ are a set of $N$ invertible functions of the $x_i$. Then we have the following self-evident equations:
\be
\begin{split}
1&=\int \prod_i dg_i~\prod_j \delta(g_j-a_j)\\
&=\int \prod_i dx_i ~\det\left|\frac{\del g_j}{\del x_i}\right| ~\prod_j \delta\Big(g_j(x_i)-a_j\Big)
\end{split}
\label{deltaid}
\ee
Suppose someone proposed an `alternative' result that $\int \prod_i dx_i ~\prod_j\delta\Big(g_j(x_i)-a_j\Big)=1$, we would object right away: this equation is not true since we have omitted the Jacobian determinant in the second line above. In special cases where this determinant is constant this does not matter much: we get a constant factor outside the path integral, but this drops out when we consider correlation functions. However a generic Jacobian determinant that varies with $x$ crucially affects the integral and without it, the second line would just be wrong. This is the analogue of the problem we face in \eref{naivegaugefix} when we try to implement gauge-fixing by inserting a $\delta$-function which involves not $A_\mu$ directly, but $\del_\mu A^\mu$: such an implementation is not correct.

We now see what is the solution to this problem. Considering the gauge orbit $\tA_\mu(x)$ defined above, we have the identity:
\be
\int [d\alpha]~ \prod_x\delta\bigg(\del_\mu \tA^\mu(x)-\omega(x)\bigg)~\det_{x,y}\left(\frac{\del \big[\delta(\del_\mu \tA^\mu(x) -\omega(x))\big]}{\del \alpha(y)}\right)=1
\label{deltaidentity}
\ee
It should be clear that this is a function-space analogue of the second line of \eref{deltaid}.
Inserting the definition of $\tA_\mu$ above, one easily sees that the Jacobian determinant is:
\be
\det_{x,y}\left(\frac{\del \big[\delta(\del_\mu \tA^\mu(x) -\omega(x))\big]}{\del \alpha(y)}\right)
=\frac{\delta \left(\frac{1}{e}\del_\mu \del^\mu\alpha(x)\right)}{\delta\alpha(y)}=\det_{x,y}\left(\square_x\,\delta^4(x-y)\right)
\label{abJacobian}
\ee
where in the last step we have dropped a constant coming from the $\frac{1}{e}$ factor, because it is independent of $A_\mu(x)$ and $\alpha(x)$ and will not affect amplitudes\footnote{Recall that any constant or function that does not depend on the integration variable $A_\mu(x)$ is irrelevant in the path integral. The basic reason is that correlation functions (amplitudes) are normalised by the partition function and any field-independent factors will cancel out.}.

Now, it is the infinite factor $\int[d\alpha]$, the volume of the gauge orbit,  that we wish to drop. So we insert the identity \eref{deltaidentity} into $\int [dA]e^{iS[A]}$ and then drop $\int [d\alpha]$. Finally, we notice that $[dA]$ and $S[A]$ are both gauge-invariant. Thus we can replace $\tA_\mu$ in the $\delta$-function by $A_\mu$. As a result, the corrected version of \eref{naivegaugefix} is:
\be
\int [dA]e^{iS[A]}\quad \to \quad \int [dA]e^{iS[A]}~\delta\big(\del_\mu A^\mu(x) -\omega(x)\big)~\det_{x,y}\big(\square_x\,\delta^4(x-y)\big)
\label{correctgaugefix}
\ee
The remaining steps are (i) we notice that the determinant in \eref{correctgaugefix} is independent of $A_\mu$. Hence it can be dropped in the same way that we drop constants (infinite or otherwise) that are independent of the integration variable, (ii) we have no particular preference for any given function $\omega(x)$, so we choose to average over all $\omega(x)$ with the weight:
\be
[d\omega]~e^{-\frac{i}{2\xi}\int d^4x~ \omega(x)^2 }
\ee
This process is again independent of $A_\mu$ so it can be inserted without changing any amplitudes. Performing the $\omega$ integral, we finally recover the $R_\xi$ gauge-fixing rule \eref{gaugefixterm}.
As we have already observed, once the gauge-fixing term is added, the gauge field has a well-defined (and $\xi$-dependent) propagator, given in \eref{covprop}.

All this has been done in the context of Abelian gauge theory to keep the notation simple, but also to explain why in earlier chapters of these notes, we used a gauge-fixing prescription  without worrying about determinants. The reason is that the determinant was independent of $A_\mu$.  As we will now show, in non-Abelian gauge theories the determinant depends non-trivially on $A_\mu$. Hence it has to be retained, and contributes to loop amplitudes in an important way. 

To see this, we generalise the identity \eref{deltaidentity} by choosing the delta-function to be $\delta(\del_\mu \tA^{\mu\,a}(x)-\omega^a(x))$. Next we assign the Gaussian weight:
\be
\prod_{a=1}^{N^2-1}~[d\omega^a]~e^{-\frac{i}{2\xi}\int d^4x~ \omega^a(x)^2}
\ee
to the functions $\omega^a(x)$. Finally we use the fact that the infinitesimal non-Abelian gauge transformation, analogous to \eref{abeliangt}, is\footnote{It must be stressed that this is only valid for infinitesimal $\alpha^a$, unlike \eref{abeliangt} which is also valid at finite $\alpha$.}:
\be
\tA_\mu^a(x)=A_\mu^a(x)+\frac{1}{g}\left(D_\mu\alpha\right)^a\equiv 
A_\mu^a + \left(\del_\mu \alpha^a(x)+f^{abc}A_\mu^b \alpha^c\right)
\ee
Thus, the non-Abelian version of the Jacobian determinant in \eref{abJacobian} is:
\be
\det_{x,y}\left(\frac{\del \big[\delta(\del_\mu \tA^{\mu\,a}(x) -\omega(x))\big]}{\del \alpha^b(y)}\right)
=\frac{\delta \left(\frac{1}{e}\del_\mu \left(D^{\mu}\alpha\right)^a(x)\right)}{\delta\alpha^b(y)}
=\det_{\buildrel{x,y}\over {\scriptscriptstyle a,b}}\big(\del_\mu D^{\mu\,ab}_x\,\delta^4(x-y)\big)
\label{nonabJacobian}
\ee

Retaining this determinant and dropping the integral over $\alpha^a$ (which is the infinite volume of the gauge orbits), we finally get the Yang-Mills path integral:
\be
\int \big[dA^{a}\big]~ \det\big(\partial_{\mu}D^\mu\big)~ e^{iS[A]-\frac{i}{2\xi}\int d^4x(\partial^{\mu} A_{\mu}^a)^2}
\ee
where we have used a simplified notation for the operator appearing in the determinant. We cannot take the determinant outside the path integral now, because the covariant derivative $D_\mu$ depends on the integration variable $A_\mu$. 

Handling a path integral that contains a determinant insertion seems rather complicated, since this means none of the usual Feynman rules can be applied. Happily there is a trick due to Fadeev and Popov that converts the determinant into a new term in the Lagrangian. For this we introduce a set of fermionic (anti-commuting) scalar fields $c^a$ and $\bar{c}^{\,a}$, and use the identity:
\be
\det\big(\partial^{\mu}D_{\mu}\big)=\int [dc][d\bar{c}] e^{-i\int d^4x~\bar{c}^{\,a}(x)\partial_{\mu}D^{ab}_{\mu}c^{b}(x)}
\ee
This is a standard result for a path integral over fields with fermionic statistics. The fields $c,\cbar$ are called ``Faddeev-Popov ghost fields''. They are not a part of the original theory. Indeed, they are not associated to elementary particles, but only propagate inside loops in Feynman diagrams. Note that the index $a$ is in the adjoint representation of the gauge group, which means that the number of ghost fields is equal to the number of gauge fields.

Although these fields are fermionic in nature, they are not spinor fields. Instead they are complex spinless (and massless) scalars. In principle they are present even in Abelian theories like QED, but because the determinant is independent of the gauge fields, the ghosts decouple from the rest of the theory and can be safely ignored. In the non-Abelian case, ghosts couple to the gauge fields via a cubic vertex coming from the covariant derivative. These contributions must be included when we compute Feynman diagrams, where ghosts propagating in loops cancel out contributions from certain modes of the gauge field which are redundant due to gauge invariance. 

To summarise, the Yang-Mills path integral including the ghost fields is given by: \begin{equation}
\int [dA_{\mu}]\,[dc]\,[d\bar{c}]~e^{i S_{\rm gauge-fixed}}
\end{equation}
and the gauge-fixed Lagragian is:
\begin{equation}
\begin{split}
L_{\rm gauge-fixed}&=-\frac{1}{4} F^{a}_{\mu \nu}F^{a\mu \nu}-\frac{1}{2\xi}\Big(\partial^{\mu}A^{a}_{\mu}\Big)^2+\bar{c}^{\,a}\partial^{\mu}\Big(\partial_{\mu}\delta^{ac}+g f^{abc} A_{\mu}^{b}\Big)c^{c}\\[2mm] &\qquad + \bar{\psi_i}\big(i\delta_{ij} i \slashed \partial - g \slashed A^a t^a_{ij} -m\delta_{ij}\big)\psi_j
\end{split}
\label{Lgf}
\end{equation} 

As expected, the above Lagrangian is not gauge-invariant. However, it is supposed to correctly describe the dynamics of a gauge-invariant underlying theory. How can we be sure of this? Of course our manipulations were all correct, as they consisted of inserting a delta-function integral that is equal to 1, and then dropping the infinite volume of gauge orbits. Both of these are well-justified. Still, it is a little un-nerving to be left with no visible evidence of gauge invariance at the end. 

It turns out that the gauge-fixed Lagrangian does possess a beautiful but unexpected symmetry which contains the information that the original theory was gauge-invariant. This is known as Becchi-Rouet-Stora-Tyutin or BRST symmetry.
Because it captures information about the original gauge invariance, BRST symmetry is crucial to prove the gauge invariance of the quantum amplitudes in the final theory. The symmetry transformations are easy enough to write down and verify. We introduce an anti-commuting scalar constant $\theta$, taken to be infinitesimal, and write:
\begin{equation}
\begin{split}
\delta A_{\mu}^a &= \frac{1}{g}  \theta D_{\mu}c^{a}\\
\delta c^{a}&=-\frac{\theta}{2}f^{abd} c^{b} c^{d}\\
\delta \bar{c}^{\,a}&=-\frac{1}{g\, \xi}\theta\, \partial^{\nu}A^{a}_{\nu}\\
\delta \psi_{i} &=i \theta\, c^{a} t^{a}_{ij} \psi_{j}
\end{split}
\end{equation}

\exercise{Verify by inserting the BRST transformations into the Lagrangian \eref{Lgf}, that they are a symmetry of this Lagrangian. Observe that this holds for any value of the gauge fixing parameter $\xi$.}

Since the parameter $\theta$ is constant, the BRST symmetry is a global symmetry of the gauge-fixed Lagrangian. 

Notice that the BRST variations of the gauge field and the physical fermion $\psi_i$ are formally the same as the gauge transformations of the same fields but with the ghost-dependent parameter $\theta\, c^a(x)$ in place of the function $\alpha^a(x)$. This is a hallmark of the transformation. BRST symmetry gives a precise relation between the unphysical gauge boson polarization states and the ghosts and anti-ghosts. This property ensures that the $S$-matrix of the resulting theory is unitary, as the processes mediated by the unphysical degrees of freedom exactly cancel each other. 

\subsection{Feynman rules for gauge-fixed Yang-Mills theory}

The Feynman rules for the gauge-fixed Yang-Mills Lagrangian are easily obtained. They include the standard 3-point and 4-point self-couplings of the vector field (which we henceforth refer to as the ``gluon'') , as well as a 3-point fermion-vector coupling and a 3-point ghost-vector field coupling. Note that ghosts do not couple directly to the fermionic matter fields. The propagator for the vector field  is the same as for the Abelian theory (since it is derived from the free, quadratic part of the Lagrangian, which is just $N^2-1$ copies of the free Abelian Lagrangian). The ghost propagator is:
\be
\langle c^a(x)\cbar^{\,b}(y)\rangle =\delta^{ab}\int \frac{d^4k}{(2\pi)^4}\frac{i}{k^2+i\epsilon}e^{-ik\cdot(x-y)}
\ee
which is just that of a complex massless scalar.

The complete momentum-space Feynman rules are given pictorially as follows:

\begin{figure}[H]
\begin{center}
\hspace*{-4mm}
\begin{tikzpicture}
\begin{feynman}
\vertex (a) {\(\nu,b\)};
\vertex [right=of a, xshift=12mm] (b) {\(\mu,a\)};
\diagram* {
(a) -- [gluon, momentum=\(p\) ] (b)
};
\end{feynman}
\end{tikzpicture}
\hspace*{1.5cm}
\begin{tikzpicture}
\begin{feynman}
\vertex (a) {\(b\)};
\vertex [right=of a, xshift=4mm] (b) {\(a\)};
\diagram* {
(a) -- [ghost, momentum=\(p\) ] (b)
};
\end{feynman}
\end{tikzpicture}
\hspace*{2.1cm}
\begin{tikzpicture}
\begin{feynman}
\vertex (a) {\(j\)};
\vertex [right=of a, xshift=4mm] (b) {\(i\)};
\diagram* {
(a) -- [fermion, momentum=\(p\) ] (b)
};
\end{feynman}
\end{tikzpicture}\\[2mm]
\hspace*{4mm}
Gluon propagator
\hspace*{1.4cm}
Ghost propagator
\hspace*{1.4cm}
Fermion propagator\\[2mm]
\hspace*{-0.8cm}
$i \frac{(-\eta^{\mu \nu}+(1-\xi)\frac{p^{\mu}p^{\nu}}{p^2})\delta^{ab}}{p^2+i\epsilon}$
\hspace*{2.5cm}
$\frac{i \delta^{ab}}{p^2+i\epsilon}$
\hspace*{3.5cm}
$\frac{i \delta_{ij}}{\slashed{p}-m+i \epsilon}$\\[4mm]
\end{center}
\caption{Feynman rules for Yang-Mills theory: Propagators}
\label{feynYMprop}
\end{figure}
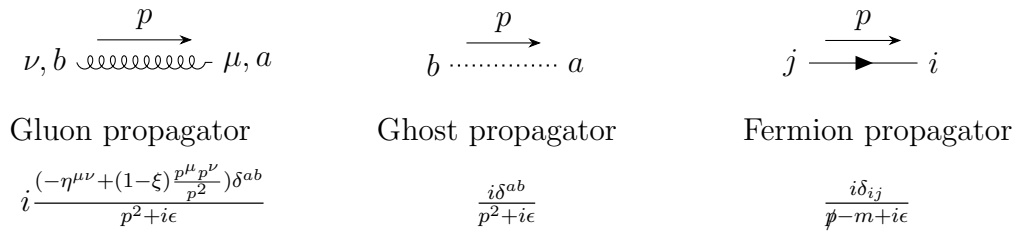

\begin{figure}[H]
\begin{center}
\begin{tikzpicture}
\begin{feynman}
\vertex (a) {\(\nu,b\)};
\vertex [right=of a, xshift=4mm] (b) ;
\vertex [above right=of b] (c) {\(\mu,a\)};
\vertex [below right=of b] (d) {\(\rho,c\)};
\diagram*{
(a) -- [gluon, momentum={[arrow shorten=0.1mm]\(p\)}] (b);
(c) -- [gluon, momentum={[arrow shorten=0.1mm]\(k\)}] (b);
(d) -- [gluon, momentum={[arrow shorten=0.1mm]\(q\)}] (b)
};
\end{feynman}
\end{tikzpicture}
\hspace*{2.5cm}
\begin{tikzpicture}
\begin{feynman}
\vertex (a);
\vertex [above left=of a] (b) {\(\mu,a\)};
\vertex [above right=of a] (c) {\(\nu,b\)};
\vertex [below left=of a] (d) {\(\rho,c\)};
\vertex [below right=of a] (e) {\(\sigma,d\)};
\diagram*{
(a) -- [gluon] (b);
(a) -- [gluon] (c);
(a) -- [gluon] (d);
(a) -- [gluon] (e)
};
\end{feynman}
\end{tikzpicture}\\
Three-gluon vertex
\hspace*{36mm}
Four-gluon vertex\\[2mm]
\hspace*{15mm}
\begin{minipage}{3cm}
$gf^{abc}\big[\eta^{\mu \nu}(k-p)^{\rho}\\[2mm]
~~~~~ + \eta^{ \nu \rho}(p-q)^{\mu}\\[2mm]
~~~~~~ +\,\eta^{\rho \mu}(q-k)^{\nu}\big]$
\end{minipage}
\hspace*{23mm}
\begin{minipage}{6cm}
$-ig^2\Big[f^{abe}f^{cde}\big(\eta^{\mu \rho}\eta^{\nu \sigma}-\eta^{\mu \sigma} \eta^{\nu \rho}\big)\\ 
~~~~~ +f^{ace}f^{bde} \big(\eta^{\mu \nu}\eta^{\rho \sigma}-\eta^{\mu \sigma}\eta^{\rho \nu}\big)\\
~~~~~ + f^{ade}f^{bce} \big(\eta^{\mu\nu}\eta^{\rho\sigma}-\eta^{\mu\rho}\eta^{\sigma\nu} \big)\Big]$
\end{minipage}
\\[4mm]
\begin{tikzpicture}
\begin{feynman}
\vertex (a) {\(\mu,a\)};
\vertex [right=of a, xshift=4mm] (b) ;
\vertex [above right=of b] (c) {\(i\)};
\vertex [below right=of b] (d) {\(j\)};
\diagram*{
(a) -- [gluon] (b);
(b) -- [fermion] (c);
(d) -- [fermion] (b)
};
\end{feynman}
\end{tikzpicture}
\hspace*{25mm}
\begin{tikzpicture}
\begin{feynman}
\vertex (a) {\(\mu,b\)};
\vertex [right=of a, xshift=4mm] (b) ;
\vertex [above right=of b] (c) {\(a\)};
\vertex [below right=of b] (d) {\(c\)};
\diagram*{
(a) -- [gluon] (b);
(b) -- [ghost, with arrow=8mm, momentum=\(p\)] (c);
(d) -- [ghost, with arrow=7mm] (b)
};
\end{feynman}
\end{tikzpicture}\\
Vector field-fermion vertex
\hspace*{25mm}
Vector field-ghost vertex\\[2mm]
\hspace*{15mm}
\begin{minipage}{3cm}
$ig\gamma^\mu t^a_{ij}$
\end{minipage}
\hspace*{40mm}
\begin{minipage}{3cm}
$-gf^{abc}p^\mu$
\end{minipage}
\end{center}
\caption{Feynman rules for Yang-Mills theory: Vertices}
\label{feynYMvert}
\end{figure}

\subsection{Renormalisation constants for Yang-Mills theory}

We are now in a position to define the renormalised fields and constants (with subscript $R$) starting from the bare fields and constants (with subscript 0):
\be
\begin{split}
A_{\mu,0} &=\sqrt{Z_3}\,A_{\mu,R},\qquad \psi_0 =\sqrt{Z_2}\, \psi_R\\
c^a_0 &=\sqrt{Z_{3c}}\, c^a_R,\qquad\quad {\bar c}^{\,a}_0=\sqrt{Z_{3c}}\, {\bar c}^{\,a}_R\\
g_0&=Z_g\, g_R,\qquad\qquad m_0=Z_m\, m_R
\end{split}
\ee
Next we write down the Yang-Mills Lagrangian in terms of renormalised quantities, thereby defining the starting point for renormalised perturbation theory:
\begin{equation}
\begin{split}
\mathcal{L}&=
-\frac{1}{4}Z_{3}\Big[\big(\partial_{\mu} A^{a}_{\nu R}-\partial_{\nu}A^{a}_{\mu R}\big)^2+\frac{1}{2\xi}\big(\partial^{\mu}A_{\mu R} \big)^2\Big]
+Z_2\, \bar{\psi}_{iR}\big(i\slashed{\partial}-Z_{m}m_{R}\big)\psi_{iR}\\
&\qquad 
+Z_{3c}\, \partial_{\mu}\bar{c}^{\,a}_{R}\partial^{\mu}c^{a}_{R}
-Z_g (Z_3)^\frac32 \,g_{R}\, f^{abc}\partial^{\mu}A^{\nu a}_{R}A_{\mu R}^{b}A^{c}_{\nu R}\\[2mm]
&\qquad -Z_g^2Z_3^2\, \frac{g_{R}^{2}}{4} f^{eab}f^{ecd}A^{\mu a}_{R}A^{\nu b}_{R}A_{\mu R}^{c}A^d_{\nu R} \\[2mm]
&\qquad  + Z_gZ_2\sqrt{Z_3}\,g_{R}\,A^{a}_{\mu R}\bar{\psi}_{i}\gamma^{\mu}(t^a)_{ij}\psi_{j}+Z_g Z_{3c}\sqrt{Z_3}\, g_{R}\,f^{abd}\partial_{\mu}\bar{c}^{a}_{R}A^{\mu b}_{R}c^{d}_{R}
\end{split}
\label{YMrpt}
\end{equation}

There is an assumption in what we have done above, namely that the coupling constant $g$ appearing in the cubic vector field term, the quartic vector field term, the vector-fermion coupling and the vector-ghost coupling, are all renormalised in the same way. It is not immediately obvious why this should be true. After all we can always start with a Lagrangian having multiple coupling constants, and set them equal to each other in the bare theory. Unless that relation is maintained by some symmetry, they will flow differently under renormalisation and cease to be equal to each other. Note that this was not an issue in QED, where there is a single coupling appearing in the vector-fermion term\footnote{The issue can arise in multi-species QED, where the photon couples to various different particles.}. 

And here we encounter the first application of BRST symmetry: it ensures that $g$ is renormalised in the same way wherever it appears. It is easily verified that if different terms in the above Lagrangian were renormalised differently then BRST invariance would fail to hold. But since this is a symmetry of the classical (gauge-fixed) theory, even after regularisation, it must remain a symmetry of the full quantum theory. Therefore the renormalised Lagrangian is the one we have written above. 

Recall that in QED we had in principle four independent renormalisation constants, which can be chosen as $Z_2,Z_3,Z_e,Z_m$ (there was also $Z_1$ which was determined to be equal to $Z_eZ_2\sqrt{Z_3}$). However the relation $Z_1=Z_2$, arising from gauge invariance, implied that $Z_e\sqrt{Z_3}=1$. Thus in the end there were three independent renormalisation constants. Now the analogue of $Z_1=Z_2$ does not hold in non-Abelian gauge theory, and also there is a new field renormalisation for the ghosts. As a result, in Yang-Mills theory coupled to a fermion species there are five independent renormalisation constants: $Z_2,Z_3,Z_{3c},Z_g,Z_m$. It is also useful to define the dependent quantities:
\be
\begin{split}
Z_1 &\equiv Z_gZ_2\sqrt{Z_3}\\
Z_{1c} &\equiv Z_gZ_{3c}\sqrt{Z_3}\\
Z_{A^3} &\equiv Z_g(Z_3)^\frac32\\
Z_{A^4} &\equiv Z_g(Z_3)^2
\end{split}
\label{dependentZ}
\ee
Inspection of the Lagrangian \eref{YMrpt} makes it clear why we chose to define these quantities. They correspond to the overall renormalisation constants for, respectively, the fermion-vector coupling, the ghost-vector coupling, the cubic vector self-coupling and the quartic vector self-coupling terms.

Henceforth we work in Feynman-'tHooft gauge, where $\xi=1$. In principle we should classify the different kinds of graphs which have a superficial degree of divergence $\geq$ 0. However this exercise is too lengthy to be carried out here and we restrict our attention to a single process where we can discuss some details of the computation. This process is the one-loop vacuum polarisation for the gluon. Our approach this time will be to write down the Feynman integrals in full, but instead of going into the computation (which presents no new features beyond those we have already discussed in the Abelian case) we will directly quote the answer for each diagram. The actual computations can be found in several excellent references. 

There are four such diagrams, one with a fermion loop, two involving a vector field loop and one with a ghost loop. All of them are divergent. But there is also a counterterm. As we saw in previous examples, this term cancels the leading $\frac{1}{\epsilon}$ divergence coming from the one-loop diagrams.
We will now compute the one-loop diagrams, and choose a counterterm of the form (in momentum space):
\begin{equation}
-i \delta_{3} (\eta_{\mu \nu}p^2-p_{\mu}p_{\nu})
\end{equation}
where the value of $\delta_{3}=Z_3-1$ is to be determined. 

The four one-loop diagrams we need to compute are the following: 
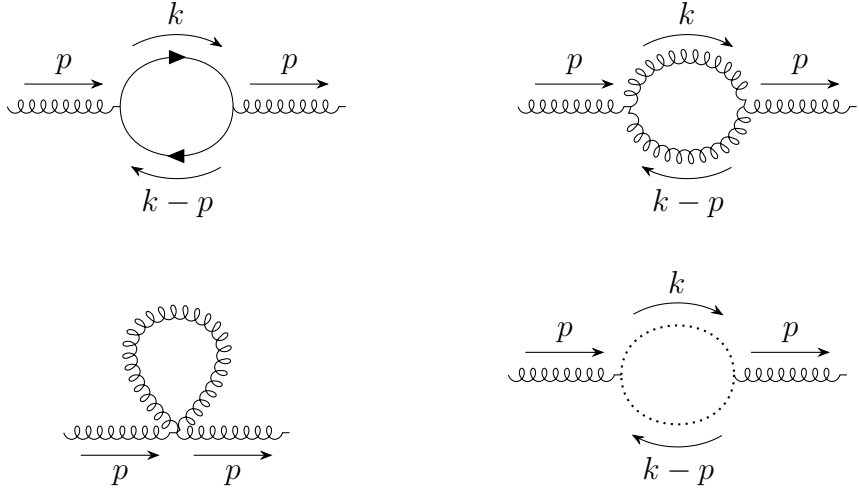
\begin{figure}[H]
\begin{center}
\begin{tikzpicture}
\begin{feynman}
\vertex (a) ;
\vertex [right=of a] (b);
\vertex [right=of b] (c);
\vertex [right=of c] (d);
\diagram* {
(a) -- [gluon, momentum=\(p\)] (b) -- [fermion, half left, momentum={[arrow shorten=0.1mm]\(k\)}] (c);
(c) -- [fermion, half left, momentum={[arrow shorten=0.1mm]\(k-p\)}] (b);
(c) -- [gluon, momentum=\(p\)] (d)
};
\end{feynman}
\end{tikzpicture}
\hspace*{2cm}
\begin{tikzpicture}
\begin{feynman}
\vertex (a) ;
\vertex [right=of a] (b);
\vertex [right=of b] (c);
\vertex [right=of c] (d);
\diagram* {
(a) -- [gluon, momentum=\(p\)] (b) -- [gluon, half left, momentum={[arrow shorten=0.1mm]\(k\)}] (c);
(c) -- [gluon, half left, momentum={[arrow shorten=0.1mm]\(k-p\)}] (b);
(c) -- [gluon, momentum=\(p\)] (d)
};
\end{feynman}
\end{tikzpicture}\\[4mm]
\begin{tikzpicture}
\begin{feynman}
\vertex (a) ;
\vertex [right=of a] (b);
\vertex [right=of b] (c);
\diagram* {
(a) -- [gluon, momentum'=\(p\)] (b); 
(b) -- [gluon, out=135, in=45, loop, min distance=3cm] (b)
(b) -- [gluon, momentum'=\(p\)] (c)
};
\end{feynman}
\end{tikzpicture}
\hspace*{2cm}
\begin{tikzpicture}
\begin{feynman}
\vertex (a) ;
\vertex [right=of a] (b);
\vertex [right=of b] (c);
\vertex [right=of c] (d);
\diagram* {
(a) -- [gluon, momentum=\(p\)] (b) -- [ghost, half left, momentum={[arrow shorten=0.1mm]\(k\)}] (c);
(c) -- [ghost, half left, momentum={[arrow shorten=0.1mm]\(k-p\)}] (b);
(c) -- [gluon, momentum=\(p\)] (d)
};
\end{feynman}
\end{tikzpicture}
\end{center}
\caption{The four one-loop contributions to gluon self-energy}
\end{figure}

Before embarking on the calculation, let us acknowledge an important fact about non-Abelian gauge theories: all fields come in multiplets. There are $N^2-1$ distinct vector fields of each polarisation in SU(N) gauge theory, and we typically couple them to fermions in the fundamental representation so there are $N$ fermions. The number of ghosts  and anti-ghosts is again equal to $N^2-1$. Thus, in addition to computing momentum integrals, we have to use certain identities following from representation theory that arise due to the presence of $f^{abc}$ and $t^a_{ij}$ in the Lagrangian. For example in QCD, the contribution from $N=3$ colours, associated to the index $i=1,2,3$ on the quark field $\psi_i$, is found using the identities we are about to work out.

Beyond that, there is the issue of number of species (flavours) $n_F$. For example in the Standard Model there will be $n_F=6$ quark flavours. Because flavours are identical copies with the same coupling to gauge fields, the loop acquires a simple multiplicative factor of $n_F$. 

Let us write out the identities we will need in terms of Hermitian generators $T^a_{IJ}$ of an arbitrary irreducible representation of SU(N). Later we will specialise to the fundamental and adjoint representations. We have:
\be
\begin{split}
\tr T^a T^b &=T_R\, \delta^{ab}\\
\sum_a T^a T^a &=C_R{\bold 1}
\end{split}
\label{Tident}
\ee
where $T_R$ and $C_R$ are real numbers associated to the representation, the latter being called the {\em quadratic Casimir} of the representation.

It is important to be clear about notation here. The first relation involves taking a matrix product of $T^a$ with $T^b$ and then taking the trace of the result, while the second involves the same matrix product but now with the upper indices equated and summed, but no trace. One should also be aware that changing the normalisation of the $T^a$ will scale both relations above. Typically the first relation is used to fix the normalisation, and then the second relation gives us information within that normalisation.

For the fundamental representation of SU(N) we take $T_F=\half$. This is familiar for SU(2), where the representation matrices are normally chosen to be $\half\sigma^a$ with $\sigma^a$ being the Pauli matrices. In this normalisation, we find $C_F=\frac{N^2-1}{2N}$. This can again be verified for SU(2) where one can easily show that:
\be
\sum_{a=1}^3 \left(\shalf\sigma^a\right)\left(\shalf\sigma^a\right)=\sfrac34\, {\bold 1}
\ee 

We also need identities for the structure constants $f^{abc}$, but these are in fact a special case of the above when we choose the adjoint representation. Viewed as matrices in the last two indices, the $f^{abc}$ are real and anti-symmetric. Thus to get a Hermitian representation we must write $T^a=i\left(f^a\right)_{bc}$. Now the second relation in \eref{Tident} gives:
\be
\sum_{a,c} i\left(f^a\right)_{bc}i\left(f^a\right)_{cd}=C_A\,\delta_{bd}
\ee
Using total antisymmetry to rearrange indices, the above equation becomes:
\be
\sum_{c,d}f^{acd}f^{bcd}=C_A\delta^{ab}
\label{casim}
\ee
The structure constants are already normalised by requiring the Lie algebra relations to hold in standard form. In this normalisation, for SU(N) one finds that  $T_A=C_A=N$. Notice that by rearranging indices again, \eref{casim} can also be interpreted as the first relation in \eref{Tident}. 

Armed with these results, let us consider the first diagram with a fermion loop. 
Its contribution is given by:
\begin{equation}
i\mathcal{M}^{ab\,\mu\nu}_{F}=-n_F\tr t^a t^b ~(ig_{R})^2 \int \frac{d^4k}{(2\pi)^4}\frac{i}{(p-k)^2-m_{R}^2}\frac{i}{k^2-m_{R}^2} tr\Big[\gamma^{\mu}(\slashed{k}-\slashed{p}+m)\gamma^{\nu}(\slashed{k}+m)\Big]
\end{equation}
Apart from the pre-factors, this was already computed for QED. In dimensional regularisation and for $p^2 \gg m_{R}^2$, the above quantity reduces to:
\begin{equation}
\mathcal{M}_F^{ab\,\mu\nu}=-\delta^{ab}n_{F}T_R^f\Big( \frac{g_{R}^2}{16\pi^2} \Big)\big( \eta^{\mu \nu}p^2-p^{\mu}p^{\nu} \big)\times \Big[ \frac{8}{3\epsilon}+\frac{4}{3}\ln\Big(\frac{\tilde{\mu}^2}{-p^2}\Big) \Big]
\label{gpropfermiloop}
\end{equation}
Here $n_{f}$ is the number of quark flavors and $T_R^f$ is the normalisation of the generators in the representations in which the fermions transform\footnote{The superscript ``$f$'' means ``fermion''. Also to avoid cluttering the notation, we assume there is a single fermion representation, but clearly if there are many then we will have a different $n_f$ for each one.}. For the most common case of fermions in the fundamental representation we have $T_F^f=\frac{1}{2}$, but we prefer to retain an arbitrary $T_R^f$ in the above formula so that it holds for any fermion representation. As before $\tilde{\mu}$ is the mass scale introduced during the renormalization procedure. Notice that the coefficient of the finite logarithm is half of the coefficient of the $\frac{1}{\epsilon}$ term. This will be relevant in what follows.

The gluon loop contribution from two cubic vertices is given by:
\begin{equation}
i\mathcal{M}_{3}^{ab\,\mu \nu}=\frac{g_{R}^2}{2}\int \frac{d^4k}{(2\pi)^4}\frac{-i}{k^2}\frac{-i}{(k-p)^2}f^{ace}f^{bdf}\delta^{cf}\delta^{ed}N^{\mu \nu},
\end{equation}
where the factor $\frac{1}{2}$ is the symmetry factor of the diagram, and \begin{equation}
\begin{split}
N^{\mu \nu}=\Big[ \eta^{\mu \alpha} & (p+k)^{\rho}+\eta^{\alpha \rho}(p-2k)^{\mu}+\eta^{\rho \mu}(k-2p)^{\alpha} \Big]\eta_{\alpha \beta}\eta_{\rho \sigma} \\& \times \Big[\eta^{\nu \beta}(p+k)^{\sigma}-\eta^{\beta \sigma}(2k-p)^{\nu}-\eta^{\sigma \nu}(2p-k)^{\beta} \Big].
\end{split}
\end{equation}
Notice that $f^{ace}f^{bdf}\delta^{cf}\delta^{ed}=f^{acd}f^{bcd}=C_A\delta^{ab}$. A similar factor of the Casimir will arise in the following diagrams.

The gluon loop contribution from the quartic vertex is given by, \begin{equation}
\begin{split}
i\mathcal{M}_{4}^{ab\,\mu \nu}=&-\frac{ig_{R}^2}{2}\int \frac{d^4k}{(2\pi)^4}
\frac{-i\eta^{\rho \sigma}\delta^{cd}}{k^2} \Bigg[f^{abe}f^{cde}\Big( \eta^{\mu\rho}\eta^{\nu\sigma}-\eta^{\mu\sigma}\eta^{\nu\rho}\Big)\\ & \qquad 
+f^{ace}f^{bde}\Big( \eta^{\mu\nu}\eta^{\rho\sigma}-\eta^{\mu\sigma}\eta^{\nu\rho}\Big)+f^{ade}f^{bce}\Big( \eta^{\mu\nu}\eta^{\rho\sigma}-\eta^{\mu\rho}\eta^{\nu\sigma}\Big) \Bigg]
\end{split}
\end{equation}
The momentum integral is quadratically divergent and it vanishes in dimensional regularization. 

Finally the ghost loop contribution is given by:
\begin{equation}
i\mathcal{M}_{ghost}^{ab\,\mu \nu}=(-1)(-g_{R})^2\int \frac{d^4k}{(2\pi)^4}\frac{i}{(k-p)^2}\frac{i}{k^2}\,f^{cad}f^{dbc}k^{\mu}(k-p)^{\nu},
\end{equation}
where the $(-1)$ arises from the anticommuting property of the ghosts. 

The above expressions were all written in $d=4$. To use dimensional regularisation we need to rewrite them in $d$ dimensions (among other things, this requires us to introduce the factor $\mu^{4-d}$ which is familiar from many previous examples). Combining the graphs involving gluon and ghost loops, we get:
\begin{equation}
\begin{split}
\mathcal{M}_{\rm gluon-ghost}^{ab\,\mu\nu}&= g_{R}^2\ \delta^{ab}\,C_A \,\frac{\mu^{4-d}}{(4\pi)^{\frac{d}{2}}}\int_{0}^{1}dx~
\Delta^{-\epsilon}\,\Gamma\left(\frac{\epsilon}{2}\right)\\
&\qquad\times \Bigg[ \eta^{\mu \nu}p^2\Big[(-2x^2+3x-1)d+x(4x-5)+\sfrac72\Big]\\
&\qquad\quad + p^{\mu}p^{\nu}\Big[\frac{d}{2}(1-2x)^2-4x^2+4x-3\Big]\Bigg].
\end{split}
\end{equation}
where $\Delta=x(x-1)p^2$.

If we naively worked in 4 dimensions, and carried out the $x$-integrals explicitly, it would seem from the above that we have  very different coefficients for the $p^2\eta^{\mu\nu}$ and $p^{\mu\nu}$ terms. But this is not true! At $d=4-\epsilon$, there is a power of $\Delta$ in the above expression. We must evaluate the $x$ integrals away from $d=4$ and then take the limit $\epsilon\to 0$ keeping both the divergent and finite parts. The result is:
\begin{equation}
\mathcal{M}_{gluon-ghost}^{ab\,\mu\nu}=\frac{g_{R}^2}{16\pi^2}\ \delta^{ab}C_A(g^{\mu \nu}p^2-p^{\mu}p^{\nu})\Bigg[ \frac{10}{3\epsilon}+\frac{5}{3}\ln\bigg(\frac{\tilde{\mu}^2}{-p^2}\bigg)+\frac{31}{9}\Bigg].
\end{equation}
It is pleasing to see that the the relative coefficient of  $p^2\eta^{\mu\nu}$ and $p^{\mu\nu}$ is precisely $-1$, so that the transverse form of the propagator is retained.

Finally we combine the above result with the fermion loop contribution in \eref{gpropfermiloop}, to get:
\begin{equation}
\mathcal{M}_{total}^{ab\,\mu\nu}=
\frac{g_{R}^2}{16\pi^2}\ \delta^{ab}\bigg(g^{\mu \nu}p^2-p^{\mu}p^{\nu}\bigg)\Bigg[C_A\Big(\frac{10}{3\epsilon}+\frac{5}{3}\ln\frac{\tilde{\mu}^2}{-p^2}\Big)-n_{f}T_{R}^f\Big( \frac{8}{3\epsilon}+\frac{4}{3}\ln\frac{\tilde{\mu}^2}{-p^2}\bigg) \Bigg].
\end{equation}

In the MS scheme, we only cancel the $\frac{1}{\epsilon}$ divergence. This fixes the counterterm such that: \begin{equation}
\delta_{3}=\frac{g_{R}^2}{16\pi^2\epsilon}\left(\frac{10}{3}C_A-\frac{8}{3}n_{f}T_{R}^f \right)
\end{equation}

So far, our calculation was valid for any gauge group and fermion representation. If now we specialise to SU(N) with fermions in the fundamental representation, we get:
\be
\delta_{3}({\rm SU(N)})=\frac{g_{R}^2}{16\pi^2\epsilon}\left(\frac{10}{3}N-\frac{4}{3}n_f\right)
\ee

Having calculated $\delta_3$ we now write down, without derivation, the remaining independent counterterms $\delta_{i}=(1-Z_{i})$ at 1-loop in the $\xi=1$ gauge, where $i=2,3,3c,g,m$:
\be
\begin{split}
\delta_{2} &=\frac{g_{R}^2}{16\pi^2\epsilon}\left(-2C_{F}^f \right)\\
\delta_{3c} &=\frac{g_{R}^2}{16\pi^2\epsilon}C_A\\
\delta_g &= \frac{g_{R}^2}{16\pi^2\epsilon}\left(-\frac{11}{3}C_A+\frac43 n_fT_R^f\right)\\
\delta_{m} &=\frac{g_{R}^2}{16\pi^2\epsilon}\left(-6C_{F}^f \right)
\end{split}
\label{deltavalues}
\ee
Here $C_{R}^f$ is the quadratic Casimir for the fermions. From the above, using \eref{dependentZ} one can easily work out the dependent $\delta_i$'s, namely those with $i=1,1c,A^3,A^4$. It turns out that almost all the $\delta_i$ are gauge-dependent. Calculations for arbitrary $\xi$ give rise to additional terms in the $\delta_i$ that are proportional to $(1-\xi)$. This gives us another reason why $\xi=1$ was such a convenient choice. However, one can verify that $\delta_g$ is independent of $\xi$. This will have important consequences.

Before concluding the section we note that $Z_g$ is the same for all quarks at 1-loop, independent of their flavour. Thus, for quarks there is a universal renormalized colour charge $g_{R}$. This is the analogue of the $Z_{1}=Z_{2}$ relation of QED.

Also notice that the same gauge charge $g_{R}$ appears in front the quark-gluon interaction as well as the gluon self-interactions. All factors of $g_{R}$ will be renormalised in the same way only if the following relations hold: \begin{equation}
\frac{Z_1}{Z_2}=\frac{Z_{A^3}}{Z_3}=\sqrt{\frac{Z_{A^4}}{Z_3}}=\frac{Z_{1c}}{Z_{3c}}.
\label{samecharge}
\end{equation}
At one loop, we observe that (for the gauge choice $\xi=1$): 
\begin{equation}
\delta_{1}-\delta_{2}=\delta_{1c}-\delta_{3c}=\delta_{A^3}-\delta_{3}=\frac{1}{2}\big(\delta_{A^4}-\delta_{3} \big)=-\frac{g_{R}^2}{8\pi^2\epsilon}C_{A}
\end{equation}
 which obeys \eref{samecharge} to order $g_{R}^2$.

\subsection{The $\beta$-function of Yang-Mills theory}

Given the renormalisation constant $\delta_g$ for Yang-Mills theory, we can use exactly the same procedure that was employed before to deduce the $\beta$-function and the consequent renormalisation group equations which tell us how $g_R$ and other physical quantities effectively depend on the scale.

Thus, again we demand that the bare gauge coupling is independent of the arbitrary scale $\mu$ introduced in dimensional regularisation. We have:
\begin{equation}
g_0=\mu^{\frac{\epsilon}{2}}Z_g\,g_R
\end{equation} 
where recall that $\epsilon=4-d$, and the power of $\mu$ has been introduced to ensure that the bare coupling has no mass dimension. It follows that:
\begin{equation}
\mu \frac{d g_0}{d\mu}=\mu \frac{d}{d\mu}\left(\mu^{\frac{\epsilon}{2}} Z_g\,g_{R}\right)=0
\end{equation}
At one-loop order, inserting $Z_g=1+\delta_g$ we find:
\begin{equation}
\beta(g_{R})=\mu \frac{d g_R}{d\mu}=\Bigg(-\frac{\epsilon}{2}-\mu \frac{d\delta_g}{d\mu}\Bigg)g_R
+\cdots
\end{equation}
Now $\delta_g$ depends on $\mu$ implicitly through $g_{R}$, and from \eref{deltavalues} we find, to one loop order:
\be
\mu \frac{d g_R}{d\mu}
=-\frac{\epsilon}{2}g_R-\frac{ g_R}{8\pi^2\epsilon}\left(-\frac{11}{3}C_A+\frac43 n_f T_R^f\right)\mu\frac{dg_R}{d\mu}
\ee
Solving, we get:
\be
\beta(g_R)=-\frac{\epsilon}{2}g_{R}+\frac{g_{R}^{3}}{16\pi^2} \ \bigg(-\frac{11}{3}C_{A}+\frac{4}{3}n_{f}T_{R}^f \bigg)
\label{YMbeta}
\ee
Recall that $\delta_g$ was (at least to one-loop order) independent of the gauge-fixing parameter $\xi$. It follows that the $\beta$-function we have just calculated is gauge-invariant. Also at this point we may drop the term of order $\epsilon$ since we can go back to 4 dimensions.

Let us now reflect on the physical significance of this extremely fundamental result. From our earlier discussions, we understand that a negative sign for the $\beta$-function tells us that the coupling constant decreases as we flow to higher energies. This is called asymptotic freedom. The $\beta$-function we have found can have either a negative or a positive sign, depending on the values of $C_A,n_f,T_R^f$. To compare with nature, we specialise to SU(N) gauge theory with fermions in the fundamental representation. Then, inserting $C_{A}=N$ and $T_{R}^f=\frac{1}{2}$ and taking the $\epsilon \rightarrow 0$ limit we have:
\begin{equation}
\mu \frac{d}{d\mu}g_{R}= \frac{g_{R}^{3}}{16\pi^2}\bigg(-\frac{11}{3}N+\frac{2}{3}n_{f} \bigg)
\end{equation}
We see that, given a value of the number of colours $N$, asymptotic freedom is preserved if the number of fermions obeys an upper bound:
\be
n_f <\frac{11}{2}N
\ee
In the Standard Model the number of colours is 3, so we should have no more than 16 fermion flavours. In fact the Standard Model has 6 flavours, so we are well within the asymptotically free regime\footnote{One should incorporate scalars, but the Standard Model does not have any strongly interacting scalars so the present result still holds.}.

So far we only considered gauge fields coupled to fermions. It turns out that a complex scalar gives a contribution that is $\frac14$ times that of a Dirac fermion. This can be verified by explicit computation. Thus as a practical recipe, we may write the $\beta$ function for a theory that includes scalars as:
\be
\beta(g_R)=\frac{g_{R}^{3}}{16\pi^2} \ \bigg(-\frac{11}{3}C_{A}+\frac{4}{3}n_{f}T_{R}^f + \frac13 n_s T_R^s \bigg)
\label{YMbetascalars}
\ee
where $n_s$ counts the number of complex scalars. To simplify future formulae we define the one-loop constant:
\be
b_0\equiv \frac{11}{3}C_{A}-\frac{4}{3}n_{f}T_{R}^f - \frac13 n_s T_R^s
\ee
and the analogue of the fine-structure constant for strong interactions by:
\be
\alpha_s\equiv \frac{g_R^2}{4\pi}
\ee
Then, we can write:
\be
\mu\frac{d}{d\mu}\alpha_s=-\frac{b_0\,\alpha_s^2}{2\pi}
\ee
Let us mention here that the contribution from Majorana fermions and real scalars is half that of Dirac fermions and complex scalars respectively. Thus one can simply set $n_f=\half$ or $n_s=\half$ to cover these cases.

The $\beta$-function equation in \eref{YMbetascalars} is easily integrated to give the $\mu$-dependence of the renormalised coupling constant: 
\begin{equation}
\alpha_s(\mu)=\frac{\alpha_s(\mu_{0})}{1+\frac{b_0\,\alpha_s(\mu_0)}{2\pi}
\ln\frac{\mu}{\mu_{0}}}.
\label{alpharunning}
\end{equation}
This determines the value of $\alpha_s$ at any scale $\mu$, given its value at another scale $\mu_0$. Again we see explicitly that when there is asymptotic freedom, the denominator increases with increasing $\mu$, as a result of which $\alpha_s$ decreases. Observe that the decrease of $\alpha_s$ is logarithmic and therefore {\em extremely slow}. In a  physical system of this type we will observe an apparently scale-independent coupling constant over small energy ranges, but as we probe larger ranges we will detect a slow variation. This phenomenon, called ``logarithmic scaling violation'', was experimentally observed. The fact that theory and experiment were found to agree about the presence of such violations was an early confirmation of the gauge theory description of strong interactions.

We can obtain additional insight by asking what is the scale at which the asymptotically free interaction becomes ``strong''. This can be estimated by formally setting the denominator on the right hand side of \eref{alpharunning} to zero so that $\alpha_s$ diverges\footnote{As before, one must be careful about the physical interpretation of this procedure. We cannot allow $\alpha_s$ to diverge in a perturbative formula, since perturbation theory breaks down for values of $\alpha_s\sim {\cal O}(1)$, long before it can diverge. What we are instead doing is to {\em define} an energy in this way, and simply substitute it back into the original formula. The scale so obtained will never be measured as the energy where $\alpha_s$ diverges, but rather by fitting the resulting $\Lambda$-dependent formula to experiment.} and denoting the corresponding mass scale by $\Lambda$. Thus we have:
\be
1+\frac{b_0 \alpha_s(\mu_0)}{2\pi}\ln\frac{\Lambda}{\mu_0}=0
\ee
Using this, one finds that \eref{alpharunning} can be rewritten as: 
\be
\alpha_s(\mu)=\frac{2\pi}{b_0\ln\frac{\mu}{\Lambda}}
\label{alpharunlambda}
\ee
This is a remarkable equation, because all mention of {\em both} $\mu_0$ and $\alpha(\mu_0)$ have disappeared! The running of the coupling is determined entirely by a constant $b_0$ that we have calculated, and a scale $\Lambda$ that cannot be calculated even in principle, but must be determined from experiment. There is no dimensionless coupling constant left in the theory! This feature is called {\em dimensional transmutation}.

As anticipated in the footnote, \eref{alpharunlambda} cannot be taken seriously all the way down to $\mu=\Lambda$, which would violate the smallness of $\alpha_s$ that justifies perturbation theory. Rather we consistently keep $\mu\gg \Lambda$ and deduce $\Lambda$ by measuring $\alpha_s$ at one such value of $\mu$. Another illuminating relation is obtained by solving \eref{alpharunlambda} for $\Lambda$:
\be
\Lambda = \mu\, e^{-\frac{2\pi}{b_0\alpha_s(\mu)}}
\label{Lambdaform}
\ee
Let us examine the rate of variation of the coupling constant and the value of $\Lambda$ for Quantum Chromodynamics, which is a Yang-Mills theory with an SU(3) colour group. In this case $C_A=N=3$ and $n_f=6,T_R^f=\half$ (if we want to use the formula below the top quark mass then we should take $n_f=5$). Then we find $b_0=7$. 

Now, it is known that a measurement at $\mu=91.2$ GeV gives us a value $\alpha_s=0.1184$. So we can ask how much $\alpha_s$ varies if, for example, we double the energy scale, increasing $\mu$ from 91.2 GeV to 182.4 GeV. It follows from \eref{alpharunning} that by the time we reach this higher energy, $\alpha_s$ has decreased by around 10 percent from 0.1184 to 0.1084. Similarly we may use \eref{Lambdaform} to find that
$\Lambda\sim 50$ MeV. However this value does change significantly if we incorporate higher-loop corrections to the $\beta$-function. The currently accepted value, based on knowing the $\beta$-function up to four loops, is closer to $\Lambda_{\rm QCD}\sim 220$ MeV. As we remarked above, the formula \eref{alpharunlambda} is to be used only for scales substantially higher than $\Lambda_{\rm QCD}$, and indeed it is generally agreed that we should keep $\mu\gsim 5$ GeV, the mass of the bottom quark.

\newpage

\section{Renormalisation of gauge theories in the Higgs phase}
\label{smoneloop}

In this final chapter we discuss how the renormalisation procedure described above works in gauge theories in the Higgs phase, with a focus on the special features in this case. These theories are sometimes known as ``spontaneously broken'' gauge theories. The electroweak theory, which is a part of the Standard Model provides, a prime motivation to address this issue. The key insight in this subject was that, although gauge invariance seems to be broken, it is not actually broken in Higgs theories. It is merely realised in a different way with resulting massive particles (in the Standard Model these are the $W,Z$ and Higgs bosons). However, the simplifications and identities that we highlighted in the previous chapters,which arise from gauge invariance, are still present. This is ultimately the reason why gauge theories in the Higgs phase are renormalisable.

\subsection{Abelian gauge theory and gauge-fixing}

To set notation and also highlight an important conceptual point, we will start with Abelian gauge theory coupled to a scalar field. The Lagrangian is:
\be
{\cal L}=-\frac14 F_{\mu\nu}F^{\mu\nu} +D_\mu\phi^* D^\mu\phi - V(\phi^*,\phi)
\label{sedlag}
\ee
where:
\be
V(\phi^*,\phi)=-m^2\phi^*\phi + \frac{\lambda}{2} (\phi^*\phi)^2
\ee
and:
\be
D_\mu\phi = \del_\mu\phi +ieA_\mu\phi
\ee
As is familiar, the Lagrangian above is invariant under the gauge transformations:
\be
A_\mu \to A_\mu -\frac{1}{e}\del_\mu\alpha(x),\qquad \phi\to e^{i\alpha(x)}\phi
\label{sedgauge}
\ee

The minimum of the potential is at all values satisfying:
\be
|\phi|=
 \sqrt{\frac{m^2}{\lambda}}\equiv \frac{v}{\sqrt2}
\label{modphivev}
\ee
The vacuum expectation value of the field $\phi$ should lie at a minimum. Now the above equation specifies a circle of distinct values in field space, since the phase of $\phi$ is left arbitrary. Usually one makes the choice:
\be
\langle\phi\rangle =\frac{v}{\sqrt2}
\label{vevchoice}
\ee
which means the real part of the complex field gets a vev. This vacuum expectation value is not gauge invariant, but by a gauge transformation (the finite version of \eref{sedgauge}) we can take it to any other expectation value satisfying \eref{modphivev}.

It is convenient to define the real components of the complex field:
\be
\phi=\frac{1}{\sqrt2}(\chi_1+i\chi_2)
\ee
The gauge transformation on the real and imaginary components is:
\be 
\delta\chi_1=-\alpha\chi_2, \quad \delta\chi_2=\alpha\chi_1
\ee
In view of the choice we made in \eref{vevchoice}, the vacuum expectation values of these fields are $\langle\chi_1\rangle=v$ and $\langle \chi_2\rangle =0$. We now perform the shift $\chi_1(x)=v+\eta(x)$. 
Then the Lagrangian becomes:
\be
{\cal L}=-\frac14 F_{\mu\nu}F^{\mu\nu} + \half (\del_\mu \eta-eA_\mu\chi_2)^2 +\half \big(\del_\mu\chi_2+eA_\mu(v+\eta)\big)^2 -V\bigg(\phi=\frac{1}{\sqrt2}(v+\eta+i\chi_2)\bigg)
\label{shiftlag}
\ee
By expanding the potential $V$, it is easily verified that the field $\eta$ has also acquired a mass term, namely $-m^2\eta^2$. This is of the correct sign (unlike the original mass term in $V$ before shifting) and tells us that the mass of the $\eta$ is $\sqrt2\,m$. By expanding the third term above, we see that the vector field also acquires a mass term:
\be
\frac{(ev)^2}{2}A_\mu A^\mu
\ee
which shows that it has acquired a mass $m_A= ev=\sqrt{\frac{2}{\lambda}}\,em$.

Now let us focus only on quadratic terms in the above Lagrangian. This will suffice to discuss gauge-fixing and propagators. One should of course work out all the terms, including interaction terms, to understand the theory completely. The quadratic terms are:
\be
{\cal L}_{\rm quad}=-\frac14 F_{\mu\nu}F^{\mu\nu} + \half (\del_\mu\eta)^2 -m^2\eta^2 +\half \big(\del_\mu\chi_2+evA_\mu\big)^2 
\ee
The only problem with this Lagrangian as it stands is that it has a linear term in $A_\mu$:
\be
e\,v A_\mu\del^\mu\chi_2
\label{linearterm}
\ee
as one can see by expanding the last term. We can get rid of this by performing the transformation:
\be
A_\mu\to A_\mu-\frac{1}{ev}\del_\mu\chi_2
\ee
Although this shift resembles a gauge transformation, we should not think of it that way, but rather as a redefinition of the field $A_\mu$. It is not accompanied by a variation of any other field. However, it has the same form as a gauge transformation where the parameter $\alpha(x)$ has been replaced by the field $\frac{1}{v}\chi_2(x)$. Since all gauge transformations leave the gauge kinetic term invariant, this field redefinition too leaves it invariant. But it has the effect of removing the term linear in $A_\mu$. And remarkably it does one more thing: it removes the kinetic term of $\chi_2$ as well. As a result, $\chi_2$ is no longer a physical field. It has been traded for the longitudinal component of the massive gauge field $A_\mu$.

The above considerations were classical. But we want to study this theory at the quantum level. For this purpose, instead of making the above field redefinition, we add the gauge-fixing term:
\be
-\frac{1}{2\xi}\left(\del_\mu A^\mu-\,ev\,\xi\,\chi_2\right)^2
\label{Higgsgaugefix}
\ee
This is a variant of the familiar $R_\xi$ gauge which includes an extra term proportional to the scalar field $\chi_2$. The cross-term in the above expression is:
\be
 ev\,\del_\mu A^\mu\,\chi_2
\ee
Combining this with \eref{linearterm}, we see that the result is a total derivative -- which will drop out from the action. Clearly this was possible because of the precise choice of coefficient of $\chi_2$ in the gauge-fixing term. Now that we have cancelled the unwanted linear term, we can proceed to extract the spectrum of the theory. Squaring the first term in \eref{Higgsgaugefix} gives us the usual gauge-fixing term that we had in unbroken gauge theory (compare \eref{gaugefix}), allowing us to define a propagator for the gauge field much as we did in the case without spontaneous symmetry breaking. Note, however, that the kinetic term for $\chi_2$ has {\em not} disappeared. Moreover, the square of the second term inside the bracket in \eref{Higgsgaugefix} gives us a ($\xi$-dependent) mass term for the $\chi_2$-field. Thus the part of the Lagrangian \eref{shiftlag} quadratic in $\chi_2$ is:
\be
\half(\del_\mu\chi_2) -\half \xi\,(ev)^2\chi_2^2
\ee
We see that the mass of $\chi_2$ is dependent on the gauge-fixing parameter: $m_{\chi_2}=\sqrt{\xi} ev=\sqrt\xi\,m_A$. 
Given that final results should be gauge-invariant, we expect this $\xi$-dependence to drop out when we calculate physical quantities. Indeed, we know that $\chi_2$ is actually decoupled from the theory, since we could have made its kinetic term disappear by a field redefinition of $A_\mu$. However, allowing it to remain in the Lagrangian has certain advantages.

Now that we have fixed the gauge, we need to compute the Faddeev-Popov determinant. Just as we did in the unbroken case, we first calculate the variation of the gauge-fixing term with respect to a gauge variation. Recalling that:
\be
\delta A_\mu = -\frac{1}{e}\del_\mu\alpha,\delta \chi_2=\alpha(v+\eta)
\ee
we have:
\be
\frac{\del}{\del \alpha(x)}\big(\del_\mu A^\mu - ev\,\xi\chi_2\big)=-\frac{1}{e}\big(\square +  \xi e^2v(v+\eta)\big)
\ee
Taking the determinant and exponentiating using a pair of anti-commuting ghosts $c$ and $\cbar$, we can read off the ghost Lagrangian:
\be
\del_\mu\cbar\, \del^\mu c -\xi m_A^2\,\left(1+\frac{\eta}{v}\right)\cbar\, c
\ee
As in the unbroken case, here too we have dropped an overall $\frac{1}{e}$ factor in order to have a canonically normalised kinetic term for the ghosts that is independent of the gauge coupling. 

There are two key features of this result: (i) the ghosts have a mass $\sqrt\xi\,m_A$, exactly the same mass as the $\chi_2$ field, (ii) the ghosts couple to the Higgs field $\eta(x)$ through a cubic coupling $\eta \,{\bar c}c$. Point (i) offers a useful hint: since the role of ghosts is to subtract degrees of freedom due to their opposite statistics, it is reasonable to expect that they will play the role of decoupling the $\chi_2$ field. This indeed turns out to be the case. Point (ii) tells us that in the Higgs phase, ghosts cannot be ignored even in the Abelian theory. Though they do not couple to the gauge field, they do couple to the Higgs field. 

In the Higgs phase, the gauge field is massive so it already has a well-defined propagator. This is modified in the presence of the gauge-fixing term: 
\be
\langle 0|{\tilde A}_\mu(k){\tilde A}_\nu(0)|0\rangle =-i\frac{\eta_{\mu\nu}-(1-\xi)\frac{k_\mu k_\nu}{k^2-\xi m_A^2}}{k^2-m_A^2+i\epsilon}
\label{covpropHphase}
\ee
In the limit $m_A\to 0$ it reduces to the gauge-fixed massless propagator in \eref{covprop}, while in the limit of $\xi\to\infty$ (where the gauge fixing term disappears) with $m_A$ fixed, we find the propagator:
\be
\langle 0|{\tilde A}_\mu(k){\tilde A}_\nu(0)|0\rangle =-i\frac{\eta_{\mu\nu}-\frac{k_\mu k_\nu}{m_A^2}}{k^2+i\epsilon}
\ee
which is appropriate for a gauge field with an explicit mass term.

With these preliminaries out of the way, we can turn to the analogous discussion in the non-Abelian context.

\subsection{Non-abelian gauge theory: a simplified model}

In this section we will examine some basic features of the Higgs mechanism in a non-Abelian gauge theory, and summarise how renormalisation is carried out in such theories. The model we will consider has SU(2) gauge fields coupled to a complex scalar field in the doublet representation. This can be thought of as a sector of the Standard Model, while the full Standard Model contains an additional U(1) gauge field corresponding to hypercharge symmetry.

The Lagrangian of our simplified model is given, as in \eref{YMscalars}, by:
\be
{\cal L}=-\frac14 F_{\mu\nu}^a F^{\mu\nu\,a}+ (D_\mu\phi)^{\dagger}_i(D^\mu\phi)_i-V(\phi_i,\phi^\dagger_i)
\ee
with:
\be
(D_\mu\phi)_i=\del_\mu\phi_i-igA_\mu^a T^a_{ij}\phi_j
\ee
Since our gauge group is SU(2), we choose $T^a_{ij}=\frac{\sigma^a_{ij}}{2}$ where $\sigma^a_{ij}$ are the Pauli matrices. This defines the doublet representation of SU(2).

Now we choose the potential to be of the form that spontaneously breaks the global SU(2) symmetry:
\be
V(\phi_i,\phi^\dagger_i)=-m^2 \phi^\dagger_i\phi_i+\frac{\lambda}{2}(\phi^\dagger_i\phi_i)^2
\ee
This potential has minima at:
\be
|\phi|\equiv\sqrt{\phi^\dagger_i\phi_i}= \sqrt{\frac{m^2}{\lambda}}\equiv \frac{v}{\sqrt2}
\ee

We may now choose the vacuum expectation value to be:
\be
\begin{pmatrix}
\langle \phi_1\rangle \\ \langle \phi_2\rangle 
\end{pmatrix}
=
\frac{1}{\sqrt2}\begin{pmatrix}
 0 \\ v
\end{pmatrix}
\label{vevchoice.nonab}
\ee
We must expand the field around this vev. A standard parametrisation is given in terms of the four real fields $\chi_1,\chi_2,\chi_3,\eta$ defined by:
\be
\begin{split}
\phi_1 &=\frac{1}{\sqrt2}(-i\chi^1-\chi^2)\\
\phi_2 &=\frac{1}{\sqrt2}(v+\eta+i\chi^3)
\end{split}
\label{su2.expand}
\ee
Inserting this parametrisation, one finds  after a small calculation that the Lagrangian has the form:
\be
{\cal L}=-\frac14 F_{\mu\nu}^a F^{\mu\nu\,a}+\half \left(\del_\mu \chi^a+\shalf gv A_\mu^a\right)^2+ \cdots
\ee
where $\chi^a, a=1,2,3$ are the three real scalar fields defined in \eref{su2.expand} that were not given a vev, and we have not written out the cubic and higher terms. As before, we see that there is a term linear in $A_\mu^a$:
\be
\half gv A_\mu^a\del^\mu \chi^a
\label{nonablin}
\ee
As in the Abelian case, at the classical level it is natural to perform a field redefinition on the gauge field:
\be
\delta A_\mu^a = \frac{1}{g}D_\mu\alpha^a
\ee
where $\alpha^a=-\frac{2}{gv}\chi^a$. However there are some important differences from the Abelian case: this redefinition has the form of a gauge transformation on $A_\mu^a$ only for infinitesimal $\alpha^a$, and moreover due to the covariant derivative it contains not just linear terms but also terms quadratic in fields. Thus we can only conclude that if $\chi^a$ are infinitesimal, this field redefinition removes their kinetic terms, at the cost of introducing complicated nonlinear terms. This information, however, is enough to conclude that sufficiently close to the point where $\chi^a=0$, i.e. for infinitesimal $\chi^a$, their kinetic terms can be removed -- along with the unwanted cross terms involving $A_\mu^a$. The three $\chi^a$ are those components of the original doublet scalar field that, in the absence of a gauge field, would have been Goldstone bosons. Due to the presence of gauge fields, these bosons instead decouple from the theory and their degrees of freedom become the longitudinal polarisations of the massive gauge fields. A better treatment of this problem can be carried out by making a different parametrisation of the gauge field instead of \eref{su2.expand}, and making use of the geometry of the SU(2) gauge group and its orbits, to show that even for finite values of the corresponding fields there is a field redefinition of $A_\mu^a$ that renders them non-dynamical. Once the field redefinition has been carried out, we see right away that, as expected, all three components $A_\mu^a$ become massive with a common mass $m_A=\frac{gv}{2}$. This is as expected, in view the fact that the vev in \eref{vevchoice} is not preserved by any element of SU(2) and thus ``breaks'' the symmetry completely. 

Again, in the quantum theory we follow a different procedure to remove linear terms in the gauge field, namely we define:
\be
G^a\equiv\del^\mu\! A_\mu^a- \shalf\xi gv \chi^a 
\ee
and add to the Lagrangian the gauge-fixing term:
\be
-\frac{1}{2\xi}\left(G^a\right)^2
\ee
The cross term from this is:
\be
\half gv\, \del^\mu \! A_\mu^a\,\chi^a
\ee
and we readily see that, when added to \eref{nonablin}, it forms a total derivative. As in the Abelian case, this leaves us with the massive gauge field of mass $m_A$ as well as extra fields $\chi^a$ of mass $\sqrt{\xi}m_A$. Thus the $R_\xi$ gauge choice has retained spurious fields that will eventually be cancelled out by the ghosts.

It remains to write down the Faddeev-Popov determinant and from it the ghost action, and conclude by writing the propagators of the various fields in the theory. Following the procedure expounded above, we need the variation of the fields $\chi^a$ under gauge transformations. These are easily found to be:
\be
\delta \chi^a=-\half\Big(\alpha^a(v+\eta) + \epsilon^{abc}\alpha^b\chi^c\Big)
\ee
We can now compute the gauge variation of $G^a$ to get:
\be
\delta G^a=\frac{1}{g}\del^\mu D_\mu\alpha^a+\sfrac14 \xi gv\Big(\alpha^a(v+\eta) + \epsilon^{abc}\alpha^b\chi^c\Big)
\ee
After an overall scaling, the ghost action then comes out to be:
\be
\del_\mu {\bar c}^a D^\mu c^a- \sfrac14 \xi (gv)^2 \,\cbar^a c^a -\sfrac14 g^2v\,\cbar^a c^a\eta
-\sfrac14 \xi g^2v\,\epsilon^{abc} {\bar c}^a c^b\chi^c
\ee

Thus to calculate physical quantities in Higgs-phase gauge theories, one has to include the ghost vertices that arise from the above terms. At the end one has a unitary, renormalisable and gauge-invariant theory.

\section{Bibliography and Further Reading}

As mentioned in the Introduction, two very useful sources to understand the renormalisation of QFT in the spirit presented here are:

\begin{itemize}
\item
M.E. Peskin and D.V. Schroeder, ``An Introduction to Quantum Field Theory''\cite{Peskin:1995ev}.

\item
M.D. Schwartz, ``Quantum Field Theory and the Standard Model''\cite{Schwartz:2013pla}.
\end{itemize}
The above textbooks contain several details beyond those provided in the present notes. Also, I have tried as far as possible to follow the notation of \cite{Peskin:1995ev}.

Many more aspects of this fascinating subject can be uncovered in the following books, reviews and papers. An influential early review article on the renormalisation of gauge theories was written by Abers and Lee in 1973 \cite{Abers:1973qs}. This review uses the path integral formalism. It also contains references to earlier works on perturbative renormalisation in Quantum Field Theory. Significant conceptual and technical material about perturbative renormalisation and its application to gauge theories can be found in the textbooks by Jean Zinn-Justin \cite{zinnjustin:2002ru} and by George Sterman \cite{sterman:1994ce}. The lecture notes on ``Secret Symmetry'' by Sidney Coleman in \cite{Coleman:1985rnk} provide helpful insights into the renormalisation of gauge theories in the Higgs phase.  A pedagogical article by David Gross on the Renormalisation Group in the context of QCD can be found in \cite{Balian:1976vq}. An entire textbook on renormalisation has been written by John Collins \cite{Collins:1984xc}.

Beyond all this, the student wishing to go deeper into the renormalisation of non-Abelian gauge theories should definitely read some of the original classic works by 'tHooft \cite{tHooft:1971akt} and by 'tHooft and Veltman \cite{tHooft:1973mfk}. Valuable insights into the history of the subject can be found in \cite{tHooft:1995wad}. 

Finally, a new chapter in the understanding of renormalisation was opened by Ken Wilson and the student interested in this story should read \cite{Wilson:1970ag, Wilson:1971bg, Wilson:1971dh}, as also Polchinski's seminal contribution \cite{Polchinski:1983gv}.

\bibliography{renorm}{}
\bibliographystyle{unsrt}

\end{document}